\pdfoutput=1
\documentclass[11pt,a4paper]{article}

\usepackage[utf8]{inputenc}
\usepackage[T1]{fontenc}
\usepackage{lmodern}
\usepackage{microtype}
\usepackage[margin=1.85cm]{geometry}
\usepackage{amsmath,amssymb,amsfonts}
\usepackage{booktabs}
\usepackage{array}
\usepackage{graphicx}
\graphicspath{{./}{../outputs/figures/}}
\usepackage{float}
\providecommand{\FloatBarrier}{}
\usepackage[table]{xcolor}
\usepackage{colortbl}
\usepackage[round,authoryear]{natbib}
\usepackage[colorlinks=true,linkcolor=blue!60!black,citecolor=blue!60!black,urlcolor=blue!60!black]{hyperref}
\usepackage{caption}
\definecolor{tabhead}{HTML}{D6E4F0}
\definecolor{rowodd}{HTML}{FFFFFF}
\definecolor{roweven}{HTML}{F3F6FA}
\definecolor{colodd}{HTML}{F0F5FA}
\definecolor{coleven}{HTML}{E3ECF5}
\definecolor{rulecol}{HTML}{B8C8D8}

\newcommand{\co}{\cellcolor{colodd}}
\newcommand{\ce}{\cellcolor{coleven}}

\renewcommand{\arraystretch}{1.24}
\arrayrulecolor{rulecol}
\title{\textbf{Does Increasing Government Spending on Tertiary Education Provide a Return on Investment? A Bayesian Causal Inference Analysis}}

\author{
  \textbf{Andrew C. Parnell}\\[0.3em]
  \small School of Mathematics and Statistics, University College Dublin\\
  \small \texttt{andrew.parnell1@ucd.ie}
}

\date{8 November 2026}

\begin{document}

\maketitle

\begin{abstract}
\noindent When a national government increases public spending on tertiary (university) education, does that investment provide a positive fiscal return on investment once general economic growth is separated out? Cross-country data make this question difficult to answer because faster-growing economies naturally collect more tax revenue, international indicators divide per-student spending by gross domestic product (GDP) per person and student headcount, and university cohorts take several years to graduate and enter higher tax brackets. We assemble a harmonised five-year panel covering 121 countries from 1985 to 2024 and fit continuous-treatment Multivariate Bayesian Causal Forests that model four interrelated tax changes simultaneously whilst adjusting for pre-treatment country conditions, predicted funding levels, and follow-up macroeconomic growth. Within the central policy range around each income group's benchmark, a $+10$ percentage-point (pp) increase in per-student university spending as a share of GDP per person shifts the tax mix toward direct income, profit, and capital gains taxation by $+0.36$ pp after 5 years ($P = 0.93$, where $P$ denotes the posterior probability of a positive effect) and $+0.71$ pp after 10 years ($P = 0.96$). Overall tax revenue as a share of GDP takes a full decade to respond as cohorts graduate and enter formal careers, with 10-year total non-resource tax revenue rising by $+0.24$ pp of GDP ($P = 0.89$, or $+0.31$ pp with $P = 0.94$ across four Markov chains, retaining the conservative single-chain baseline throughout) and direct income tax revenue rising by $+0.12$ pp of GDP ($P = 0.90$, or $+0.15$ pp with $P = 0.93$ across four chains) per $+10$ pp of per-student spending, by $+0.37$ pp ($P > 0.99$) following five-year within-country funding increases, and by $+0.22$ pp in total tax ($P = 0.95$) and $+0.15$ pp in direct income tax ($P = 0.98$) per $+25\%$ increase in real purchasing power parity dollars per student. Translating these 10-year causal slopes into a single-cohort comparison between the multi-year degree cost of educating one entering cohort and that cohort's cumulative tax payments across a 40-year working career reveals a positive expected return on investment across all income groups. In high-income economies, a $+10$ pp per-student funding boost generates a typical three- to five-fold expected career return in total non-resource tax revenue, though with wide long-horizon credible intervals reflecting macroeconomic noise in aggregate tax ratios (such as an illustrative \texteuro2.4 billion cohort input in Ireland returning roughly \texteuro9.7 billion in total tax, 95\% credible interval \texteuro$-10$ billion to \texteuro$+32$ billion, with posterior probability of net profit $0.77$, and \texteuro2.9 billion in direct income tax alone, or \texteuro11.6 billion in the United Kingdom returning roughly \texteuro58 billion, 95\% credible interval \texteuro$-60$ billion to \texteuro$+190$ billion) and a four- to twelve-fold return in middle- and low-income economies, whereas expanding the overall university budget without protecting spending per student fails to generate a fiscal return.

\vspace{0.5em}
\end{abstract}

\section{Introduction}
\label{sec:intro}

Public debates over higher education finance often centre on a simple, practical question. Does government spending on tertiary (university) education provide a positive return on investment to the public purse, or is it a net fiscal cost? At the individual level, decades of labour economics research show that university graduates earn higher wages, experience lower unemployment, and work in larger, more formalised firms than non-graduates \citep{carneiro2011estimating, zimmerman2014returns, psacharopoulos2018returns, deming2022four}. Empirical studies in the United States also show that public investments in college access and per-student instructional resources can recoup a substantial share of their upfront cost over a graduate's working career through higher income tax payments \citep{trostel2010fiscal, chakrabarti2020state, hendren2020unified}. As university-educated cohorts enter the labour market and raise firm productivity \citep{moretti2004estimating, valero2019economic}, those private wage gains and corporate spillovers should show up at the national level as stronger personal income, payroll, consumption, and corporate profit tax receipts that pay back the initial university outlay.

Yet working out whether national government spending on tertiary education actually improves a country's tax take is surprisingly tricky. First, if a country's tax revenue rises in the decade after an increase in university funding, we must determine whether universities specifically strengthened the formal tax base or whether the whole economy simply grew richer. If an analyst constructs a national tax-per-person variable by multiplying the tax-to-GDP ratio by GDP per person, any positive effect of education on GDP automatically inflates the tax outcome by mathematical identity. Isolating a genuine fiscal return therefore requires modelling tax shares and tax-to-GDP ratios whilst controlling for starting economic conditions and follow-up economic growth \citep{imai2010general, acharya2016explaining, vanderweele2016mediation}. Second, the standard international indicator compiled by the United Nations Educational, Scientific and Cultural Organisation (UNESCO) Institute for Statistics\footnote{UNESCO Institute for Statistics (UIS) education finance database: \url{https://data.uis.unesco.org/}.} and the World Bank\footnote{\raggedright World Bank World Development Indicators (WDI) portal: \url{https://datatopics.worldbank.org/world-development-indicators/}.} measures public expenditure per university student as a percentage of GDP per person. Because this ratio places both student headcount and GDP per person in the denominator, a low-income country with a tiny elite university sector (such as Malawi or Ethiopia, enrolling under 2\% of young adults) can appear to spend over 150\% of GDP per person per student, whereas a high-income country enrolling 60\% of young adults (such as Denmark or South Korea) appears to spend 27\% despite spending nearly three times as many real purchasing power parity (PPP) dollars per student (where PPP adjusts US dollars for differences in local price levels).

A third obstacle is that university investments are continuous policy choices that take time to mature and face diminishing marginal returns. Treating government university spending as a simple binary on/off switch would discard variation in funding intensity and imply that a tiny spending nudge across an arbitrary cutoff produces the same tax return as a major fiscal expansion. At the same time, governments cannot keep increasing university spending indefinitely and expect linear tax gains forever. In our causal framework, we capture non-linearity and diminishing returns in three ways. First, we evaluate local spending changes within a central $\pm 15$ percentage-point (pp) policy window around each income group's typical funding level and adjust for each country's predicted spending level $\hat{z}(X_{i,t}) = \mathbb{E}[Z_{i,t} \mid X_{i,t}]$ given its starting covariates $X_{i,t}$. Second, our regression trees allow the causal slope per $+10$ pp of funding to vary with enrolment and national income, showing that aggregate tax-to-GDP gains taper off as university systems mature, and third, we verify our results using concave logarithmic funding specifications where each additional dollar of spending yields a smaller marginal return.

\begin{figure}[htbp]
  \centering
  \includegraphics[width=\textwidth]{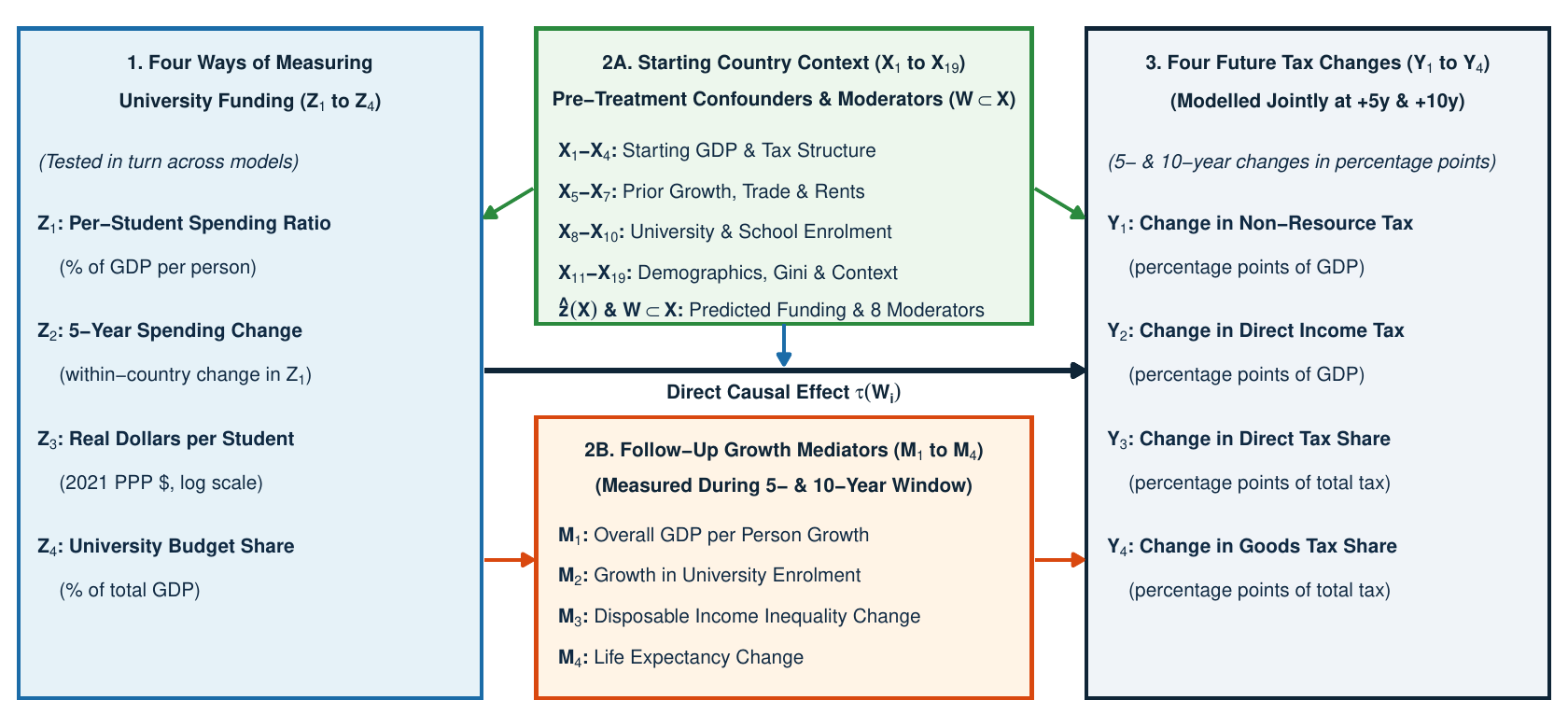}
  \caption{Overview of the study design across 121 countries from 1985 to 2024. Box 1 shows the four continuous university funding measures ($Z_1$ to $Z_4$, tested in turn in separate models), Box 2A shows the 19 pre-treatment country covariates ($X_1$ to $X_{19}$), predicted funding $\hat{z}(X)$, and the eight treatment-effect moderators ($W \subset X$) that modify the direct causal slope $\tau(W_i)$, Box 2B shows the four follow-up macroeconomic growth mediators ($M_1$ to $M_4$), and Box 3 shows the four percentage-point tax revenue changes ($Y_1$ to $Y_4$, modelled jointly in every run) over $+5$ and $+10$ years.}
  \label{fig:schematic}
\end{figure}
\FloatBarrier

Figure~\ref{fig:schematic} summarises our study design for addressing these challenges across 121 countries from 1985 to 2024. On the outcome side (Box 3 of Figure~\ref{fig:schematic}), we jointly model four percentage-point tax changes ($Y_1, Y_2, Y_3, Y_4$) in a single multivariate system at both $+5$ and $+10$ years. Specifically, $Y_1$ is the change in total non-resource tax revenue as a share of GDP (measuring the full Exchequer return across all tax streams, including payroll, consumption, property, and income taxes), $Y_2$ is the change in direct income, profit, and capital gains tax revenue as a share of GDP (measuring the conservative direct income-tax floor), $Y_3$ is the change in direct income and profit taxes as a share of total tax revenue, and $Y_4$ is the change in indirect goods and services (sales and value-added) taxes as a share of total tax revenue. On the treatment side (Box 1 of Figure~\ref{fig:schematic}), rather than relying on a single spending definition, we estimate separate models for four continuous university funding variables ($Z_1, Z_2, Z_3, Z_4$) in turn. Here $Z_1$ is our primary per-student spending ratio (per $+10$ pp of GDP per person per student within each World Bank income group), $Z_2$ is the five-year within-country change in per-student funding (per $+10$ pp increase), $Z_3$ is real 2021 PPP dollars per student on a logarithmic scale (per $+25\%$ relative increase, removing GDP from the denominator), and $Z_4$ is the public university budget as a percentage of GDP (per $+0.25$ pp of GDP, removing student headcount from the denominator). In every model, we control for 19 pre-treatment country characteristics $X_1$ to $X_{19}$ (Box 2A of Figure~\ref{fig:schematic}, of which $W \subset X$ moderate the treatment slope $\tau(W_i)$) and four follow-up macroeconomic growth mediators $M_1$ to $M_4$ (Box 2B); throughout the paper, pp denotes percentage points, CI denotes a Bayesian posterior credible interval, and $P = \Pr(\tau > 0 \mid \text{data})$ denotes the posterior probability that the causal slope $\tau$ is positive given the data.

We believe this is the first global study to use modern causal inference techniques to test whether public university funding provides a national fiscal return on investment. Existing microeconomic studies estimate fiscal returns by tracking individual college cohorts in a single country such as the United States \citep{trostel2010fiscal, chakrabarti2020state, hendren2020unified}, which provides detailed lifecycle tax estimates but cannot capture economy-wide corporate spillovers or cross-country differences in tax enforcement. Conversely, sub-national and cross-country macroeconomic studies examine how universities affect GDP per person \citep{valero2019economic} or how employment structure affects tax collection \citep{besley2014pillars, jensen2022employment} using linear panel regressions that assume constant slopes and analyse one outcome at a time. Unlike single-country lifecycle accounting or linear fixed-effects regressions, our continuous-treatment Multivariate Bayesian Causal Forest approach models all four interrelated tax changes ($Y_1$ to $Y_4$) simultaneously across 121 countries while allowing non-linear confounding and heterogeneous returns across development stages. By unbundling the four funding measures ($Z_1$ to $Z_4$), separating out follow-up macroeconomic mediators ($M$), and translating the resulting 10-year causal slopes into single-cohort degree inputs versus 40-year career tax outputs, our approach gives finance and education ministries concrete return-on-investment estimates in their own currency.

The remainder of the paper is organised as follows. Section~\ref{sec:lit} reviews the literature on tax capacity, university resources, and Bayesian regression tree models for continuous treatments. Section~\ref{sec:data} describes the 40-year panel dataset, our four tax outcomes ($Y_1$ to $Y_4$), the UNESCO denominator trap across our four funding measures ($Z_1$ to $Z_4$), and the country-blocked holdout design. Section~\ref{sec:methods} explains the continuous-treatment Multivariate Bayesian Causal Forest (MV-BCF) model, why we separate baseline predictors $X$ and growth mediators $M$ from treatment moderators $W$, and how we account for expected funding levels, with full mathematical derivations in Appendix~\ref{app:math}. Section~\ref{sec:results} presents the empirical results across time horizons, funding measures, country subgroups, cohort return-on-investment calculations for policy makers, and out-of-sample holdout validation, with additional benchmarks, linear-versus-logarithmic comparisons, and a 20-country policy-maker lookup table in Appendices~\ref{app:benchmarks} to \ref{app:policymakers}. Finally, Section~\ref{sec:discussion} discusses multi-cohort fiscal payback timelines, limitations, and directions for future research.

\section{Literature Review}
\label{sec:lit}

In this section we review three strands of research that have largely developed in parallel: the political economy of tax structure modernisation, the economics of per-student university resources, and Bayesian tree ensembles for causal inference. Bringing these three literatures together clarifies why university resources should affect direct income taxation more strongly than consumption taxation. It also explains why flexible nonparametric models are needed to separate continuous funding slopes from general economic growth. Readers interested in comprehensive surveys of each field may refer to \citet{besley2014pillars}, \citet{deming2022four}, and \citet{hill2020bayesian}.

\subsection{Tax Capacity, Formal Employment, and Tax Structure}

A central question in public finance is why high-income countries collect 30\% to 45\% of GDP in tax revenue, whereas low- and middle-income countries often collect less than 15\% of GDP \citep{besley2009origins, gordon2009tax, besley2014pillars}. High-income economies raise most of their revenue through direct personal income taxes, social security contributions, and corporate profit taxes, while developing economies rely heavily on indirect consumption taxes and customs duties. Early cross-country studies emphasised political institutions, property rights, and administrative state capacity \citep{besley2009origins, kaufmann2011worldwide}, alongside international accounting standards that separate domestic tax effort from volatile oil and mineral royalties \citep{prichard2016reassessing, ictd2023grdv}.\footnote{See the UNU-WIDER Government Revenue Dataset (GRD) documentation at \url{https://doi.org/10.35188/UNU-WIDER/GRD-2023}.} Whilst these institutional studies establish that fiscal capacity evolves alongside economic development, they leave open the specific channels through which public investments in skilled workers expand the collectible tax base.

More recent empirical work has focussed on the microeconomic mechanics of tax enforcement, paper trails, and formal employment \citep{kleven2016why, keen2017optimal, pomeranz2019taking}. \citet{kleven2016why} show that modern governments can enforce high direct tax compliance at low audit cost because large, formal firms act as third-party tax collectors. Complex businesses require internal accounting records shared across managers and skilled staff, which makes collusive tax evasion between employers and employees difficult to sustain. Building on this mechanism, \citet{jensen2022employment} analyses roughly 100 household surveys across development levels and shows that as the share of salaried employees rises relative to self-employed workers, governments lower income tax exemption thresholds and expand direct income tax collection. Field studies confirm that electronic tax filing and third-party information trails reduce evasion when paired with capable firms and literate taxpayers \citep{pomeranz2019taking, okunogbe2022technology}, whilst broader direct taxation can also strengthen citizen engagement with the state \citep{weigel2020participation}. By contrast, in developing economies with large informal sectors, indirect consumption taxes such as value-added tax (VAT) remain the main revenue source because formal retail chains only gradually replace informal vendors as incomes rise \citep{bachas2024informality}.

This public finance literature suggests a direct connection between university funding and the national tax base that goes beyond headline GDP growth. Universities are themselves large formal employers whose academic and administrative staff pay income tax through automatic payroll withholding. Over the medium term, universities also train the professional, managerial, scientific, and accounting graduates required by formal corporate employers \citep{kleven2016why, valero2019economic}. As those graduates enter the workforce, they shift employment into salaried jobs above national income tax exemption thresholds \citep{jensen2022employment}. When this mechanism operates at the national level, higher public spending per university student should expand direct income, profit, and capital gains taxation ($Y_2$ and $Y_3$) whilst reducing relative reliance on indirect goods and services taxes ($Y_4$).

\subsection{Returns to University Resources and Fiscal Externalities}

A large microeconometric literature documents strong private wage returns to university education across both advanced and developing economies \citep{carneiro2011estimating, zimmerman2014returns, psacharopoulos2018returns, chetty2020income, mountjoy2022community}. Reviewing this evidence, \citet{deming2022four} notes that human capital accounts for at least one-third of within-country wage variation and at least half of cross-country income differences. Cognitive and problem-solving skills play a central role in long-run growth \citep{hanushek2012better, angrist2021measuring}, even after accounting for skilled emigration from smaller developing countries \citep{beine2008brain}. Beyond individual earnings, university graduates generate productivity spillovers for surrounding workers and firms. \citet{moretti2004estimating} finds that a higher share of college graduates in a city raises the wages of both graduates and non-graduates, and \citet{valero2019economic} examine roughly 15,000 universities across 1,500 regions in 78 countries and find that university expansions raise regional GDP per person.

Importantly for our study, higher education research demonstrates that graduation rates and subsequent earnings depend heavily on per-student instructional resources rather than enrolment headcounts alone. \citet{bound2010have} and \citet{deming2017impact} show that when public colleges expand student numbers whilst state funding per student stagnates or declines (a phenomenon known as institutional crowding), degree completion rates fall and time to graduation lengthens. Conversely, \citet{chakrabarti2020state} find that increases in state funding for public universities raise bachelor's degree completion and improve students' long-term financial outcomes into their thirties. When graduates complete their degrees and enter formal careers, the fiscal return to the public purse over their working lives can be substantial. \citet{trostel2010fiscal} estimates that lifetime additional income and payroll taxes paid by college graduates exceed direct government subsidies, and \citet{hendren2020unified} show within their Marginal Value of Public Funds framework that direct investments in college access and per-student spending generate positive fiscal externalities through future tax receipts.

\subsection{Bayesian Regression Tree Ensembles for Continuous Treatments}

Separating the fiscal effect of continuous university funding from general economic development involves two statistical challenges. First, we must control for intermediate economic growth without introducing bias, and second, we must account for non-linear confounding across diverse countries. In cross-country panels, standard linear two-way fixed effects regressions can suffer from negative weighting and functional-form misspecification when treatment slopes vary across countries and over time \citep{jorda2005estimation, acemoglu2019democracy, dechaisemartin2020two, callaway2021difference}. Furthermore, as shown in the causal mediation literature \citep{imai2010general, acharya2016explaining, vanderweele2016mediation, ting2025estimating}, adjusting for follow-up macroeconomic changes (such as GDP per person growth, enrolment growth, inequality, and life expectancy) allows us to isolate the direct structural effect of university spending net of overall economic growth. We also verify this decomposition using two-stage sequential g-estimation \citep{acharya2016explaining}.

Bayesian Additive Regression Trees (BART; \citealt{chipman2010bart, hill2011bayesian, hill2020bayesian}) and Bayesian Causal Forests (BCF; \citealt{hahn2020bayesian}) provide a flexible nonparametric alternative to linear panel regressions, sitting alongside frequentist causal forests and double machine learning \citep{chernozhukov2018double, athey2019generalized}. Tree ensembles automatically capture non-linear relationships and multi-way interactions amongst covariates, with extensions supporting high-dimensional predictors \citep{linero2018bayesian}, functional smoothing \citep{starling2020bart}, count outcomes \citep{murray2021loglinear}, fast stochastic tree search \citep{krantsevich2023stochastic}, causal mediation \citep{ting2025estimating}, and low-dimensional posterior summaries \citep{woody2021model}. \citet{hahn2020bayesian} show that standard single-forest BART models can suffer from regularisation-induced confounding when strong baseline predictors of the outcome also predict treatment intensity. BCF resolves this problem by splitting the model into a prognostic forest $\mu(\cdot)$ and a separately regularised treatment-effect forest $\tau(\cdot) \cdot Z_{i,t}$, with recent extensions to multivariate outcomes and longitudinal panels \citep{mcjames2024lbcf, mcjames2024mvbcf}. Because $Z_{i,t}$ enters our structural equation in Section~\ref{sec:methods} (and Appendix~\ref{app:math}) multiplicatively as a scalar multiplier on the four-dimensional leaf vector $\tau_{l,b} \in \mathbb{R}^4$, the MV-BCF algebra extends directly to a continuous treatment variable $Z_{i,t} \in \mathbb{R}$ by replacing the binary propensity score with the expected continuous funding level $\hat{z}(X_{i,t}) = \mathbb{E}[Z_{i,t} \mid X_{i,t}]$ inside $\mu(\cdot)$, mirroring continuous-treatment residualisation in causal forests and double machine learning \citep{robinson1988root, hirano2004propensity, chernozhukov2018double, athey2019generalized}.

Readers will note that the Multivariate Bayesian Causal Forest (MV-BCF) method used in this paper \citep{mcjames2024mvbcf} was co-developed by the author with Nathan McJames, Ann O'Shea, and Yong Chen Goh. Given this connection, we want to make it clear from the outset that this paper is not written to promote one causal estimator over all others. In Section~\ref{sec:holdout}, we benchmark the out-of-sample predictive accuracy of continuous MV-BCF against single-forest BART, random forests, historical training means, ridge regression, zero-change persistence, and linear local projections. In Appendix~\ref{app:benchmarks}, we also compare its causal slope estimates directly against separate univariate BCFs, single-forest BART, linear local projections, two-way fixed effects, cross-fitted double machine learning, and frequentist orthogonal causal forests \citep{athey2019generalized}. Throughout the paper, we discuss the limitations of the BCF approach alongside its strengths.

\section{Data}
\label{sec:data}

\subsection{Global Five-Year Panel Structure and Country-Blocked Holdout Split}

Annual macroeconomic series for education spending and tax revenue contain frequent reporting gaps and short-term business-cycle fluctuations. Following standard practice in empirical macro-panel studies \citep{acemoglu2019democracy}, we average annual observations from six international databases into eight non-overlapping five-year waves spanning 40 years: \texttt{1985--1989}, \texttt{1990--1994}, \texttt{1995--1999}, \texttt{2000--2004}, \texttt{2005--2009}, \texttt{2010--2014}, \texttt{2015--2019}, and \texttt{2020--2024}. These sources are the World Bank WDI, the UNESCO Institute for Statistics, the International Monetary Fund (IMF) Fiscal Monitor,\footnote{IMF Fiscal Monitor database: \url{https://www.imf.org/en/Publications/FM}.} the Worldwide Governance Indicators (WGI), the Standardised World Income Inequality Database (SWIID version 9.6; \citealt{solt2020swiid}),\footnote{SWIID version 9.6 database and documentation: \url{https://fsolt.org/swiid/}.} and the Organisation for Economic Co-operation and Development (OECD) \emph{Education at a Glance} financing taxonomy.\footnote{OECD \emph{Education at a Glance} portal: \url{https://www.oecd.org/en/publications/education-at-a-glance.html}.} Constructing five-year waves matches the typical duration of an undergraduate degree programme and reduces serial correlation from annual budgetary carryover.

Although the full World Bank register lists 217 sovereign territories, many small island states and conflict-affected territories do not report tertiary education expenditure or disaggregated tax revenues to UNESCO and the World Bank. In addition, raw international fiscal and education archives occasionally contain reporting anomalies where currency-unit mismatches, decimal-place errors, or offshore sovereign resource funds distort headline ratios. Screening out impossible tax-to-GDP values above 60\% or below 0.5\% of GDP and near-zero tertiary reporting glitches below 0.5\% of GDP per person removes five raw reporting anomalies in the pipeline: Sudan in \texttt{1995--1999} (recorded at 629.2\% of GDP in tax due to a currency-unit mismatch), Timor-Leste in \texttt{2010--2014} (recorded at 110.0\% of non-oil GDP due to offshore petroleum fund accounting), the United Arab Emirates in \texttt{2010--2014} and \texttt{2015--2019} (recorded below 0.4\% of GDP in central federal tax), and Madagascar in \texttt{2015--2019} (recorded at 0.036\% of GDP per person per student due to a decimal-place glitch). Across the full cleaned panel before requiring forward tax changes, 808 five-year country-waves have observed university spending or tax structure data (summarised descriptively in Table~\ref{tab:eda1}). Requiring strictly observed per-student university funding ($Z_1$ or $Z_3$) and all four forward tax changes (with zero imputation on treatments or outcomes) yields 371 country-waves across 121 countries at the $+5$-year horizon ($t \to t+1$, waves 1 to 7) and 283 country-waves across 112 countries at the $+10$-year horizon ($t \to t+2$, waves 1 to 6), while requiring two consecutive waves of per-student funding for $Z_2$ yields 176 waves across 86 countries at $+10$ years and requiring aggregate tertiary budget data without student headcounts for $Z_4$ yields 403 waves across 123 countries at $+10$ years (with $Z_4$ also verified on the overlapping $N = 272$ sample in Check \texttt{RC17} of Appendix Table~\ref{tab:robust}).

To evaluate how well our continuous-treatment models generalise to unseen countries, we partition the countries using a deterministic 5-fold country-blocked split stratified by each country's modal World Bank income group (sorting countries alphabetically by ISO-3 code within each modal income group and assigning folds $1, \dots, 5$ sequentially). Every wave of a given country is assigned to the same fold so that no country ever straddles the training and test sets. Folds 1 to 4 (80\% of countries: 98 countries and 301 country-waves at $+5$ years; 90 countries and 229 country-waves at $+10$ years) form the training sample used to compute income-group funding medians and covariate centring statistics without test-set leakage, whereas Fold 5 (20\% of countries: 23 unseen countries and 70 country-waves at $+5$ years; 22 unseen countries and 54 country-waves at $+10$ years) serves as a locked out-of-sample holdout test set evaluated in Section~\ref{sec:holdout}. Occasional missing values in auxiliary covariates inside $X_{i,t}$ and $W_{i,t}$ (such as baseline gross tertiary enrolment in 39 of the 283 ten-year waves where UNESCO records per-student expenditure ratios without publishing the separate enrolment headcount) are imputed using strictly out-of-fold training medians so tree splits can be evaluated on all 283 waves, and we confirm in Check \texttt{RC4} of Appendix Table~\ref{tab:robust} that dropping all incomplete covariate rows ($N = 238$) leaves our results unchanged. In Sections~\ref{sec:results}.1 to \ref{sec:results}.3, we report posterior treatment slopes from the full 5-fold country-blocked cross-fitted panel ($N = 371$ at $+5$ years and $N = 283$ at $+10$ years, where $\hat{z}(X_{i,t})$ is predicted strictly out-of-fold across the five country folds), whilst in Section~\ref{sec:holdout} and Check \texttt{RC13} we fit both the random forest for $\hat{z}(X_{i,t})$ (using 4-fold internal cross-fitting across Folds 1 to 4) and the MV-BCF model exclusively on the 80\% training countries to generate strictly out-of-sample predictions on the 22 locked holdout countries in Fold 5.

\subsection{Tax Revenue Outcomes (\texorpdfstring{$Y_1$ to $Y_4$}{Y1 to Y4})}

To examine how university funding affects both overall tax collection and the mix of taxes without introducing mechanical GDP product identities, we distinguish baseline tax levels $L_{1,i,t}, \dots, L_{4,i,t}$ at wave $t$ from our four percentage-point forward change outcomes $Y_{k,i,t+h} = L_{k,i,t+h} - L_{k,i,t}$ ($k \in \{1, 2, 3, 4\}$) over $+5$ years ($h=1$) and $+10$ years ($h=2$). Our first outcome, $Y_{1,i,t+h} = L_{1,i,t+h} - L_{1,i,t}$, is the change in total non-resource tax revenue as a percentage of GDP ($\Delta \text{Tax/GDP}$, in pp of GDP), based on the World Bank WDI tax-to-GDP series (\texttt{GC.TAX.TOTL.GD.ZS}). Following the fiscal-capacity literature \citep{prichard2016reassessing, ictd2023grdv}, in major oil- and mineral-exporting country-years where natural resource rents (\texttt{NY.GDP.TOTL.RT.ZS}) exceed 10\% of GDP, we subtract half of the excess resource rents at the annual level before five-year wave averaging ($L_{1,i,\text{yr}} = \text{Tax/GDP}_{i,\text{yr}} - 0.5 \max(0, \text{Rents/GDP}_{i,\text{yr}} - 10)$) as a harmonised proxy so that volatile commodity royalties do not masquerade as domestic tax effort. Whilst the UNU-WIDER Government Revenue Dataset provides consolidated general-government non-resource tax series for a subset of waves, we verify in Appendix~\ref{app:benchmarks} (Check \texttt{RC1}) that using raw, unadjusted WDI tax-to-GDP yields nearly identical slopes.

Our second outcome, $Y_{2,i,t+h} = L_{2,i,t+h} - L_{2,i,t}$, is the change in direct income, profit, and capital gains tax revenue as a percentage of GDP ($\Delta \text{IncomeTax/GDP}$, in pp of GDP), which captures revenue collected directly from individual wages, salaries, and corporate profits. Our third outcome, $Y_{3,i,t+h} = L_{3,i,t+h} - L_{3,i,t}$, is the change in direct income, profit, and capital gains taxes as a percentage of total tax revenue (\texttt{GC.TAX.YPKG.ZS}, $\Delta \text{IncomeTax/Rev}$, in pp of revenue), which measures structural modernisation toward progressive direct taxation \citep{besley2014pillars, jensen2022employment}. Our fourth outcome, $Y_{4,i,t+h} = L_{4,i,t+h} - L_{4,i,t}$, is the change in indirect goods and services taxes (such as VAT, sales taxes, and excise duties, \texttt{GC.TAX.GSRV.RV.ZS}) as a percentage of total tax revenue ($\Delta \text{GoodsTax/Rev}$, in pp of revenue; \citealt{bachas2024informality}). Readers should note that in levels at any wave $t$, direct income tax as a share of GDP satisfies $L_{2,i,t} = (L_{1,i,t} \times L_{3,i,t}) / 100$, so for a marginal funding change the first-difference slopes satisfy $\tau_{Y_2} \approx (\bar{L}_1 \tau_{Y_3} + \bar{L}_3 \tau_{Y_1}) / 100$. Thus, $Y_1$, $Y_3$, and $Y_4$ capture the two underlying policy dimensions (whether overall tax collection rises and whether the tax mix shifts from consumption taxes toward income taxes), whilst $Y_2$ measures their combined impact on the national budget, and our Multivariate Bayesian Causal Forest models all four change outcomes ($Y_1$ to $Y_4$) simultaneously through a joint $4 \times 4$ covariance matrix $\Sigma$.

\subsection{The UNESCO Denominator Trap and Four University Funding Measures (\texorpdfstring{$Z_1$ to $Z_4$}{Z1 to Z4})}
\label{sec:unesco_trap}

\begin{figure}[htbp]
  \centering
  \includegraphics[width=0.96\textwidth]{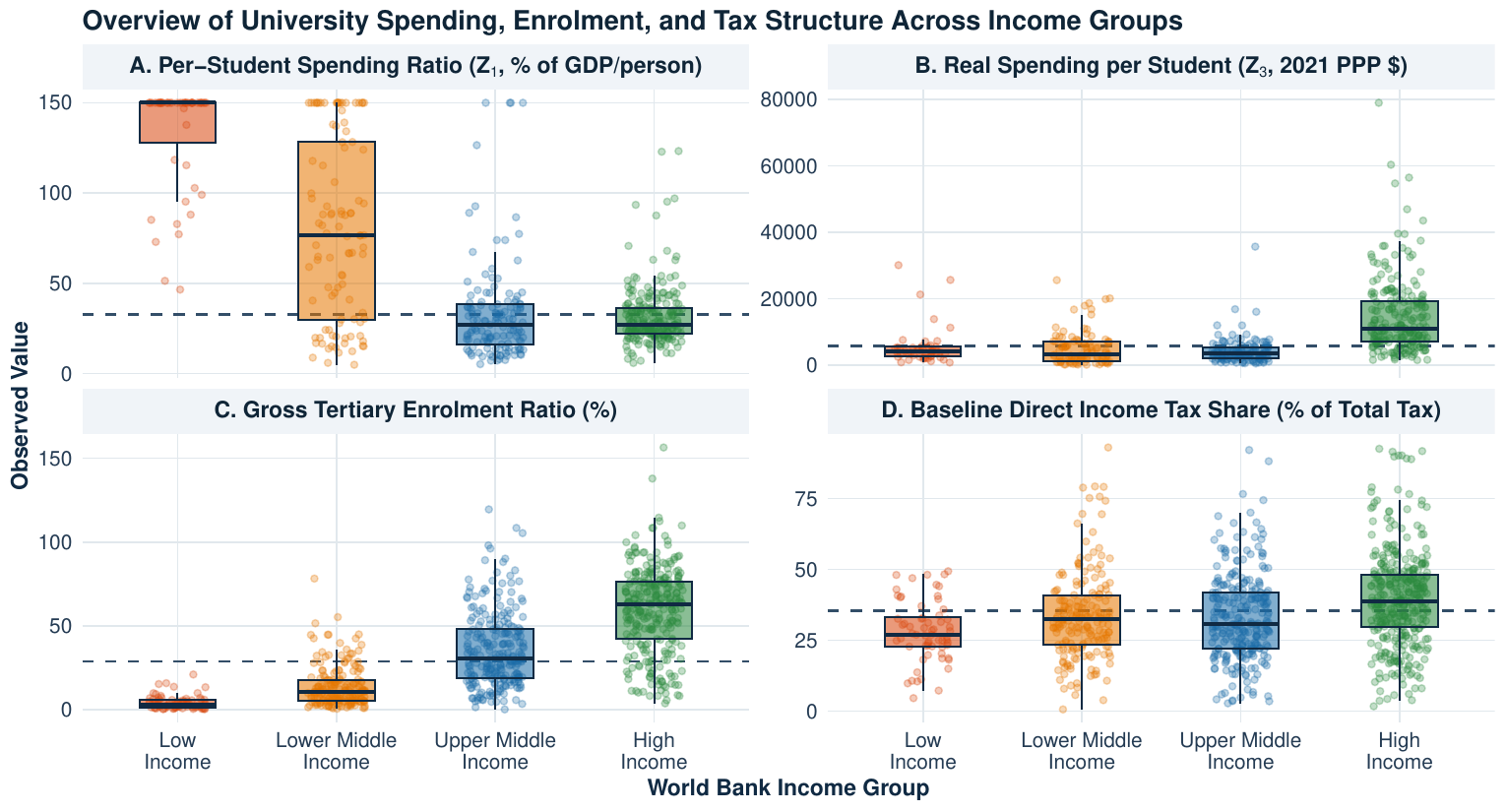}
  \caption{Observed distributions of (A) the UNESCO per-student spending ratio (\% of GDP per person, used for $Z_1$ and winsorised at 150\% for display), (B) real spending per university student in 2021 PPP dollars (used for $Z_3$), (C) gross tertiary enrolment ratio (\%), and (D) baseline direct income and profit tax share (\% of revenue, $Y_3$) across the four World Bank income groups. Horizontal dashed lines mark the global median (or mean in Panel D).}
  \label{fig:data_overview}
\end{figure}

The standard international indicator of per-student university spending compiled by UNESCO and the World Bank divides public spending per university student by GDP per person. As shown in Equation~\eqref{eq:unesco}, this ratio is algebraically identical to dividing the public university budget (\% of GDP) by the number of university students per person:
\begin{equation}
  \text{UNESCO Ratio}_{i,t} = \frac{\text{University Spend per Student}_{i,t}}{\text{GDP per Person}_{i,t}} \times 100 = \frac{\text{University Budget (\% GDP)}_{i,t}}{\text{Students per Person}_{i,t}}.
  \label{eq:unesco}
\end{equation}
Because both GDP per person and student headcount appear in the denominator, countries with very different university systems can record similar headline ratios for completely different reasons. To see the raw data before any statistical modelling, Figure~\ref{fig:data_overview} plots the observed distributions of the UNESCO per-student ratio (used to construct $Z_1$), real 2021 PPP dollars per student (used to construct $Z_3$), gross tertiary enrolment, and the direct income tax share ($Y_3$) across the four World Bank income groups. Table~\ref{tab:eda1} reports the corresponding group summaries.

\begin{table}[htbp]
  \centering
  \caption{University spending indicators, enrolment, and baseline direct income tax share ($L_3$) across World Bank income groups in the full cleaned descriptive panel ($N = 808$ five-year waves across 185 country-by-income-group appearances prior to requiring forward tax changes and student headcounts; individual countries that move across World Bank income thresholds over the 40-year window appear in more than one row). Group medians are reported for skewed spending and enrolment variables and group means for the baseline direct income tax share.}
  \label{tab:eda1}
  \small
  \setlength{\tabcolsep}{4.8pt}
  \begin{tabular}{l|c|c|c|c|c}
    \toprule
    \rowcolor{tabhead}
    \textbf{Income Group} & \textbf{UNESCO} & \textbf{Real PPP \$} & \textbf{Univ. Budget} & \textbf{Tertiary} & \textbf{Direct Income} \\
    \rowcolor{tabhead}
    \textbf{(Countries, Waves)} & \textbf{Ratio (\%)} & \textbf{/ Student} & \textbf{(\% GDP)} & \textbf{Enrol. (\%)} & \textbf{Tax Share (\%)} \\
    \midrule
    \rowcolor{rowodd}
    \textbf{Low income} (21, 80) & \co \textbf{205\%} & \$3,957 & \co 0.47\% & 2.3\% & \co 29.2\% \\
    \rowcolor{roweven}
    \textbf{Lower middle} (41, 172) & \ce 76.5\% & \$3,258 & \ce 0.65\% & 7.4\% & \ce 30.8\% \\
    \rowcolor{rowodd}
    \textbf{Upper middle} (54, 219) & \co 26.8\% & \$3,521 & \co 0.64\% & 25.5\% & \co 33.1\% \\
    \rowcolor{roweven}
    \textbf{High income} (69, 337) & \ce 26.9\% & \textbf{\$11,148}& \ce \textbf{0.94\%} & \textbf{58.3\%} & \ce \textbf{38.4\%} \\
    \bottomrule
  \end{tabular}
\end{table}
\FloatBarrier

Figure~\ref{fig:data_overview} and Table~\ref{tab:eda1} show clearly how the UNESCO denominator trap works in practice. In low-income countries (such as Uganda, Malawi, or Ethiopia), only 2.3\% of young adults attend university. Dividing a small university budget (0.47\% of GDP) by a tiny student headcount and a low GDP per person produces a raw median UNESCO ratio of 205\%, which we cap at 150\% in Figure~\ref{fig:data_overview}A to make the plot easier to read. That tiny elite student body also explains why median real PPP spending per student in low-income countries (\$3,957, concentrated in one or two national universities) is slightly higher than in lower-middle-income countries (\$3,258), where enrolment is over three times higher (7.4\% median) and budgets are stretched across many more students. Once we unbundle the denominators in Table~\ref{tab:eda1}, high-income countries (such as Denmark, Norway, or South Korea) spend nearly three times as much per student in real purchasing power (\$11,148 vs \$3,957) and twice as much of their national income (0.94\% vs 0.47\% of GDP), whilst collecting a larger share of their tax revenue from direct income and profit taxes (38.4\% vs 29.2\%).

Because comparing raw UNESCO ratios across income groups would conflate small elite enrolment with high per-student funding, we construct our primary continuous funding variable $Z_1$ (which we refer to as Scaling A) within each World Bank income group. Specifically, we centre each country-wave's winsorised per-student spending ratio on the training-set median of its own income group ($m_{g(i,t)} = 150.0\%$ for low income after the $150\%$ pre-winsorisation cap vs $183.8\%$ raw, $68.9\%$ for lower-middle income, $27.6\%$ for upper-middle income, and $26.4\%$ for high income) and divide by $10$ percentage points. To ensure our causal slope reflects realistic policy changes rather than linear extrapolation to extreme ratios, we winsorise $Z_1$ at $\pm 1.5$ units ($\pm 15$ pp around the group median) so that $Z_{1,i,t} \in [-1.5, +1.5]$ is centred at zero, where $Z_{1,i,t} = 0$ represents the income-group median benchmark and a unit change ($\Delta Z_1 = +1.0$ or $-1.0$, $\text{SD}(Z_1) = 1.05$) represents a $+10$ pp increase or $-10$ pp decrease in per-student funding (we also compare $Z_1$ against a logarithmic $+25\%$ transformation of the same ratio, termed Scaling B, in Appendix~\ref{app:scaling_comparison}). Readers should note that because more than half of low-income country-waves exceed the $150\%$ pre-winsorisation cap ($m_{\text{Low}} = 150.0\%$), $Z_{1,i,t}$ in the low-income stratum varies over $[-1.5, 0]$ below the cap, so we verify in Check \texttt{RC1b} (omitting the $150\%$ cap), Scaling B (using raw $m_{\text{Low}} = 183.8\%$), and $Z_3$ (real PPP dollars) that preserving upper-tail variation yields virtually identical conclusions.

On this zero-centred $Z_1$ scale, Nordic and continental European high-income countries with generous public university funding sit near the top of the distribution (Norway, Sweden, and Denmark at $+1.2$ to $+1.5$, and Austria at $+0.9$ to $+1.0$). Countries near their group median sit near $0$ (New Zealand and Iceland at $-0.03$), whereas countries relying more heavily on private tuition sit below $0$ (Spain at $-0.40$, the United States at $-0.59$, Chile at $-0.69$, and South Korea at $-1.15$). Ireland illustrates a striking chronological trajectory along $Z_1$: it sat above the high-income median in \texttt{1995--1999} ($28.2\%$, $Z_1 = +0.18$), dipped slightly below in \texttt{2000--2004} ($25.6\%$, $Z_1 = -0.08$), recovered above the median in \texttt{2005--2009} ($28.4\%$, $Z_1 = +0.20$) and \texttt{2010--2014} ($27.6\%$, $Z_1 = +0.12$), and then dropped sharply to $17.3\%$ ($Z_1 = -0.91$, or $9.1$ pp below the median) in \texttt{2015--2019} as student numbers grew much faster than state grants after the 2008 financial crisis. Alongside $Z_1$, we construct three additional zero-centred continuous funding variables tested in turn in Section~\ref{sec:unbundling} (each winsorised at $\pm 1.5$ units so $Z \in [-1.5, +1.5]$): $Z_2 = \text{clip}(\Delta \text{Spend}_{i,t} / 10, -1.5, +1.5)$ measures within-country five-year per-student funding changes centred at zero change (per $+10$ pp expansion), $Z_3 = \text{clip}\bigl(\log(\text{PPP\$}_{i,t} / m^{\text{PPP}}_{g(i,t)}) / \log(1.25), -1.5, +1.5\bigr)$ measures log real 2021 PPP dollars per student centred on the training income-group median (per $+25\%$ relative dollar increase), and $Z_4 = \text{clip}\bigl((\text{BudgetGDP}_{i,t} - m^{\text{Bud}}_{g(i,t)}) / 0.25, -1.5, +1.5\bigr)$ measures the total public university budget as a share of GDP centred on the training income-group median (per $+0.25$ pp of GDP).

\section{Methodology}
\label{sec:methods}

In an ideal experiment, we would randomly assign different countries to increase per-student university funding by varying amounts and then track their tax receipts over the next decade. In observational data, however, governments set their university budgets in response to economic conditions, demographics, and existing tax capacity. The relationships among these macroeconomic variables are also strongly non-linear \citep{hill2011bayesian, hill2020bayesian}. To separate the causal slope of continuous university funding from confounding macroeconomic trends whilst modelling all four interrelated tax outcomes ($Y_1$ to $Y_4$) simultaneously, we use the continuous-treatment Multivariate Bayesian Causal Forest (MV-BCF) model \citep{hahn2020bayesian, mcjames2024mvbcf}. We present the intuitive setup of the model here and provide full mathematical derivations in Appendix~\ref{app:math}.

The Multivariate Bayesian Causal Forest \citep{hahn2020bayesian, mcjames2024mvbcf} splits the prediction of future tax changes into two distinct regression tree ensembles \citep{chipman2010bart}. The first ensemble estimates a baseline prognostic trajectory $\mu(\cdot)$ (what would have happened to a country's tax take naturally when $Z_{i,t} = 0$, given its 19 pre-treatment country characteristics $X_{i,t}$, its predicted funding level $\hat{z}(X_{i,t})$, and its four follow-up macroeconomic growth mediators $M_{i,t\to t+h}$), whilst the second ensemble estimates a causal treatment slope $\tau(W_{i,t})$ that measures the effect of a unit change in university funding $Z_{i,t}$ and can vary with a vector of eight pre-treatment structural moderators $W_{i,t} = (W_{1,i,t}, \dots, W_{8,i,t})^\top \subset X_{i,t}$ (a complete data dictionary defining every variable in $Y$, $Z$, $X$, $W$, and $M$ is provided in Appendix~\ref{app:dictionary}, Table~\ref{tab:dictionary}). For country $i$ at wave $t$ and follow-up horizon $h \in \{1, 2\}$ ($+5$ and $+10$ years), the four tax changes $Y_{i, t+h} = (Y_{1,i,t+h}, Y_{2,i,t+h}, Y_{3,i,t+h}, Y_{4,i,t+h})^\top \in \mathbb{R}^4$ are modelled jointly as:
\begin{equation}
  Y_{i, t+h} = \underbrace{\mu\bigl(X_{i,t}, M_{i,t\to t+h}, \hat{z}(X_{i,t})\bigr)}_{\text{Baseline Trajectory at } Z_{i,t}=0} + \underbrace{\tau\bigl(W_{i,t}\bigr)}_{\text{Causal Slope}} \cdot Z_{i,t} + \varepsilon_{i, t+h}, \quad \varepsilon_{i, t+h} \stackrel{\text{i.i.d.}}{\sim} \mathcal{N}_4(0, \Sigma).
  \label{eq:mvbcf}
\end{equation}

To estimate the baseline trajectory $\mu\bigl(X_{i,t}, M_{i,t\to t+h}, \hat{z}(X_{i,t})\bigr)$, we use an ensemble of 75 decision trees, where each terminal leaf outputs a vector of four values corresponding to our four tax outcomes ($Y_1$ to $Y_4$). Whether these 75 decision trees are shallow or deep is not fixed in advance, but is learned directly from the data by the model during Markov chain Monte Carlo (MCMC) sampling. The baseline forest receives $X_{i,t}$ comprising 19 pre-treatment country characteristics measured at or before wave $t$ (starting log real GDP per person in PPP dollars, starting total non-resource tax-to-GDP ratio $L_1$, starting direct income tax share $L_3$, starting goods and services tax share $L_4$, disposable income Gini inequality, life expectancy, gross tertiary and secondary enrolment ratios, total education spending as a share of GDP, working-age population share, log population, unemployment rate, natural resource rents, trade openness, lagged five-year GDP per person growth, World Bank income group, geographic region, OECD membership, and wave time index; starting direct income tax-to-GDP $L_2 = L_1 L_3 / 100$ is implicitly captured by tree splits on $L_1$ and $L_3$; see Table~\ref{tab:dictionary}). In our primary specification (\texttt{M2a} at $+5$ years and \texttt{M2b} at $+10$ years), $\mu(\cdot)$ also receives four follow-up macroeconomic mediators $M_{i,t\to t+h}$ over $[t, t+h]$ (forward GDP per person growth, university enrolment expansion, income inequality changes, and life expectancy changes) to isolate the controlled direct fiscal effect $\tau^{\text{dir}}(W_{i,t})$ net of overall economic growth under sequential ignorability of $Z_{i,t}$ and $M_{i,t\to t+h}$ \citep{imai2010general, acharya2016explaining, vanderweele2016mediation}. Because holding follow-up GDP growth $M_{i,t\to t+h}$ fixed in a single stage could introduce collider bias if an unobserved 10-year economic boom simultaneously raises both GDP growth and tax receipts (slightly attenuating the estimated university funding slope), we treat \texttt{M2a/M2b} ($+0.24$ pp on $Y_1$ and $+0.12$ pp on $Y_2$ at $+10$ years) as a conservative lower bound alongside the unmediated total causal effect $\tau^{\text{tot}}(W_{i,t})$ omitting $M_{i,t\to t+h}$ entirely (\texttt{M1a/M1b}: $+0.30$ pp on $Y_1$ and $+0.16$ pp on $Y_2$ in Appendix Table~\ref{tab:benchmarks}) and two-stage sequential g-estimation (\texttt{RC9}: $+0.46$ pp on $Y_1$ and $+0.20$ pp on $Y_2$ in Appendix Table~\ref{tab:robust}; \citealt{acharya2016explaining}).

In standard Bayesian Causal Forests \citep{hahn2020bayesian}, the treatment variable $Z$ is binary and a propensity score $P(Z = 1 \mid X)$ is included inside $\mu(\cdot)$ to control for selection bias. However, in our approach the university funding variable $Z_{i,t} \in [-1.5, +1.5]$ is continuous rather than binary, so a binary propensity score no longer applies. Instead, following the partially linear and orthogonalised forest literature \citep{robinson1988root, chernozhukov2018double, athey2019generalized}, we estimate each country-wave's predicted continuous funding level $\hat{z}(X_{i,t}) = \mathbb{E}[Z_{i,t} \mid X_{i,t}]$ strictly from the 19 pre-treatment covariates $X_{i,t}$ (excluding post-treatment mediators $M_{i,t\to t+h}$) using a 500-tree random forest (\texttt{ranger}; \citealt{breiman2001random, wright2017ranger}) with 5-fold country-blocked cross-fitting, and we include $\hat{z}(X_{i,t})$ as a covariate inside $\mu(\cdot)$. Under conditional mean independence of treatment variation around $\hat{z}(X_{i,t})$ (a first-moment analogue of the generalised propensity score of \citealt{hirano2004propensity}), including $\hat{z}(X_{i,t})$ guards against regularisation-induced confounding \citep{hahn2020bayesian} and identifies $\tau(W_{i,t})$ from local funding deviations around what a country with similar history is predicted to spend. In Appendix~\ref{app:diagnostics}, we confirm that adjusting for $\hat{z}(X_{i,t})$ reduces the mean absolute correlation between $Z_1$ and the 19 starting covariates by more than 60\% (from $0.157$ to $0.061$).

To allow the fiscal slope per unit of university funding to vary across countries, we model $\tau(W_{i,t})$ using a second ensemble of 30 decision trees that split on the eight pre-treatment moderators $W_{i,t} = (W_{1,i,t}, \dots, W_{8,i,t})^\top \subset X_{i,t}$: gross tertiary enrolment ($W_1$), log real GDP per person ($W_2$), gross secondary enrolment ($W_3$), starting direct income tax share ($W_4$), working-age population share ($W_5$), World Bank income group ($W_6$), OECD membership ($W_7$), and wave time index ($W_8$). Again, the exact tree structures and depths in $\tau(W_{i,t})$ are chosen by the data during MCMC sampling, but following \citet{hahn2020bayesian} we place stronger regularisation priors on the $\tau$ tree shapes than on the $\mu$ tree shapes so that the treatment forest shrinks noisy subgroup differences toward the overall average slope (full hyperparameter details are provided in Appendix~\ref{app:math}). Because $Z_{i,t} \in [-1.5, +1.5]$ is centred at zero (the income-group median benchmark for $Z_1, Z_3, Z_4$, or zero change for $Z_2$), $\mu\bigl(X_{i,t}, M_{i,t\to t+h}, \hat{z}(X_{i,t})\bigr)$ directly represents the expected baseline tax trajectory when $Z_{i,t} = 0$, allowing us to adjust funding $Z_{i,t}$ up ($Z_{i,t} > 0$) or down ($Z_{i,t} < 0$) symmetrically around zero (Check \texttt{RC16} in Appendix Table~\ref{tab:robust} confirms that shifting location by $+1.5$ onto $[0, 3.0]$ yields virtually identical slopes). All four tax outcomes ($Y_1$ to $Y_4$) are linked through the $4 \times 4$ residual covariance matrix $\Sigma$, which receives a conjugate Inverse-Wishart prior; whilst Equation~\eqref{eq:mvbcf} uses a working i.i.d. $\mathcal{N}_4(0, \Sigma)$ likelihood across country-waves, we verify that accounting for within-country serial correlation and overlapping two-wave horizons via country-clustered posterior standard errors (reported in Appendix Table~\ref{tab:mcmc}), non-overlapping 10-year decadal waves (\texttt{RC2}), and within-country fixed-effect demeaning (\texttt{RC3a/RC3b}) leaves our conclusions intact.

An important modelling question is why we feed all 19 starting covariates $X_{i,t}$ (plus the four follow-up growth controls $M_{i,t\to t+h}$) into the baseline forest $\mu(\cdot)$, while feeding the targeted subset of eight structural moderators $W_{i,t} \subset X_{i,t}$ into the treatment forest $\tau(\cdot)$ (see Table~\ref{tab:dictionary} in Appendix~\ref{app:dictionary} for the side-by-side mapping of $X$ and $W$). In Bayesian causal forest modelling \citep{hahn2020bayesian, mcjames2024mvbcf}, separating prognostic covariates $X$ from effect moderators $W$ is recommended when the sample size is moderate ($N = 283$ to $371$ country-waves). The 75 baseline trees in $\mu(X_{i,t}, M_{i,t\to t+h}, \hat{z}_{i,t})$ need all 19 starting variables and growth controls to absorb complex macroeconomic confounding, whereas restricting $\tau(W_{i,t})$ to the eight core institutional, educational, and demographic moderators prevents the 30 treatment trees from splitting on noisy short-term nuisance variables (such as trade openness or resource rents). However, one could reasonably argue that both parts of the model should be given the exact same set of covariates ($W_{i,t} = X_{i,t}$) so that the data alone decide which variables moderate the treatment slope. To test whether this choice affects our findings, we run a dedicated sensitivity check in Appendix Table~\ref{tab:robust} (Check \texttt{RC15}) where $\mu(\cdot)$ and $\tau(\cdot)$ use the identical covariate matrix ($W_{i,t} = X_{i,t}$); doing so yields virtually identical 10-year posterior slopes ($+0.33$ pp on total tax/GDP $Y_1$, $P = 0.97$; $+0.12$ pp on direct income tax/GDP $Y_2$, $P = 0.85$; and $+0.51$ pp on the direct income tax share $Y_3$, $P = 0.91$).

The main advantage of the continuous-treatment MV-BCF framework is that it estimates smooth funding slopes, captures non-linear confounding and diminishing returns across country tiers, and shares information across correlated tax outcomes whilst providing full Bayesian posterior distributions. The main disadvantage is that tree ensembles rely on MCMC sampling and conditional ignorability assumptions rather than linear fixed-effect differencing, and assume locally linear funding slopes within terminal leaves over the central $\pm 15$ pp range $Z_{1,i,t} \in [-1.5, +1.5]$. Across our 68-model pipeline, all baseline MV-BCF models are fitted in C++ for 2,000 MCMC iterations (750 burn-in, 1,250 retained posterior draws), which we retain throughout our headline figures, tables, and return-on-investment calculations as the conservative benchmark because pooling four dispersed chains (8,000 retained draws; Section~\ref{sec:holdout}) produces slightly larger $+10$-year slopes on $Y_1$ and $Y_2$ and strengthens every conclusion. We report multi-chain convergence, residual, and out-of-sample holdout validation diagnostics in Section~\ref{sec:holdout} and Appendices~\ref{app:diagnostics} and \ref{app:scaling_comparison}.

\section{Results}
\label{sec:results}

\subsection{Continuous Funding Slopes After 5 and 10 Years}

\begin{figure}[htbp]
  \centering
  \includegraphics[width=\textwidth]{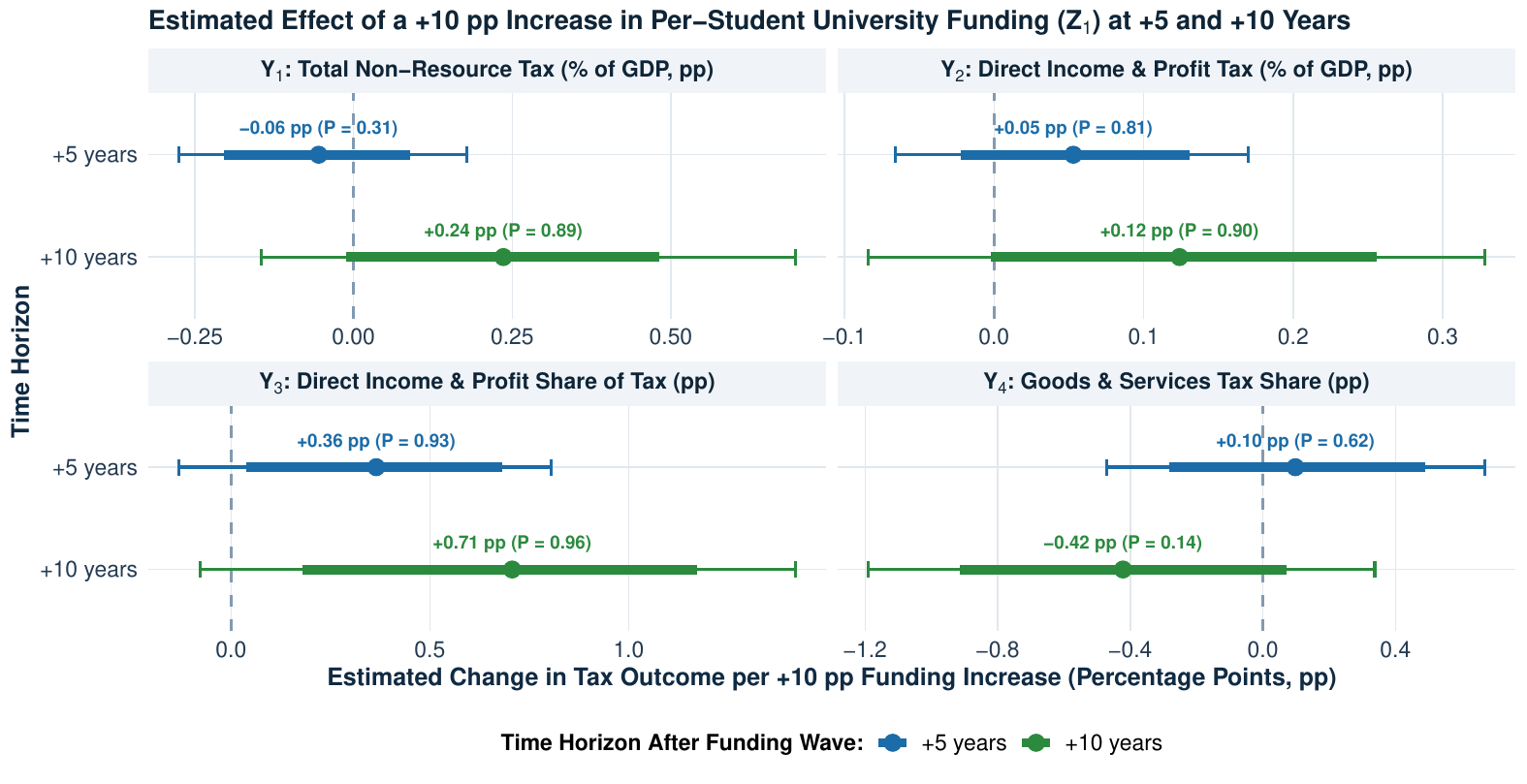}
  \caption{Posterior mean treatment slopes (in percentage points, pp, per $+10$ pp increase in per-student university spending $Z_1$) across the four jointly modelled tax outcomes ($Y_1$ to $Y_4$) at $+5$ years ($N = 371$, blue) and $+10$ years ($N = 283$, green). Thick horizontal bars show 80\% Bayesian posterior credible intervals, thin horizontal lines show 95\% credible intervals, and $P = \Pr(\tau > 0 \mid \text{data})$ denotes the posterior probability that the causal slope is greater than zero given the observed data (baseline single-chain estimates shown as the conservative choice; pooling four dispersed chains in Section~\ref{sec:holdout} strengthens $+10$-year slopes to $+0.31$ pp, $P = 0.94$ on $Y_1$ and $+0.15$ pp, $P = 0.93$ on $Y_2$).}
  \label{fig:posteriors}
\end{figure}

We begin by examining our primary question using funding measure $Z_1$. When a country increases its per-student university spending ratio by $+10$ percentage points of GDP per person ($\Delta Z_1 = +1.0$ unit), what happens to its four tax outcomes ($Y_1$ to $Y_4$) after 5 and 10 years once starting country conditions, predicted funding $\hat{z}(X_{i,t})$, and follow-up economic growth are controlled for? Figure~\ref{fig:posteriors} plots the posterior mean slopes, 80\% and 95\% credible intervals, and posterior probabilities $P = \Pr(\tau > 0 \mid \text{data})$ across all four tax outcomes at $+5$ years ($N = 371$, 121 countries) and $+10$ years ($N = 283$, 112 countries).

\FloatBarrier

Figure~\ref{fig:posteriors} reveals a clear maturation pattern across the four panels from $Y_1$ to $Y_4$, with the $+5$-year effect shown at the top of each panel and the $+10$-year effect shown below it. Starting in the top-left panel ($Y_1$, total non-resource tax revenue as a share of GDP), a $+10$ pp increase in per-student university spending ($Z_1$) has a slightly negative effect after 5 years at $-0.06$ pp of GDP ($P = 0.31$) while students are still enrolled, before rising to $+0.24$ pp of GDP after 10 years ($P = 0.89$, or $+0.31$ pp with $P = 0.94$ across four dispersed MCMC chains in Section~\ref{sec:holdout} and $+0.30$ pp with $P = 0.95$ when follow-up growth mediators are omitted in Appendix Table~\ref{tab:benchmarks}; we retain the single-chain $+0.24$ pp baseline throughout as the conservative choice). In the top-right panel ($Y_2$, direct income, profit, and capital gains tax revenue as a share of GDP), the slope is $+0.05$ pp of GDP after 5 years ($P = 0.81$) and more than doubles to $+0.12$ pp of GDP after 10 years ($P = 0.90$, or $+0.15$ pp with $P = 0.93$ across four chains and $+0.16$ pp with $P = 0.96$ for the total effect) as university cohorts graduate and advance into higher-salaried careers. In the bottom-left panel ($Y_3$, direct income and profit taxes as a share of total tax revenue), higher per-student funding produces a strong and sustained shift in the tax mix toward direct taxation, rising by $+0.36$ pp after 5 years ($P = 0.93$) and by $+0.71$ pp after 10 years ($P = 0.96$). Finally, in the bottom-right panel ($Y_4$, indirect goods and services taxes as a share of total tax revenue), the slope is flat after 5 years at $+0.10$ pp ($P = 0.62$) and shifts downward after 10 years by $-0.42$ pp ($P = 0.14$, or $-0.65$ pp with $P = 0.04$ for the total effect) as direct income taxation expands its share of the revenue pie.

To put these continuous slopes into concrete national currency terms, we hold each country's latest population and 2021 PPP GDP per person constant so that demographic growth does not inflate the calculation. At the 10-year mark, an annual direct income tax gain ($Y_2$) of $+0.12$ to $+0.16$ pp of GDP (and $+0.24$ to $+0.30$ pp in total non-resource tax $Y_1$) per $+10$ pp funding increase corresponds to roughly \texteuro$0.23$ billion to \texteuro$0.30$ billion per year in additional direct income tax receipts (and \texteuro$0.44$ billion to \texteuro$0.55$ billion per year in total tax) for a median-sized economy with a GDP of \texteuro$183$ billion. In a medium-sized high-income economy such as Ireland, the Year-10 annual gain from a $+10$ pp funding increase is \texteuro$0.36$ billion per year in direct income tax and \texteuro$1.21$ billion per year in total non-resource tax, and in a large economy such as the United Kingdom it is \texteuro$2.52$ billion per year in direct income tax and \texteuro$7.29$ billion per year in total non-resource tax. Because our data are averaged into five-year waves, a $+10$ pp increase in $Z_1$ represents an annual spending commitment (averaging $0.40\%$ of GDP per year across below-median countries, or \texteuro$0.73$ billion per year in a median economy and \texteuro$2.37$ billion per year in Ireland) sustained over the five-year wave while a student cohort is at university. Comparing a single year of tuition costs for the entire student body against a single year of tax receipts at Year 10 (when only the first two five-year graduate cohorts have entered the workforce) can make the annual funding cost look larger than the Year-10 annual return in high-enrolment countries, even though each cohort only requires university funding during its degree and then pays higher taxes across a 40-year working career; in Section~\ref{sec:roi} and Appendix~\ref{app:policymakers}, we translate these 10-year causal slopes directly into a single-cohort return on investment across a 40-year career.

\subsection{Comparing the Four Ways of Measuring University Funding (\texorpdfstring{$Z_1$ to $Z_4$}{Z1 to Z4})}
\label{sec:unbundling}

\begin{figure}[htbp]
  \centering
  \includegraphics[width=\textwidth]{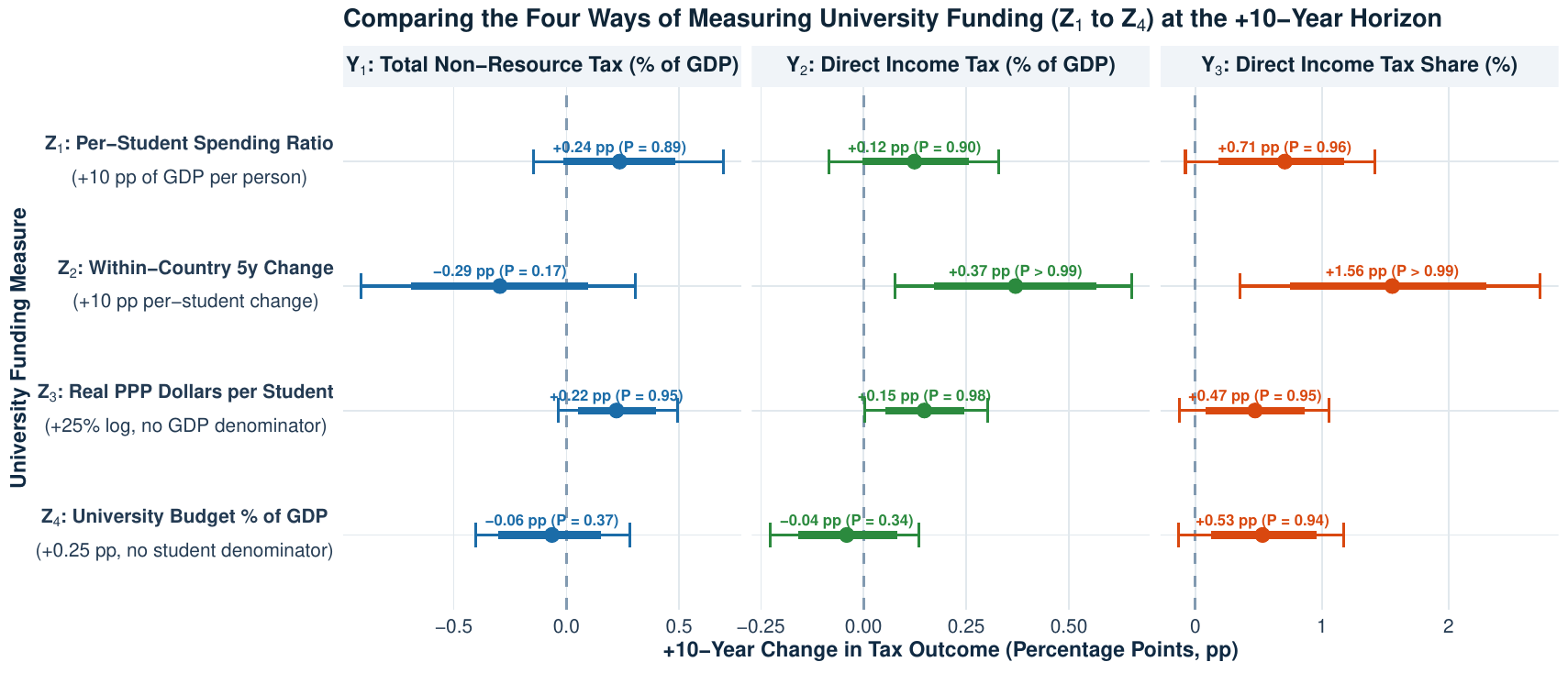}
  \caption{Comparison of $+10$-year posterior mean slopes (in percentage points, pp) across the four continuous university funding measures ($Z_1$ to $Z_4$, estimated in separate models) for total non-resource tax/GDP ($Y_1$), direct income tax/GDP ($Y_2$), and the direct income tax share of revenue ($Y_3$). Thick horizontal bars show 80\% Bayesian credible intervals, thin lines show 95\% credible intervals, and $P = \Pr(\tau > 0 \mid \text{data})$ denotes the posterior probability of a positive effect. Protecting real resources per student ($Z_1$, $Z_2$, and $Z_3$) raises 10-year direct income tax revenue, whereas expanding the aggregate budget without accounting for student numbers ($Z_4$) yields near-zero tax-to-GDP gains.}
  \label{fig:horserace}
\end{figure}

Whilst Figure~\ref{fig:posteriors} compared the $+5$-year and $+10$-year time horizons for our primary per-student spending ratio ($Z_1$), a natural follow-up question is how the $+10$-year fiscal slope changes when we unbundle the UNESCO ratio into its separate components. In Figure~\ref{fig:horserace}, we hold the time horizon fixed at $+10$ years and compare our four continuous university funding variables ($Z_1$ to $Z_4$), each estimated in its own four-outcome MV-BCF model. Specifically, we compare (1) the per-student spending ratio $Z_1$ (per $+10$ pp of GDP per person per student, $N = 283$, 112 countries), (2) observed within-country five-year funding changes $Z_2$ (per $+10$ pp five-year expansion, $N = 176$, 86 countries), (3) real 2021 PPP dollars per student $Z_3$ (per $+25\%$ relative logarithmic increase, removing GDP per person from the denominator, $N = 283$, 112 countries), and (4) the public university budget as a percentage of GDP $Z_4$ (per $+0.25$ pp of GDP, removing student headcount from the denominator, $N = 403$, 123 countries). Comparing $Z_1$ to $Z_4$ on the same 10-year horizon tests whether future tax gains are driven by real instructional resources per student or merely by aggregate budget size.

\FloatBarrier

Figure~\ref{fig:horserace} shows a simple pattern across the four ways of measuring university funding. The three measures that track resources \emph{per student} ($Z_1$, $Z_2$, and $Z_3$) all produce positive 10-year gains in direct income tax revenue ($Y_2$) and in the direct income tax share ($Y_3$). Removing GDP from the denominator by measuring real PPP dollars per student ($Z_3$) raises 10-year total tax revenue ($Y_1$) by $+0.22$ pp ($P = 0.95$) and direct income tax revenue ($Y_2$) by $+0.15$ pp ($P = 0.98$). Focusing on five-year within-country funding increases ($Z_2$) produces the largest gains in direct income taxation ($+0.37$ pp on $Y_2$ and $+1.56$ pp on $Y_3$, $P \ge 0.99$), even though total tax-to-GDP ($Y_1$) dips slightly because reforming countries often reduce trade and sales taxes at the same time (see Appendix~\ref{app:benchmarks} for the algebraic decomposition). By contrast, expanding the overall university budget without accounting for student numbers ($Z_4$) leaves 10-year tax-to-GDP slightly negative ($-0.06$ pp on $Y_1$, $P = 0.37$, and $-0.04$ pp on $Y_2$, $P = 0.34$, which holds on the overlapping $N = 272$ sample in Check \texttt{RC17} of Appendix Table~\ref{tab:robust}), confirming that funding must keep pace with enrolment to avoid institutional crowding \citep{bound2010have, deming2017impact, chakrabarti2020state}.

\subsection{Subgroup Moderation, Negative Scenarios, and Cohort Maturation Trajectories}

\begin{figure}[htbp]
  \centering
  \includegraphics[width=\textwidth]{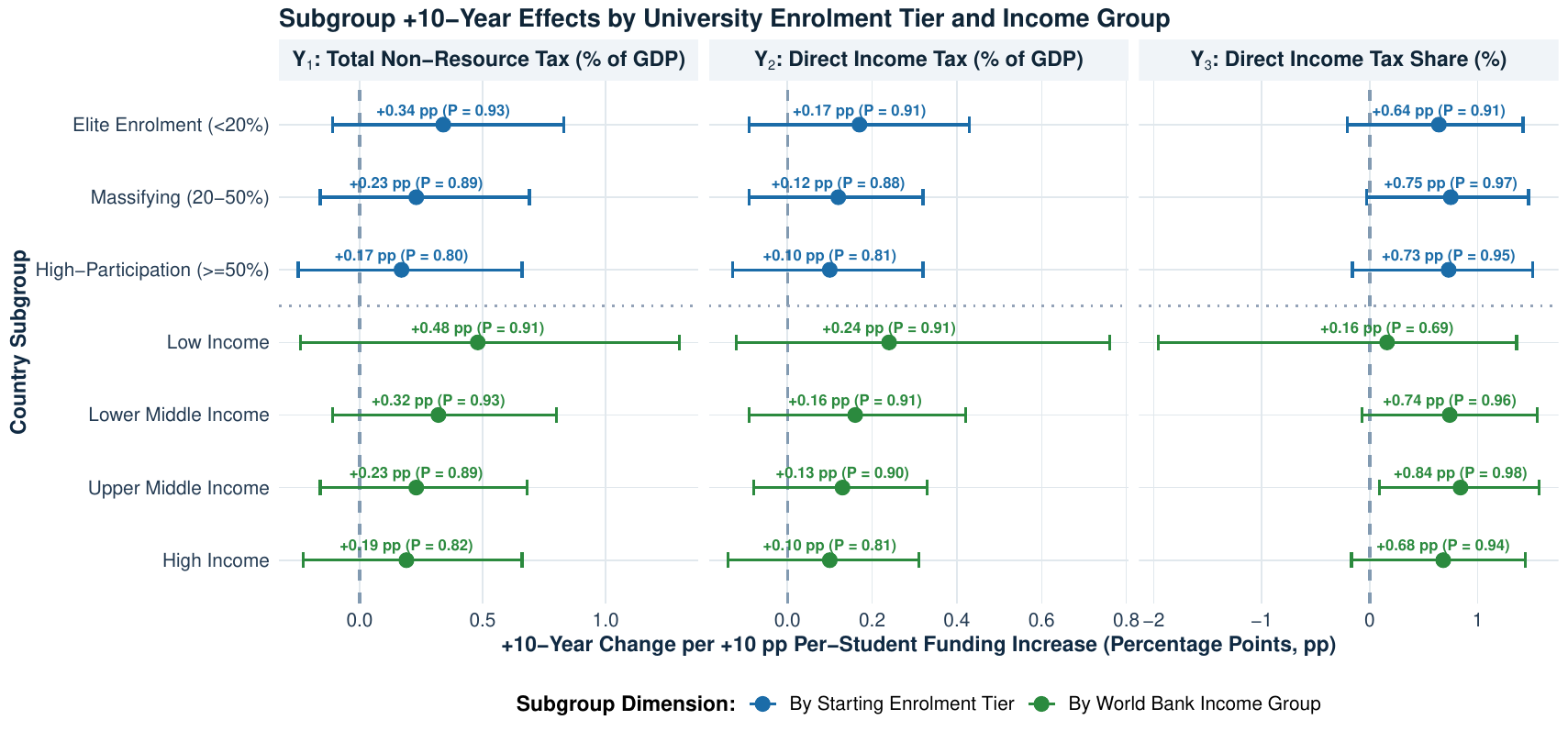}
  \caption{Subgroup $+10$-year posterior mean slopes (in percentage points, pp per $+10$ pp per-student funding increase $Z_1$) across starting university enrolment tiers and World Bank income groups for total non-resource tax/GDP ($Y_1$), direct income tax/GDP ($Y_2$), and the direct income tax share ($Y_3$). Horizontal bars show 95\% Bayesian credible intervals, and $P = \Pr(\tau > 0 \mid \text{data})$ gives the posterior probability of a positive effect.}
  \label{fig:moderation}
\end{figure}

How do the causal slopes per $+10$ pp of per-student funding ($Z_1$) vary across countries at different stages of university enrolment and economic development, and how does the model capture diminishing marginal returns? Figure~\ref{fig:moderation} plots the $+10$-year subgroup posterior mean slopes and 95\% credible intervals across starting university enrolment tiers and World Bank income groups for $Y_1$, $Y_2$, and $Y_3$ (with full numerical breakdowns, including our five-group OECD financing regime sub-model, reported in Appendix Table~\ref{tab:subgroups}). As shown in Figure~\ref{fig:moderation}, the two fiscal dimensions vary in contrasting ways across development stages. On the tax-mix dimension ($Y_3$, right panel), the shift toward direct income and profit taxation rises as economies develop a formal corporate and salaried sector capable of hiring university graduates. The $Y_3$ slope is smallest in low-income economies ($+0.16$ pp, $P = 0.69$), averages $+0.64$ pp ($P = 0.91$) in elite systems with under 20\% enrolment, rises to $+0.74$ pp ($P = 0.96$) in lower-middle-income economies, $+0.75$ pp ($P = 0.97$) in massifying systems with 20--50\% enrolment, and $+0.84$ pp ($P = 0.98$) in upper-middle-income economies, and remains strongly positive at $+0.68$ to $+0.73$ pp ($P = 0.94\text{--}0.95$) in high-income and high-participation systems ($\ge 50\%$ enrolment, and $+1.05$ to $+1.14$ pp across all five OECD financing regimes). By contrast, on the aggregate tax-to-GDP dimensions ($Y_1$ and $Y_2$, left and middle panels), the treatment trees capture clear diminishing returns as university systems expand: the largest gains per $+10$ pp of funding occur in low-income ($+0.48$ pp on $Y_1$ and $+0.24$ pp on $Y_2$, $P = 0.91$), lower-middle-income ($+0.32$ pp and $+0.16$ pp, $P = 0.91\text{--}0.93$), and elite-enrolment ($+0.34$ pp and $+0.17$ pp, $P = 0.91\text{--}0.93$) economies, tapering off to $+0.19$ pp ($P = 0.82$) and $+0.10$ pp ($P = 0.81$) in high-income economies.

\FloatBarrier

\begin{figure}[htbp]
  \centering
  \includegraphics[width=\textwidth]{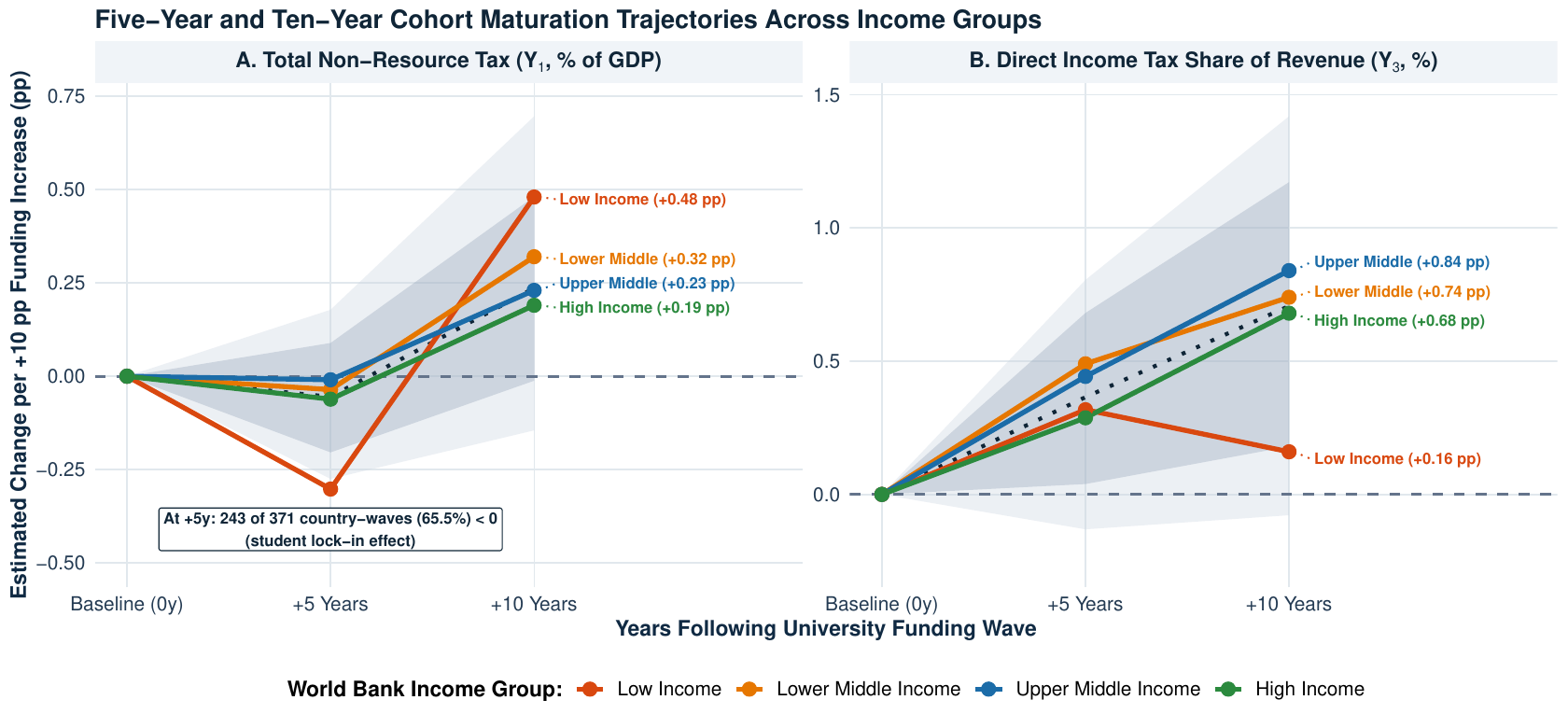}
  \caption{Five-year ($N = 371$) and 10-year ($N = 283$) cohort maturation trajectories of the causal slopes per $+10$ pp university spending increase ($Z_1$) across World Bank income groups for (A) total non-resource tax revenue as a percentage of GDP ($Y_1$, pp) and (B) the direct income and profit tax share of revenue ($Y_3$, pp). Coloured lines show income-group mean trajectories (with the callout in Panel A noting that 243 of 371 country-waves, or 65.5\%, are negative at $+5$ years due to student lock-in), the dotted line marks the global posterior mean, and the shaded bands show the global 80\% (dark grey) and 95\% (light grey) posterior credible fan.}
  \label{fig:trajectories}
\end{figure}

Both the historical data and our model estimates also highlight three specific scenarios where increasing university funding is associated with a \emph{decrease} in the tax take. First, as annotated in Panel A of Figure~\ref{fig:trajectories}, in the short run ($+5$ years) 243 out of 371 country-waves ($65.5\%$) have negative point estimates on total tax-to-GDP $Y_1$ (down to $-0.96$ pp of GDP per $+10$ pp funding increase, with an overall mean of $-0.06$ pp and $-0.30$ pp in low-income economies). This short-run dip reflects a temporary ``student lock-in'' and cohort-gestation effect: even when follow-up enrolment headcount is controlled for, higher per-student funding supports full-time study persistence and degree completion so that young adults spend more time in lecture theatres rather than working full time during those initial five years. Second, at the $+10$-year horizon in very low-income economies with tiny elite enrolment below 2\% and extreme UNESCO ratios above 150\% (such as Ethiopia in \texttt{1995--1999} at $-0.94$ pp and Malawi in \texttt{2010--2014} at $-0.68$ pp on $Y_3$), pouring additional subsidies into a tiny elite university sector without formal-sector job creation or protection against skilled emigration \citep{beine2008brain} fails to modernise the domestic income tax share. Third, when we fit a separate sub-model restricted exclusively to the 119 OECD country-waves where total tax revenue already exceeds 35\% of GDP (Rows 8 to 12 of Appendix Table~\ref{tab:subgroups}, where within-OECD $Y_1$ slopes average $-0.14$ to $-0.17$ pp per $+10$ pp increase, compared with $+0.23$ pp when evaluated inside the pooled $N = 283$ global model), expanding public university budgets within mature welfare states does not raise aggregate tax-to-GDP further even as the income tax share of revenue ($Y_3$) rises by $+1.05$ to $+1.14$ pp ($P = 0.93\text{--}0.95$).

To show how these effects evolve from the initial $+5$-year window to the $+10$-year window across income groups, Figure~\ref{fig:trajectories} plots the four World Bank income-group trajectory lines alongside the global posterior mean (dotted line) and the 80\% and 95\% posterior credible fan. In Panel A of Figure~\ref{fig:trajectories}, total non-resource tax revenue as a share of GDP ($Y_1$) dips below zero across every income group at the $+5$-year mark ($-0.30$ pp in low-income, $-0.04$ pp in lower-middle-income, $-0.01$ pp in upper-middle-income, and $-0.06$ pp in high-income economies, with 65.5\% of individual country-waves below zero) whilst students are still enrolled in university. By the $+10$-year mark in Panel A, $Y_1$ rebounds into positive territory across all four groups ($+0.48$ pp in low-income, $+0.32$ pp in lower-middle-income, $+0.23$ pp in upper-middle-income, and $+0.19$ pp in high-income economies). In Panel B of Figure~\ref{fig:trajectories}, the direct income and profit tax share of revenue ($Y_3$) already rises at $+5$ years across all four groups ($+0.29$ to $+0.49$ pp) and then diverges sharply at $+10$ years. In lower-middle-income ($+0.74$ pp), upper-middle-income ($+0.84$ pp), and high-income ($+0.68$ pp) economies where formal salaried employment can absorb new graduates, the shift toward direct income taxation roughly doubles by Year 10, whereas in low-income economies it flattens out at $+0.16$ pp. In Appendix~\ref{app:benchmarks}, we confirm that this 10-year shift toward direct income taxation holds across 17 sensitivity checks (\texttt{RC1} to \texttt{RC17}), including within-country fixed-effect demeaning, non-overlapping decadal waves, pre-trend controls, sequential g-estimation, identical covariates in $\mu$ and $\tau$ ($W = X$), zero-centred treatment scaling, and competing Bayesian and frequentist estimators.

\FloatBarrier

\subsection{Cohort Return on Investment for Policy Makers}
\label{sec:roi}

To answer whether increasing government spending on tertiary education provides a positive return on investment, we must compare the public cost of educating a cohort of students on the exact same basis as the taxes that cohort pays across its working life. In any calendar year, a country's university system holds all $D$ year-groups of students simultaneously (for example, 1st-, 2nd-, 3rd-, and 4th-year undergraduates, or roughly 245,000 students in total in Ireland, corresponding to about 60,000 students per entering year-group). Because a single entering cohort makes up $1/D$ of the total university student body in any single year and studies for $D$ years, the total multi-year cost of funding that one cohort across all $D$ years of their degree ($D \text{ years} \times 1/D \text{ of the student body}$) is algebraically identical to one year of the system-wide $+10$ pp university budget increase (\texteuro$2.37$ billion in Ireland, or one-fifth of a five-year wave). Once that cohort graduates and works across a 40-year career (ages 22 to 62, spanning eight five-year waves), our $+10$-year causal slope ($\tau_{10}$, which measures the annual tax gain from a five-year wave of five cohorts) implies that each single cohort generates one-fifth of the Year-10 annual tax flow in every working year for 40 years. Consequently, under the benchmark accounting translation that the cohort sustains its Year-10 annual tax contribution across its working life (when tax receipts grow in line with GDP, $r = g$), the cumulative 40-year career tax output per cohort equals eight times the Year-10 annual flow ($40 \text{ working years} \div 5 \text{ cohorts} = 8$; equivalently, multiplying both the cohort input and career output by $5$ gives the five-year wave total with the exact same benefit-cost ratio and probability of a net positive return). Under alternative accounting assumptions, applying a net real treasury discount rate of $r - g = 1.5\%$ per year across years 5 to 44 after university entry scales this 40-year multiplier from $8.0\times$ to $5.7\times$ in discounted present value, whilst assuming that two consecutive five-year waves of persistent spending ($Z_1$) have entered the labour force by Year 10 provides a conservative $4.0\times$ lower bound and using the unmediated total effect (\texttt{M1b}) or five-year funding expansion impulse ($Z_2$) yields substantially higher career multipliers. We emphasise that this 40-year career calculation is inherently speculative because we are extrapolating our 10-year empirical estimates across four decades of a cohort's working life (during which graduating many cohorts could also gradually compress the graduate wage premium), so these figures should be interpreted as an illustrative order-of-magnitude benchmark rather than a guaranteed 40-year forecast.

When presenting these numbers to a national treasury, a policy maker faces a practical question: should the return on investment be judged using total non-resource tax revenue ($Y_1$) or direct income and profit tax revenue ($Y_2$)? For the bottom-line budget decision, the treasury should focus primarily on total non-resource tax revenue ($Y_1$), because universities are funded out of the general Exchequer pot and every euro of additional tax that flows back into the Exchequer counts toward paying back that outlay. When university graduates earn higher salaries, they do not only pay direct income and corporate profit taxes ($Y_2$). They and their employers also pay social insurance and payroll taxes (such as Pay Related Social Insurance in Ireland or National Insurance in the United Kingdom, which international accounts record separately from income tax), value-added and excise duties when graduates spend their higher take-home pay in the domestic economy, and property and business taxes generated by wider skilled economic activity. Looking only at $Y_2$ therefore leaves out more than two-thirds of the total tax revenue generated across a cohort's career. At the same time, because aggregate tax-to-GDP $Y_1$ is subject to broader macroeconomic noise (so out-of-sample cross-validation skill in Section~\ref{sec:holdout} is strongest on $Y_3$ and $Y_2$, and the small $N = 119$ OECD-only sub-model shrinks $Y_1$ below zero in mature welfare states), policy makers should treat $Y_1$ (the full Exchequer return from the pooled global model) and $Y_2$ paired with $Y_3$ (the conservative direct income-tax floor) as joint upper and lower bounds.

Evaluating this cohort return on investment across the 70 countries with complete \texttt{2020--2024} macroeconomic, enrolment, and fiscal data (shown in Appendix~\ref{app:policymakers}, Figure~\ref{fig:roi_scatter} and Table~\ref{tab:policymakers}, and for all 70 countries in our interactive web application) reveals positive expected 40-year career returns across all income groups, though because 10-year national tax-to-GDP changes carry substantial macroeconomic noise ($R^2_{\text{OOS}} = -0.07$ on $Y_1$ in Section~\ref{sec:holdout}), we round headline currency figures in the text to avoid implying false precision. In Ireland, where current university spending is $17.3\%$ of GDP per person (USD 15,150 in 2021 PPP dollars), raising per-student funding by $+10$ pp requires a cohort degree input of roughly \texteuro$2.4$ billion ($0.43\%$ of GDP). Across that cohort's 40-year working career, this \texteuro$2.4$ billion input is estimated to generate around \texteuro$+9.7$ billion in total non-resource tax output $Y_1$ (95\% CI $[\text{\texteuro}-10\text{B}, \text{\texteuro}+32\text{B}]$, benefit-cost ratio $\text{BCR} = 4.1\times$, posterior probability of a net positive return $P_{\text{net}} = 0.77$, though because out-of-sample 95\% predictive coverage on $Y_1$ drops to $88.9\%$ on unseen countries in Section~\ref{sec:holdout}, $P_{\text{net}} = 0.77$ is likely slightly optimistic and closer to $0.70\text{--}0.72$ after empirical interval recalibration; or roughly \texteuro$+6.9$ billion, $\text{BCR} = 2.9\times$ when discounted at $r - g = 1.5\%$, and \texteuro$+4.8$ billion, $\text{BCR} = 2.0\times$ under the conservative $4\times$ two-wave bound) and around \texteuro$+2.9$ billion in direct income tax output $Y_2$ alone (95\% CI $[\text{\texteuro}-8\text{B}, \text{\texteuro}+13\text{B}]$, $\text{BCR} = 1.2\times$, $P_{\text{net}} = 0.54$), alongside a $+0.73$ pp shift toward direct income taxation ($P = 0.96$). Similarly, in the United Kingdom, a cohort input of roughly \texteuro$11.6$ billion ($0.35\%$ of GDP) generates an estimated \texteuro$+58$ billion in 40-year career total tax output $Y_1$ (95\% CI $[\text{\texteuro}-60\text{B}, \text{\texteuro}+190\text{B}]$, $\text{BCR} = 5.1\times$, $P_{\text{net}} = 0.79$; or $\text{BCR} = 3.6\times$ at a $1.5\%$ net discount rate) and roughly \texteuro$+20$ billion in direct income tax output $Y_2$ (95\% CI $[\text{\texteuro}-47\text{B}, \text{\texteuro}+86\text{B}]$, $\text{BCR} = 1.7\times$, $P_{\text{net}} = 0.60$). Across high-enrolment OECD economies, the expected career return on total tax $Y_1$ is typically $3\times$ to $5\times$ the cohort input ($P_{\text{net}} = 0.68\text{--}0.81$) while the conservative direct income-tax floor $Y_2$ is $0.6\times$ to $2.9\times$ ($P_{\text{net}} = 0.38\text{--}0.68$). In middle- and low-income economies with smaller university cohorts (such as Colombia, Malaysia, Mexico, Bangladesh, and Rwanda), raw benefit-cost ratios are much larger ($3.9\times$ to $41.4\times$ on $Y_1$ and $1.8\times$ to $25.5\times$ on $Y_2$) because a $+10$ pp per-student funding boost costs a tiny fraction of national GDP when enrolment is small ($0.07\%$ of GDP in Rwanda and $0.19\%$ in Bangladesh) while the macro slope $\tau_{10}(W_i)$ captures wider formalisation gains across GDP; scaling those developing-country returns proportionally by their tertiary enrolment rate relative to the high-income benchmark places their cohort benefit-cost ratios in the same $2.0\times$ to $4.5\times$ range as high-income economies.

\subsection{MCMC Convergence and Out-of-Sample Country Validation}
\label{sec:holdout}

Because cross-country macro panels have modest sample sizes, it is important to check both multi-chain MCMC convergence and out-of-sample generalisation to unseen countries. In our baseline single-chain runs (1,250 post-burn-in draws; Appendix~\ref{app:diagnostics}, Table~\ref{tab:mcmc}), split-chain $\hat{R}$ values lie between $0.999$ and $1.030$ with effective sample sizes (ESS) of $65$ to $120$, which implies a Monte Carlo standard error of roughly $0.03$ to $0.04$ on tail probabilities near $0.90$ (so single-chain probabilities such as $P = 0.89$ and $P = 0.90$ are indistinguishable within Monte Carlo error). Running four independent chains from dispersed starting seeds ($4 \times 3,000$ iterations, $8,000$ retained draws)\footnote{\label{fn:multichain}\raggedright Four-chain MCMC convergence table in our GitHub repository: \url{https://github.com/andrewcparnell/education-tax-bcf/blob/main/outputs/tables/tableA5_multichain_mcmc_diagnostics.csv}.} yields between-chain Gelman-Rubin $\hat{R}$ between $1.000$ and $1.006$, combined $\text{ESS}$ between $767$ and $1,932$ ($\text{MCSE}_P \le 0.011$), and slightly stronger $+10$-year pooled posterior slopes of $+0.31$ pp on $Y_1$ ($P = 0.94$), $+0.15$ pp on $Y_2$ ($P = 0.93$), $+0.71$ pp on $Y_3$ ($P = 0.96$), and $-0.43$ pp on $Y_4$ ($P = 0.15$); we retain the baseline single-chain figures ($+0.24$ pp on $Y_1$ and $+0.12$ pp on $Y_2$) throughout our tables and return-on-investment calculations as the conservative choice, noting that the better-sampled multi-chain run strengthens rather than weakens every conclusion. In-sample 95\% posterior predictive coverage is $95.8\%$ to $97.9\%$ at $+10$ years ($95.4\%$ to $97.9\%$ across both $+5$ and $+10$ years in Appendix~\ref{app:diagnostics}) and LOESS smooths of standardised residuals across $Z_1 \in [-1.5, +1.5]$ are flat (Appendix~\ref{app:scaling_comparison}, Figure~\ref{fig:scaling_ab}B), although because residual dispersion ($\sigma_1 = 2.51$ pp) is an order of magnitude larger than the treatment slope ($0.24$ pp), flat residual curves rule out gross mean misspecification rather than subtle non-linearity in $\tau(\cdot)$. Additional residual audits in Appendix~\ref{app:diagnostics} show no common wave shocks across five-year blocks spanning the 2008 financial crisis and 2020 COVID-19 window ($+10$-year wave ANOVA $p = 0.22\text{ to }>0.99$), near-zero within-country lag-1 residual autocorrelation at the non-overlapping $+5$-year horizon ($\hat{\rho}_1 = -0.14\text{ to }-0.03$) versus $+0.28\text{ to }+0.32$ from mechanical window overlap at $+10$ years (motivating Check \texttt{RC2} and country clustering), and stable $+10$-year residual standard deviations across World Bank income groups ($4.63\text{--}4.88$ pp on $Y_3$ and $1.74\text{--}2.46$ pp on $Y_1$).

\begin{figure}[htbp]
  \centering
  \includegraphics[width=\textwidth]{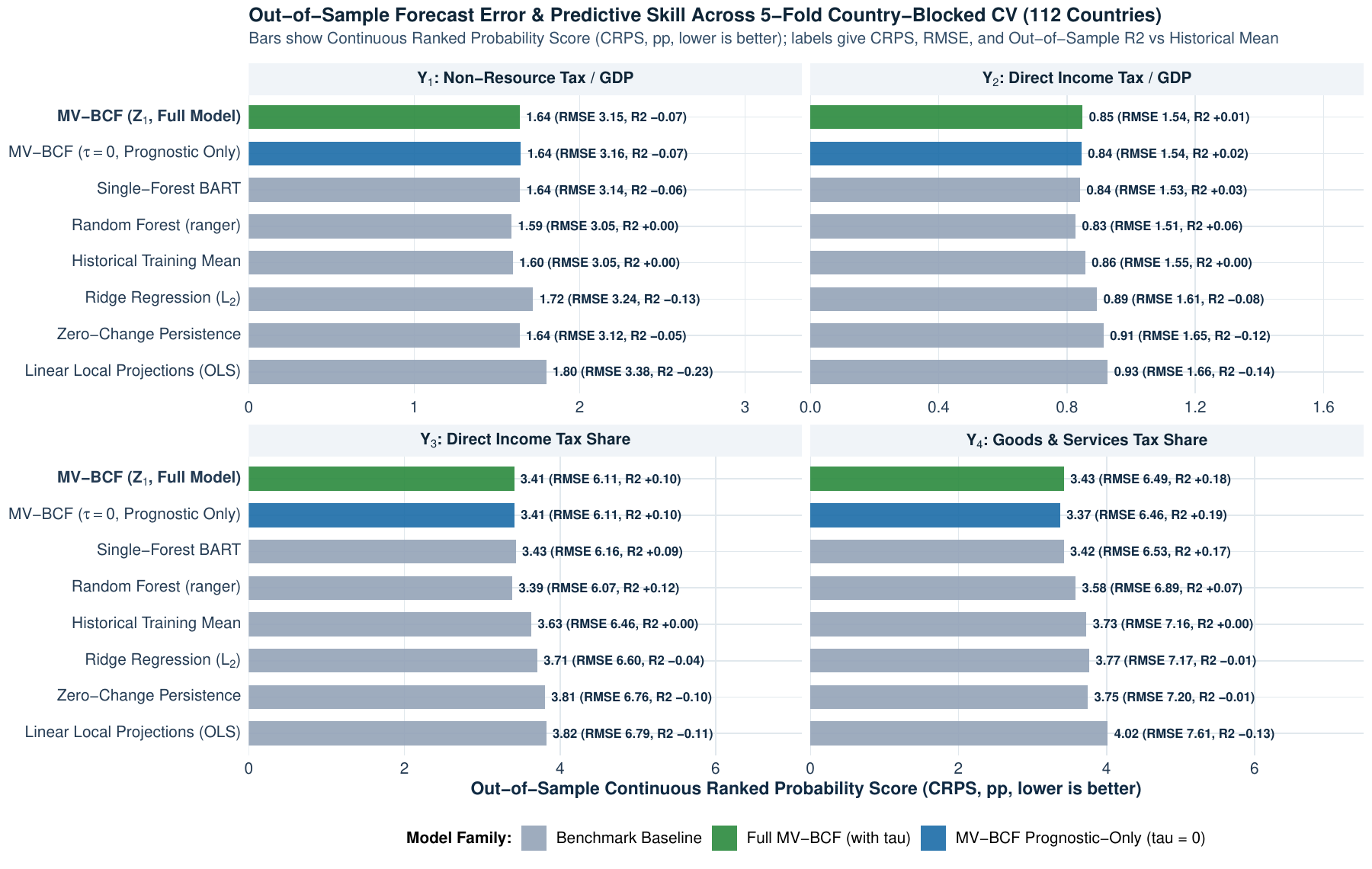}
  \caption{Out-of-sample 5-fold country-blocked cross-validation error and predictive skill across all 112 countries at the $+10$-year horizon ($N = 283$) for all four jointly modelled outcomes ($Y_1$ to $Y_4$), comparing full continuous MV-BCF ($Z_1$, green) and prognostic-only MV-BCF ($\tau \equiv 0$, treatment forest switched off, blue; on $Y_3$, the two models score $\text{CRPS} = 3.409$ vs $3.413$ pp and $\text{RMSE} = 6.112$ vs $6.111$ pp and therefore coincide at two decimal places) against single-forest BART, random forests, historical training means, ridge regression ($L_2$), zero-change persistence, and linear local projections (OLS). Horizontal bars show mean out-of-sample Continuous Ranked Probability Score (CRPS, pp, lower is better) and labels report CRPS, Root Mean Squared Error (RMSE, pp), and out-of-sample $R^2_{\text{OOS}}$ relative to the historical training mean.}
  \label{fig:holdout}
\end{figure}

When evaluating model fit, it is essential to distinguish in-sample posterior predictive fit ($R^2_{\text{in}} = 0.30\text{--}0.64$ at $+10$ years in Appendix Table~\ref{tab:mcmc}, which reflects how flexibly the 75 prognostic trees $\mu(\cdot)$ absorb within-sample country histories) from strictly out-of-sample predictive skill ($R^2_{\text{OOS}} = 1 - \text{MSE}_{\text{model}} / \text{MSE}_{\text{mean}}$ relative to the historical training mean) in unseen countries. Figure~\ref{fig:holdout} reports 5-fold country-blocked out-of-sample cross-validation continuous ranked probability scores (CRPS; \citealt{diebold1995comparing, gneiting2007strictly}), root mean squared errors (RMSE), and $R^2_{\text{OOS}}$ across all four outcomes ($Y_1$ to $Y_4$, $N = 283$, 112 countries) for eight models, including an ablation of our identical MV-BCF model with the treatment forest switched off ($\tau \equiv 0$, prognostic forest only). Averaged across all four outcomes, continuous MV-BCF achieves a mean out-of-sample CRPS of $2.33$ pp ($92.8\%$ out-of-sample 95\% coverage), outperforming historical training means ($2.45$ pp), ridge regression ($2.52$ pp), zero-change persistence ($2.53$ pp), and linear local projections ($2.64$ pp, which records negative $R^2_{\text{OOS}}$ of $-0.11$ to $-0.23$ on every outcome). However, examining individual panels in Figure~\ref{fig:holdout} reveals a sharp contrast between aggregate tax-to-GDP levels and tax-mix shares: MV-BCF has no out-of-sample predictive advantage over the historical mean on $Y_1$ ($R^2_{\text{OOS}} = -0.07$, $\text{CRPS} = 1.64$ vs $1.60$ pp, $\text{RMSE} = 3.15$ vs $3.05$ pp, where no model exceeds $R^2_{\text{OOS}} = 0.003$) and ties the mean on $Y_2$ ($R^2_{\text{OOS}} = +0.02$, $\text{CRPS} = 0.85$ vs $0.86$ pp, or $+0.12$ vs zero-change persistence), whereas it explains meaningful out-of-sample variance on the direct income tax share $Y_3$ ($R^2_{\text{OOS}} = +0.10$, $\text{CRPS} = 3.41$ vs $3.63$ pp, $\text{RMSE} = 6.11$ vs $6.46$ pp) and achieves its strongest out-of-sample skill on the goods and services tax share $Y_4$ ($R^2_{\text{OOS}} = +0.18$, $\text{CRPS} = 3.43$ vs $3.73$ pp, $\text{RMSE} = 6.48$ vs $7.16$ pp, beating random forests at $+0.07$). Comparing full MV-BCF against the $\tau \equiv 0$ prognostic-only forest in Figure~\ref{fig:holdout} ($R^2_{\text{OOS}} = -0.08, +0.02, +0.10, +0.19$; mean CRPS $2.32$ pp)\footnote{\raggedright On $Y_3$, full MV-BCF ($\text{CRPS} = 3.409$ pp, $\text{RMSE} = 6.112$ pp, $R^2_{\text{OOS}} = +0.104$) and $\tau \equiv 0$ prognostic-only MV-BCF ($\text{CRPS} = 3.413$ pp, $\text{RMSE} = 6.111$ pp, $R^2_{\text{OOS}} = +0.104$) coincide when rounded to two decimal places in Figure~\ref{fig:holdout}.} shows why: a causal slope of $0.24$ pp across $Z_1$ ($\text{SD} = 1.05$) has variance $(0.24 \times 1.05)^2 \approx 0.06\text{ pp}^2$, or under $1\%$ of total 10-year $Y_1$ shock variance ($3.05^2 = 9.30\text{ pp}^2$), so causal identification relies on pre-treatment orthogonalisation via $\hat{z}(X_{i,t})$ rather than $\tau(W)Z_1$ forecasting country-wide macro shocks, while the gap between $R^2_{\text{in}} = 0.42$ and $R^2_{\text{OOS}} = -0.07$ on $Y_1$ reflects prognostic-tree overfitting of idiosyncratic country level shocks.

Finally, our cleanest test trains all models exclusively on the 90 countries in Folds 1 to 4 and predicts the 22 locked holdout countries in Fold 5 ($N = 54$ waves at $+10$ years; Appendix~\ref{app:scaling_comparison}, Table~\ref{tab:scaling_ab} Rows A6--B6 and Figure~\ref{fig:scaling_ab}C). Across those 22 unseen countries under Scaling A, per-outcome performance mirrors the cross-validation pattern: on noisy tax-to-GDP changes, MV-BCF records $\text{RMSE} = 3.19$ pp ($R^2_{\text{OOS}} = -0.34$, $\text{CRPS} = 1.73$ pp, 95\% coverage $88.9\%$) on $Y_1$ and $\text{RMSE} = 1.53$ pp ($R^2_{\text{OOS}} = -0.15$ vs mean and $0.00$ vs zero-change, $\text{CRPS} = 0.86$ pp, coverage $94.4\%$) on $Y_2$, whereas on tax-structure shares it achieves $\text{RMSE} = 5.78$ pp ($R^2_{\text{OOS}} = +0.14$, $\text{CRPS} = 3.24$ pp, coverage $94.4\%$; improving under Scaling B to $\text{RMSE} = 5.69$ pp, $R^2_{\text{OOS}} = +0.16$, $\text{CRPS} = 3.15$ pp, coverage $96.3\%$) on $Y_3$ and $\text{RMSE} = 5.48$ pp ($R^2_{\text{OOS}} = +0.02$, $\text{CRPS} = 3.08$ pp, coverage $96.3\%$) on $Y_4$. Notice in particular that while 95\% holdout interval coverage is well calibrated on $Y_2, Y_3$, and $Y_4$ ($92.6\%\text{--}96.3\%$), it falls below the nominal $95\%$ level on $Y_1$ ($88.9\%$ under Scaling A and $85.2\%$ under Scaling B), implying that unseen countries experience slightly wider aggregate fiscal shocks than captured by $\Sigma$ and that posterior net-profit probabilities attached to $Y_1$ (such as $P_{\text{net}} = 0.77$ for Ireland) are likely slightly optimistic. Regressing observed on out-of-sample predicted 10-year changes in the direct income tax share $Y_3$ across the 22 holdout countries (Figure~\ref{fig:scaling_ab}C) yields a Mincer-Zarnowitz calibration slope of $\hat{\beta}_{\text{cal}} = 0.83$ (95\% CI $[0.29, 1.37]$, $r = 0.38$) under Scaling A and $\hat{\beta}_{\text{cal}} = 1.01$ (95\% CI $[0.41, 1.61]$, $r = 0.42$) under Scaling B, alongside positive out-of-sample predicted treatment slopes ($\bar{\tau}_{\text{test}} = +0.44$ pp on $Y_1$, $+0.21$ pp on $Y_2$, and $+0.72$ pp on $Y_3$).

\FloatBarrier

\section{Conclusions and Discussion}
\label{sec:discussion}

\subsection{Cohort Maturation Timelines and Multi-Cohort Payback Arithmetic}

Our findings across 121 countries suggest that government spending on university education shifts a nation's tax structure reliably toward direct income, profit, and capital gains taxation over a 10-year horizon, alongside positive - though substantially noisier - estimated gains in aggregate tax revenue. Because our panel averages annual data into five-year waves, a university funding increase in wave $t$ (years 0 to 4) is evaluated in wave $t+1$ (years 5 to 9) and wave $t+2$ (years 10 to 14). During the first five-year window, a $+10$ pp increase in per-student university funding ($Z_1$) immediately expands formal university payrolls whilst keeping a larger cohort of young adults in full-time study, which shifts the tax mix toward direct income tax ($Y_3 = +0.36$ pp, $P = 0.93$) without raising overall tax-to-GDP ($Y_1 = -0.06$ pp). By the 10-year window (years 10 to 14), the first two five-year university cohorts have completed their degrees and moved into salaried professional, managerial, and technical roles inside formal firms \citep{kleven2016why, valero2019economic, jensen2022employment}, raising the direct income tax share ($Y_3$) by $+0.71$ pp ($P = 0.96$), direct income tax revenue ($Y_2$) by $+0.12$ to $+0.37$ pp of GDP, and total non-resource tax revenue ($Y_1$) by $+0.22$ to $+0.31$ pp of GDP. Consistent with this mechanism, out-of-sample predictive skill across unseen countries is concentrated in the tax-composition shares ($R^2_{\text{OOS}} = +0.10\text{ to }+0.18$ in 5-fold cross-validation and $+0.14\text{ to }+0.16$ on $Y_3$ in the 22-country holdout) rather than noisy aggregate tax-to-GDP changes ($R^2_{\text{OOS}} = -0.34$ on $Y_1$ and $-0.15$ on $Y_2$ in the 22-country holdout).

For finance and education ministries, our comparison of the four funding definitions ($Z_1$ to $Z_4$) in Section~\ref{sec:unbundling} carries a clear practical lesson. Expanding the overall university budget ($Z_4$) or student numbers without protecting funding \emph{per student} yields weak fiscal returns ($-0.04$ pp on $Y_2$). Countries that maintain or expand real resources per student avoid institutional crowding \citep{bound2010have, deming2017impact, chakrabarti2020state} and achieve much stronger 10-year gains in direct income tax receipts ($+0.15$ pp of GDP per $+25\%$ increase in real PPP dollars per student $Z_3$ and $+0.37$ pp of GDP per $+10$ pp within-country five-year expansion $Z_2$). Crucially, when comparing the upfront public cost of a per-student funding boost against the resulting tax return, policy makers must distinguish between a single-year snapshot at Year 10 and an illustrative lifecycle return across a student cohort's working career. Because a single university cohort only requires public tuition subsidies while enrolled in university, whereas after graduation that cohort works and pays taxes for roughly 40 years (eight five-year waves), extrapolating our 10-year estimates across a 40-year career provides a useful back-of-the-envelope lifecycle benchmark that explains why multi-decade microeconomic studies find high lifetime fiscal payback \citep{trostel2010fiscal, hendren2020unified}. Furthermore, our analysis focuses strictly on direct fiscal tax receipts and does not capture the many wider non-fiscal benefits associated with university education - including lower crime rates, improved health and life expectancy, higher civic participation, intergenerational social mobility, and non-taxed research spillovers \citep{moretti2004estimating, trostel2010fiscal, valero2019economic, chetty2020income} - meaning our tax return estimates represent a conservative lower bound on the total social return to higher education investment.

\subsection{Limitations and Future Research}

Several caveats apply when interpreting observational cross-country macro panels, and we highlight five main limitations of our study:
\begin{enumerate}
  \item \emph{Reverse causality and fiscal-capacity feedback:} A central concern in macro-fiscal studies is that growing tax capacity enables governments to spend more on higher education rather than the other way round \citep{baumol1967macroeconomics, unesco2024efw, worldbank2024taxing}. We adjust for starting tax-to-GDP, starting income and goods tax shares, lagged GDP growth, predicted continuous funding $\hat{z}(X_{i,t})$, and explicit pre-treatment fiscal trends (Appendix Table~\ref{tab:robust}). Even so, observational conditioning cannot completely rule out multi-decade fiscal feedback loops.
  \item \emph{Unobserved concurrent policy reforms:} Our causal estimates rely on conditional ignorability given the 19 baseline characteristics and four follow-up growth controls \citep{imai2010general, acharya2016explaining}. If a reformist government simultaneously modernises its electronic tax administration \citep{pomeranz2019taking, okunogbe2022technology} and increases university funding within the same five-year wave, part of the estimated slope could reflect that concurrent tax reform.
  \item \emph{Skilled emigration (brain drain):} In low- and lower-middle-income countries, a substantial share of university graduates emigrate to higher-income economies \citep{beine2008brain}. Because emigrants pay direct income taxes in their destination country whilst sending untaxed remittances home (which are spent on domestic goods subject to VAT), high skilled emigration can dampen domestic direct income tax returns in smaller developing nations.
  \item \emph{Central government reporting, five-year wave coarseness, and overlapping horizons:} World Bank fiscal series primarily record central government revenue, which can understate state or provincial income taxes in federal countries. In addition, averaging annual data into five-year waves smooths out exact year-to-year budgetary timing, and consecutive 10-year forward differences ($t \to t+2$ and $t+1 \to t+3$) share a five-year window, which we verify via non-overlapping decadal waves (\texttt{RC2}), country-clustered posterior probabilities ($P_{\text{clu}}$), and country-clustered frequentist causal forests (\texttt{B7}).
  \item \emph{Out-of-sample level noise, interval calibration, and career extrapolation assumptions:} Modelling university funding as a zero-centred continuous variable $Z_1 \in [-1.5, +1.5]$ (winsorised at $\pm 15$ pp) avoids the artificial cliff-edge of a binary cutoff, so a country sitting just $0.31$ pp below its income-group median (such as New Zealand at $26.1\%$ vs the $26.4\%$ high-income median) has a tiny gap cost (\texteuro$0.04$ billion per year) and a proportionally tiny gap return (\texteuro$0.004$ billion per year). Comparing a single entering cohort's multi-year degree input cost against its cumulative 40-year career tax output (Section~\ref{sec:roi} and Appendix~\ref{app:policymakers}) resolves the apparent mismatch between annual system-wide tuition costs and single-year Year-10 tax flows in high-enrolment countries such as Ireland (roughly \texteuro$2.4$ billion cohort input vs \texteuro$9.7$ billion career total tax output on $Y_1$, $\text{BCR} = 4.1\times$, $P_{\text{net}} = 0.77$, and \texteuro$2.9$ billion on $Y_2$, $\text{BCR} = 1.2\times$, $P_{\text{net}} = 0.54$). Even so, two strong caveats apply to those currency projections: first, on the 22 unseen holdout countries in Fold 5, out-of-sample predictive skill is negative on aggregate tax-to-GDP changes ($R^2_{\text{OOS}} = -0.34$ on $Y_1$ and $-0.15$ on $Y_2$, vs $+0.14\text{ to }+0.16$ on the tax-mix share $Y_3$), and 95\% holdout interval coverage on $Y_1$ falls to $88.9\%$ under Scaling A ($85.2\%$ under Scaling B), indicating that nominal net-profit probabilities such as $P_{\text{net}} = 0.77$ on $Y_1$ are likely somewhat optimistic. Second, our empirical panel directly observes outcomes up to the $+10$-year horizon, so multiplying the per-cohort Year-10 annual flow by 40 working years is a speculative cohort-accounting projection (which scales from $8.0\times$ undiscounted to $5.7\times$ under a $1.5\%$ net real discount rate, or $4.0\times$ under a conservative two-wave level attribution, and abstracts from long-run general-equilibrium compression of the graduate wage premium).
\end{enumerate}

These limitations point to several promising avenues for future research. First, whilst this paper has focussed on tertiary education because university cohorts enter the labour market within a 5-to-10-year window, extending continuous-treatment multivariate Bayesian causal forests to 15-to-25-year horizons would allow researchers to compare government spending on early childhood, primary, and secondary schooling \citep{hanushek2012better, hendren2020unified, angrist2021measuring} directly against university funding. Second, incorporating bilateral skilled migration data \citep{beine2008brain} directly into the treatment-effect moderators $W$ would quantify how brain drain alters the fiscal return to public university subsidies in developing countries. Third, linking university funding cohorts to administrative employer-employee tax registers within individual countries would make it possible to trace graduate earnings and corporate tax payments over a full 40-year career. To help policy makers and researchers explore our dataset and country-level estimates interactively, we provide a dedicated policy-maker guide in Appendix~\ref{app:policymakers} alongside an interactive web application in the project repository.

\section*{Declarations and Statements}

\paragraph{Dedication.}
This paper is dedicated to the memory of Robert Christopher Parnell.

\paragraph{Author Contributions (CRediT).}
\textbf{Andrew C. Parnell:} Conceptualisation, Data curation, Formal analysis, Methodology, Software, Visualisation, Writing -- original draft, Writing -- review and editing.

\paragraph{Data Availability Statement.}
All raw and processed datasets analysed in this study are derived from publicly available international databases (World Bank WDI, UNESCO UIS, IMF, WGI, SWIID, and OECD) licensed under CC-BY-4.0. The complete dataset, SHA-256 checksums, and provenance documentation are openly available in the project repository at \url{https://github.com/andrewcparnell/education-tax-bcf}.

\paragraph{Code Availability Statement.}
All R and C++ (\texttt{RcppArmadillo}) source code, data-cleaning scripts, MCMC samplers, diagnostic scripts, interactive exploration app, automated pipeline tests, and figure generation pipelines are openly available under the MIT License at \url{https://github.com/andrewcparnell/education-tax-bcf}. The entire analysis and manuscript can be reproduced end-to-end via \texttt{bash run\_all.sh}. % style-check: ignore

\paragraph{Funding and Open Access.}
Andrew Parnell's work was supported by: the UCD-Met \'Eireann Research Professorship Programme (28-UCDNWPAI); a Research Ireland, Northern Ireland's Department of Agriculture, Environment and Rural Affairs (DAERA), UK Research and Innovation (UKRI) via the International Science Partnerships Fund (ISPF) under Grant number [22/CC/11103] at the Co-Centre for Climate + Biodiversity + Water; Decarb AI, supported by AIB and Research Ireland: an Innovate for Ireland Centre (25/I4I-TC/13542); and Taighde \'Eireann - Research Ireland, under Grant number [23/RC/13506] at the Research Ireland Centre - Rinn Artificial Intelligence. For the purpose of Open Access, the authors have applied a CC BY public copyright licence to any Author Accepted Manuscript version arising from this submission.

\paragraph{Declaration of Generative AI and AI-Assisted Technologies.}
During the preparation of this work, the author used large language models (Anthropic Claude and Google Gemini within the Antigravity agentic coding environment) to assist with assembling the reproducible R and C++ data and MCMC pipeline, running multi-agent adversarial peer-review checks, building the companion interactive web application, and editing the manuscript text. All AI-generated code, statistical outputs, citations, and text were inspected, verified, and revised by the author. A complete stage-by-stage audit of the 840 LLM queries, token volumes, and estimated electrical energy and carbon footprint is provided in Appendix~\ref{app:compute} (Table~\ref{tab:compute}). The author takes full responsibility for the integrity and content of this publication.

\paragraph{Ethics and Conflict of Interest.}
This study analyses aggregate sovereign macroeconomic statistics and involves no human subjects or personally identifiable information, so institutional ethical approval is not required. As disclosed in Section~\ref{sec:lit}, the author co-developed the Multivariate Bayesian Causal Forest methodology \citep{mcjames2024mvbcf} evaluated alongside competing estimators in this study, and declares no financial or other competing interests.

\appendix

\section{Mathematical Specification of the Continuous-Treatment MV-BCF Sampler}
\label{app:math}

Let $Y_{i, t+h} = \bigl(Y_{1,i,t+h}, Y_{2,i,t+h}, Y_{3,i,t+h}, Y_{4,i,t+h}\bigr)^\top \in \mathbb{R}^4$ denote the vector of four disaggregated percentage-point fiscal changes for country $i$ over horizon $h \in \{1, 2\}$ ($+5$ and $+10$ years), and let $Z_{i,t} \in [-1.5, +1.5]$ denote the zero-centred continuous university funding variable. Following \citet{hahn2020bayesian} and \citet{mcjames2024mvbcf}, the continuous-treatment Multivariate Bayesian Causal Forest (MV-BCF) models $Y_{i, t+h}$ as:
\begin{equation}
  Y_{i, t+h} = \sum_{m=1}^{M_\mu} g_\mu\bigl(X_{i,t}, M_{i,t\to t+h}, \hat{z}_{i,t}; T^\mu_m, M^\mu_m\bigr) + Z_{i,t} \sum_{l=1}^{M_\tau} g_\tau\bigl(W_{i,t}; T^\tau_l, M^\tau_l\bigr) + \varepsilon_{i, t+h}, \qquad \varepsilon_{i, t+h} \stackrel{\text{i.i.d.}}{\sim} \mathcal{N}_4(0, \Sigma),
\end{equation}
where $\hat{z}_{i,t} = \hat{\mathbb{E}}[Z_{i,t} \mid X_{i,t}]$ is the random forest predicted funding level given the 19 pre-treatment covariates $X_{i,t}$. The prognostic forest uses $M_\mu = 75$ decision trees with depth-penalty prior hyperparameters $(\alpha_\mu = 0.95, \beta_\mu = 2.0)$, and the treatment-effect forest uses $M_\tau = 30$ decision trees with stronger regularisation priors on tree depth $(\alpha_\tau = 0.25, \beta_\tau = 3.0)$ \citep{hahn2020bayesian}. Each prognostic leaf vector $\mu_{m,b} \in \mathbb{R}^4$ and treatment slope leaf vector $\tau_{l,b} \in \mathbb{R}^4$ receives an independent diagonal multivariate Gaussian prior $\mu_{m,b} \sim \mathcal{N}_4(0, \Sigma_{\mu})$ and $\tau_{l,b} \sim \mathcal{N}_4(0, \Sigma_{\tau})$ calibrated to the empirical outcome variances $s_d^2 = \widehat{\text{Var}}(Y_{d,i,t+h})$ ($d=1,\dots,4$) and the treatment range $\Delta_Z = \max(\max_i Z_{i,t} - \min_i Z_{i,t}, 10^{-4})$:
\begin{equation}
  \Sigma_{\mu} = \text{diag}\!\left(\frac{s_1^2}{M_\mu}, \dots, \frac{s_4^2}{M_\mu}\right), \qquad \Sigma_{\tau} = \text{diag}\!\left(\frac{s_1^2}{M_\tau \Delta_Z^2}, \dots, \frac{s_4^2}{M_\tau \Delta_Z^2}\right).
\end{equation}

Let $R^{(l)}_{i} = Y_{i,t+h} - \sum_{m=1}^{M_\mu} g_\mu(X_{i,t}, M_{i,t\to t+h}, \hat{z}_{i,t}; T^\mu_m, M^\mu_m) - Z_{i,t} \sum_{k \ne l} g_\tau(W_{i,t}; T^\tau_k, M^\tau_k)$ denote the four-dimensional partial residual for treatment tree $l$ (and analogously for prognostic tree $m$). Because $Z_{i,t}$ multiplies $\tau_{l,b}$ linearly inside leaf $b$, the conjugate treatment leaf posterior is exact multivariate Gaussian with precision $\Omega_{\tau, b}^{-1} = \Sigma_{\tau}^{-1} + \bigl(\sum_{i \in b} Z_{i,t}^2\bigr)\Sigma^{-1}$ and mean $\Omega_{\tau, b} \Sigma^{-1} \sum_{i \in b} Z_{i,t} R^{(l)}_i$ (with $Z_{i,t} \equiv 1$ and $\Sigma_\mu^{-1}$ for prognostic leaves). During reversible-jump Metropolis-Hastings tree updates \citep{chipman2010bart} with equal proposal probabilities $p_{\text{grow}} = p_{\text{prune}} = 0.5$, a \texttt{GROW} move on a terminal node $\eta$ at depth $d_\eta$ (with splitting probability $p_{\text{split}}(\eta) = \alpha (1 + d_\eta)^{-\beta}$ and minimum node size $n_{\min} = 3$) is accepted with probability $\min(1, r_{\text{grow}})$:
\begin{equation}
  r_{\text{grow}} = \frac{p_{\text{prune}} \, b(T)}{p_{\text{grow}} \, w(T^*)} \cdot \frac{p_{\text{split}}(\eta)\bigl(1 - p_{\text{split}}(\eta_L)\bigr)\bigl(1 - p_{\text{split}}(\eta_R)\bigr)}{1 - p_{\text{split}}(\eta)} \cdot \frac{L(R \mid T^*, \Sigma)}{L(R \mid T, \Sigma)},
\end{equation}
where $b(T)$ is the number of terminal leaves in $T$ and $w(T^*)$ is the number of singly internal (\texttt{nog}) parent nodes whose two children are both terminal leaves in the proposed tree $T^*$ (with the inverse ratio used for \texttt{PRUNE} moves). Given full residuals $E_i = Y_{i,t+h} - \mu_i - \tau_i Z_{i,t}$, the cross-outcome covariance matrix $\Sigma$ receives a conjugate Inverse-Wishart prior $\mathcal{IW}(\nu_0, \Psi_0)$ with $\nu_0 = D + 2 = 6$ and scale matrix $\Psi_0 = \text{diag}(s_1^2, s_2^2, s_3^2, s_4^2)$, and is updated at each Gibbs iteration via $\Sigma \mid E \sim \mathcal{IW}\bigl(\nu_0 + N, \Psi_0 + \sum_{i=1}^N E_i E_i^\top\bigr)$.

\section{Data Dictionary for Outcomes \texorpdfstring{$Y$}{Y}, Funding Measures \texorpdfstring{$Z$}{Z}, Baseline Controls \texorpdfstring{$X$}{X}, Moderators \texorpdfstring{$W$}{W}, and Mediators \texorpdfstring{$M$}{M}}
\label{app:dictionary}

Table~\ref{tab:dictionary} provides the complete data dictionary for all variables entering the continuous-treatment Multivariate Bayesian Causal Forest model in Equation~\eqref{eq:mvbcf}. Specifically, it lists the four jointly modelled forward tax change outcomes ($Y_1$ to $Y_4$), the four continuous university funding measures ($Z_1$ to $Z_4$, each zero-centred on $[-1.5, +1.5]$ and tested in turn), the 19 pre-treatment country covariates ($X_1$ to $X_{19}$) that enter both the random forest model for $\hat{z}(X_{i,t}) = \mathbb{E}[Z_{i,t} \mid X_{i,t}]$ and the 75-tree baseline prognostic forest $\mu(\cdot)$, the eight pre-treatment moderators ($W_1$ to $W_8 \subset X$) that enter the 30-tree treatment-effect forest $\tau(W_{i,t})$, and the four follow-up macroeconomic growth mediators ($M_1$ to $M_4$) included inside $\mu(\cdot)$ in the mediator-removed direct-effect specifications (\texttt{M2a/M2b}). Full variable ranges and missingness statistics across all 62 raw and constructed columns are available in the online GitHub repository.\footnote{\raggedright Complete machine-readable data dictionary: \url{https://github.com/andrewcparnell/education-tax-bcf/blob/main/data/processed/data_dictionary.csv}.}

\begin{table}[htbp]
  \centering
  \caption{Complete data dictionary for the four tax change outcomes ($Y_1$ to $Y_4$), the four zero-centred continuous university funding measures ($Z_1$ to $Z_4 \in [-1.5, +1.5]$), the 19 pre-treatment baseline controls ($X_1$ to $X_{19}$), the eight treatment-effect moderators ($W_1$ to $W_8 \subset X$), and the four follow-up macroeconomic growth mediators ($M_1$ to $M_4$).}
  \label{tab:dictionary}
  \resizebox{\textwidth}{!}{%
  \renewcommand{\arraystretch}{0.96}
  \footnotesize
  \setlength{\tabcolsep}{3.8pt}
  \begin{tabular}{l|l|l|l|l}
    \toprule
    \rowcolor{tabhead}
    \textbf{Symbol} & \textbf{Code Name} & \textbf{Model Role} & \textbf{Definition \& Scaling Units} & \textbf{Primary Source} \\
    \midrule
    \rowcolor{rowodd}
    \textbf{$Y_1$} & \co \texttt{d\_t\{1,2\}\_nonres\_tax\_gdp\_pp} & $Y$ Outcome 1 ($+5\text{y}, +10\text{y}$) & \co $\Delta$ domestic non-resource tax revenue (pp of GDP) & IMF GFS / WDI / GRD \\
    \rowcolor{roweven}
    \textbf{$Y_2$} & \ce \texttt{d\_t\{1,2\}\_inc\_tax\_gdp\_pp} & $Y$ Outcome 2 ($+5\text{y}, +10\text{y}$) & \ce $\Delta$ direct income, profit \& capital gains tax (pp of GDP) & IMF GFS / WDI \\
    \rowcolor{rowodd}
    \textbf{$Y_3$} & \co \texttt{d\_t\{1,2\}\_inc\_tax\_share\_pp} & $Y$ Outcome 3 ($+5\text{y}, +10\text{y}$) & \co $\Delta$ direct income \& profit share of tax revenue (pp) & WDI (\texttt{GC.TAX.YPKG.ZS}) \\
    \rowcolor{roweven}
    \textbf{$Y_4$} & \ce \texttt{d\_t\{1,2\}\_goods\_tax\_share\_pp} & $Y$ Outcome 4 ($+5\text{y}, +10\text{y}$) & \ce $\Delta$ goods \& services (VAT/sales) share of tax revenue (pp) & WDI (\texttt{GC.TAX.GSRV.RV.ZS}) \\
    \hline
    \rowcolor{rowodd}
    \textbf{$Z_1$} & \co \texttt{Z\_cont\_10pp} & Primary Funding $Z_1$ & \co Per-student spend (\% GDP/person) centred on group median (per $+10$ pp) & UNESCO (\texttt{SE.XPD.TERT.PC.ZS}) \\
    \rowcolor{roweven}
    \textbf{$Z_2$} & \ce \texttt{Z\_cont\_delta\_10pp} & Funding Change $Z_2$ & \ce Within-country 5-year change in per-student ratio (per $+10$ pp) & UNESCO (\texttt{SE.XPD.TERT.PC.ZS}) \\
    \rowcolor{rowodd}
    \textbf{$Z_3$} & \co \texttt{Z\_cont\_real\_ppp\_log25pct} & Real PPP Funding $Z_3$ & \co Log real 2021 PPP \$ per student centred on group median (per $+25\%$) & UNESCO \& WDI PPP GDP \\
    \rowcolor{roweven}
    \textbf{$Z_4$} & \ce \texttt{Z\_cont\_budget\_gdp\_025pp} & Total Budget $Z_4$ & \ce Public tertiary budget (\% of GDP) centred on group median (per $+0.25$ pp) & UNESCO (\texttt{SE.XPD.TOTL.GD.ZS}) \\
    \hline
    \rowcolor{rowodd}
    \textbf{$X_1$ ($W_1$)} & \co \texttt{gross\_tert\_enrolment\_pct} & Control $X_1$ \& Moderator $W_1$ & \co Baseline gross tertiary enrolment ratio (\% of age cohort) & UNESCO (\texttt{SE.TER.ENRR}) \\
    \rowcolor{roweven}
    \textbf{$X_2$ ($W_2$)} & \ce \texttt{log\_gdp\_pc\_ppp} & Control $X_2$ \& Moderator $W_2$ & \ce Log baseline real GDP per person (constant 2021 PPP USD) & WDI (\texttt{NY.GDP.PCAP.PP.KD}) \\
    \rowcolor{rowodd}
    \textbf{$X_3$ ($W_3$)} & \co \texttt{gross\_seco\_enrolment\_pct} & Control $X_3$ \& Moderator $W_3$ & \co Baseline gross secondary school enrolment ratio (\% of cohort) & UNESCO (\texttt{SE.SEC.ENRR}) \\
    \rowcolor{roweven}
    \textbf{$X_4$ ($W_4$)} & \ce \texttt{income\_profit\_tax\_share\_rev\_pct} & Control $X_4$ \& Moderator $W_4$ & \ce Baseline direct income, profit \& capital gains share of tax ($L_3$, \%) & WDI (\texttt{GC.TAX.YPKG.ZS}) \\
    \rowcolor{rowodd}
    \textbf{$X_5$ ($W_5$)} & \co \texttt{working\_age\_pop\_15\_64\_pct} & Control $X_5$ \& Moderator $W_5$ & \co Baseline share of population aged 15 to 64 (\%) & WDI (\texttt{SP.POP.1564.TO.ZS}) \\
    \rowcolor{roweven}
    \textbf{$X_6$ ($W_6$)} & \ce \texttt{income\_group} & Control $X_6$ \& Moderator $W_6$ & \ce World Bank income group tier (1 = Low to 4 = High income) & World Bank Metadata \\
    \rowcolor{rowodd}
    \textbf{$X_7$ ($W_7$)} & \co \texttt{is\_oecd} & Control $X_7$ \& Moderator $W_7$ & \co Indicator for OECD member country (0 or 1) & OECD / World Bank \\
    \rowcolor{roweven}
    \textbf{$X_8$ ($W_8$)} & \ce \texttt{wave\_index} & Control $X_8$ \& Moderator $W_8$ & \ce Five-year wave time index (1 = \texttt{1985--1989} to 8 = \texttt{2020--2024}) & Constructed (5y windows) \\
    \rowcolor{rowodd}
    \textbf{$X_9$} & \co \texttt{non\_resource\_tax\_gdp\_pct} & Baseline Control $X_9$ & \co Baseline domestic non-resource tax revenue ($L_1$, \% of GDP) & IMF GFS / WDI / GRD \\
    \rowcolor{roweven}
    \textbf{$X_{10}$} & \ce \texttt{goods\_services\_tax\_share\_rev\_pct} & Baseline Control $X_{10}$ & \ce Baseline goods \& services tax share of revenue ($L_4$, \%) & WDI (\texttt{GC.TAX.GSRV.RV.ZS}) \\
    \rowcolor{rowodd}
    \textbf{$X_{11}$} & \co \texttt{edu\_spend\_total\_gdp\_pct} & Baseline Control $X_{11}$ & \co Baseline total public education spending across all levels (\% GDP) & UNESCO (\texttt{SE.XPD.TOTL.GD.ZS}) \\
    \rowcolor{roweven}
    \textbf{$X_{12}$} & \ce \texttt{gini\_disp\_harmonised} & Baseline Control $X_{12}$ & \ce Baseline harmonised disposable income Gini index (0--100) & SWIID v9.6 (\citealt{solt2020swiid}) \\
    \rowcolor{rowodd}
    \textbf{$X_{13}$} & \co \texttt{life\_expectancy\_birth\_yrs} & Baseline Control $X_{13}$ & \co Baseline life expectancy at birth (years) & WDI (\texttt{SP.DYN.LE00.IN}) \\
    \rowcolor{roweven}
    \textbf{$X_{14}$} & \ce \texttt{log\_population} & Baseline Control $X_{14}$ & \ce Natural log of baseline total national population & WDI (\texttt{SP.POP.TOTL}) \\
    \rowcolor{rowodd}
    \textbf{$X_{15}$} & \co \texttt{unemployment\_rate\_pct} & Baseline Control $X_{15}$ & \co Baseline total unemployment rate (\% of labour force) & WDI (\texttt{SL.UEM.TOTL.ZS}) \\
    \rowcolor{roweven}
    \textbf{$X_{16}$} & \ce \texttt{natural\_resource\_rents\_gdp\_pct} & Baseline Control $X_{16}$ & \ce Baseline total natural resource rents (\% of GDP) & WDI (\texttt{NY.GDP.TOTL.RT.ZS}) \\
    \rowcolor{rowodd}
    \textbf{$X_{17}$} & \co \texttt{trade\_openness\_gdp\_pct} & Baseline Control $X_{17}$ & \co Baseline sum of exports and imports of goods \& services (\% GDP) & WDI (\texttt{NE.TRD.GNFS.ZS}) \\
    \rowcolor{roweven}
    \textbf{$X_{18}$} & \ce \texttt{lag\_gdp\_pc\_growth\_pct} & Baseline Control $X_{18}$ & \ce Lagged (wave $t-1$) annual real GDP per person growth (\%) & WDI (\texttt{NY.GDP.PCAP.KD.ZG}) \\
    \rowcolor{rowodd}
    \textbf{$X_{19}$} & \co \texttt{region} & Baseline Control $X_{19}$ & \co World Bank geographic region (7 global regions) & World Bank Metadata \\
    \hline
    \rowcolor{roweven}
    \textbf{$M_1$} & \ce \texttt{d\_t\{1,2\}\_log\_gdp\_pc\_pct} & Mediator $M_1$ ($+5\text{y}, +10\text{y}$) & \ce Follow-up real GDP per person PPP log growth over $[t, t+h]$ (\%) & WDI (\texttt{NY.GDP.PCAP.PP.KD}) \\
    \rowcolor{rowodd}
    \textbf{$M_2$} & \co \texttt{d\_t\{1,2\}\_tert\_enrol\_pp} & Mediator $M_2$ ($+5\text{y}, +10\text{y}$) & \co Follow-up change in gross tertiary enrolment over $[t, t+h]$ (pp) & UNESCO (\texttt{SE.TER.ENRR}) \\
    \rowcolor{roweven}
    \textbf{$M_3$} & \ce \texttt{d\_t\{1,2\}\_gini\_disp} & Mediator $M_3$ ($+5\text{y}, +10\text{y}$) & \ce Follow-up change in disposable income Gini over $[t, t+h]$ (pts) & SWIID v9.6 (\citealt{solt2020swiid}) \\
    \rowcolor{rowodd}
    \textbf{$M_4$} & \co \texttt{d\_t\{1,2\}\_life\_exp\_yrs} & Mediator $M_4$ ($+5\text{y}, +10\text{y}$) & \co Follow-up change in life expectancy at birth over $[t, t+h]$ (years) & WDI (\texttt{SP.DYN.LE00.IN}) \\
    \bottomrule
  \end{tabular}%
  }
\end{table}
\FloatBarrier

\section{Subgroup Breakdowns, Method Benchmarks, and Robustness Checks}
\label{app:benchmarks}

In this appendix we present and interpret the detailed subgroup slopes, competing estimator benchmarks, sensitivity checks, and country-level cost-return calculations referenced in Section~\ref{sec:results}. Table~\ref{tab:subgroups} reports the $+10$-year subgroup posterior mean slopes and 95\% credible intervals per $+10$ pp funding increase ($Z_1$) across starting university enrolment tiers ($N = 244$ strictly observed enrolment rows out of $283$; the remaining 39 waves receive training-median enrolment in the tree splits and have mean slopes of $+0.24$ pp on $Y_1$ and $+0.70$ pp on $Y_3$), all four World Bank income groups ($N = 283$), and the five OECD financing regimes ($N = 119$ OECD country-waves). Rows 1 to 7 of Table~\ref{tab:subgroups} aggregate the country-wave posterior draws $\tau(W_{i,t})$ directly from our primary $N = 283$ global model (\texttt{M2b}), whereas Rows 8 to 12 come from a dedicated OECD-only sub-model ($N = 119$, 37 OECD countries, $M_\mu = 50, M_\tau = 25$) that adds the OECD public tertiary funding share to $X_{i,t}$ and $W_{i,t}$. When the global $N = 283$ model is averaged across the 119 OECD country-waves, it predicts $+0.23$ pp on $Y_1$, $+0.08$ pp on $Y_2$, and $+0.75$ pp on $Y_3$; when the sample is restricted solely to the 119 high-tax OECD waves in Rows 8 to 12, the strong regularisation prior ($\alpha_\tau = 0.25, \beta_\tau = 3.0$) shrinks all five OECD tuition regimes toward a common within-OECD slope of $-0.14$ to $-0.17$ pp on $Y_1$, $+0.04$ to $+0.09$ pp on $Y_2$, and $+1.05$ to $+1.14$ pp ($P = 0.93\text{--}0.95$) on $Y_3$.

\begin{table}[htbp]
  \centering
  \caption{Subgroup $+10$-year posterior mean continuous slopes (in percentage points, pp per $+10$ pp university spending increase $Z_1$, with 95\% CIs on the second line of each cell and posterior probabilities $P = \Pr(\tau > 0 \mid \text{data})$) across starting university enrolment tiers ($N = 244$ strictly observed rows from the $N = 283$ global model), World Bank income groups ($N = 283$ global model), and OECD financing regimes ($N = 119$ separate OECD sub-model). Subgroups with $N \le 20$ are small-sample exploratory strata.}
  \label{tab:subgroups}
  \resizebox{\textwidth}{!}{%
  \renewcommand{\arraystretch}{1.12}
  \begin{tabular}{l|c|c|c|c}
    \toprule
    \rowcolor{tabhead}
    \textbf{Subgroup ($+10$-Year Horizon, $Z_1$)} & \textbf{Waves (Countries)} & \textbf{Total Tax} & \textbf{Direct Income} & \textbf{Direct Income} \\
    \rowcolor{tabhead}
    & & \textbf{/ GDP ($Y_1$, pp)} & \textbf{Tax / GDP ($Y_2$, pp)} & \textbf{Tax Share ($Y_3$, pp)} \\
    \midrule
    \rowcolor{rowodd}
    \textbf{1. Elite Enrolment ($<20\%$)} & \co 74 (43) & \textbf{$+0.34$ ($P=0.93$)} & \co \textbf{$+0.17$ ($P=0.91$)} & \textbf{$+0.64$ ($P=0.91$)} \\
    \rowcolor{rowodd}
                                               & \co & {\scriptsize $[-0.11, +0.83]$} & \co {\scriptsize $[-0.09, +0.43]$} & {\scriptsize $[-0.21, +1.42]$} \\
    \hline
    \rowcolor{roweven}
    \textbf{2. Massifying ($20\text{--}50\%$)} & \ce 73 (42) & $+0.23$ ($P=0.89$) & \ce $+0.12$ ($P=0.88$) & \textbf{$+0.75$ ($P=0.97$)} \\
    \rowcolor{roweven}
                                               & \ce & {\scriptsize $[-0.16, +0.69]$} & \ce {\scriptsize $[-0.09, +0.32]$} & {\scriptsize $[-0.03, +1.47]$} \\
    \hline
    \rowcolor{rowodd}
    \textbf{3. High-Participation ($\ge 50\%$)}& \co 97 (44) & $+0.17$ ($P=0.80$) & \co $+0.10$ ($P=0.81$) & \textbf{$+0.73$ ($P=0.95$)} \\
    \rowcolor{rowodd}
                                               & \co & {\scriptsize $[-0.25, +0.66]$} & \co {\scriptsize $[-0.13, +0.32]$} & {\scriptsize $[-0.16, +1.51]$} \\
    \hline
    \rowcolor{roweven}
    \textbf{4. Low Income (exploratory)} & \ce 15 (9) & \textbf{$+0.48$ ($P=0.91$)} & \ce \textbf{$+0.24$ ($P=0.91$)} & $+0.16$ ($P=0.69$) \\
    \rowcolor{roweven}
                                               & \ce & {\scriptsize $[-0.24, +1.30]$} & \ce {\scriptsize $[-0.12, +0.76]$} & {\scriptsize $[-1.96, +1.36]$} \\
    \hline
    \rowcolor{rowodd}
    \textbf{5. Lower Middle Income} & \co 47 (21) & \textbf{$+0.32$ ($P=0.93$)} & \co \textbf{$+0.16$ ($P=0.91$)} & \textbf{$+0.74$ ($P=0.96$)} \\
    \rowcolor{rowodd}
                                               & \co & {\scriptsize $[-0.11, +0.80]$} & \co {\scriptsize $[-0.09, +0.42]$} & {\scriptsize $[-0.07, +1.55]$} \\
    \hline
    \rowcolor{roweven}
    \textbf{6. Upper Middle Income} & \ce 76 (35) & $+0.23$ ($P=0.89$) & \ce \textbf{$+0.13$ ($P=0.90$)} & \textbf{$+0.84$ ($P=0.98$)} \\
    \rowcolor{roweven}
                                               & \ce & {\scriptsize $[-0.16, +0.68]$} & \ce {\scriptsize $[-0.08, +0.33]$} & {\scriptsize $[+0.09, +1.57]$} \\
    \hline
    \rowcolor{rowodd}
    \textbf{7. High Income} & \co 145 (47) & $+0.19$ ($P=0.82$) & \co $+0.10$ ($P=0.81$) & \textbf{$+0.68$ ($P=0.94$)} \\
    \rowcolor{rowodd}
                                               & \co & {\scriptsize $[-0.23, +0.66]$} & \co {\scriptsize $[-0.14, +0.31]$} & {\scriptsize $[-0.17, +1.44]$} \\
    \hline
    \rowcolor{roweven}
    \textbf{8. OECD Nordic High-Public} & \ce 20 (5) & $-0.17$ ($P=0.23$) & \ce $+0.04$ ($P=0.57$) & \textbf{$+1.13$ ($P=0.94$)} \\
    \rowcolor{roweven}
                                               & \ce & {\scriptsize $[-0.62, +0.38]$} & \ce {\scriptsize $[-0.34, +0.49]$} & {\scriptsize $[-0.18, +2.51]$} \\
    \hline
    \rowcolor{rowodd}
    \textbf{9. OECD Low-Tuition Public} & \co 46 (15) & $-0.15$ ($P=0.26$) & \co $+0.05$ ($P=0.58$) & \textbf{$+1.05$ ($P=0.93$)} \\
    \rowcolor{rowodd}
                                               & \co & {\scriptsize $[-0.58, +0.36]$} & \co {\scriptsize $[-0.33, +0.47]$} & {\scriptsize $[-0.26, +2.38]$} \\
    \hline
    \rowcolor{roweven}
    \textbf{10. OECD Hybrid Public-Fee} & \ce 29 (8) & $-0.15$ ($P=0.25$) & \ce $+0.04$ ($P=0.56$) & \textbf{$+1.11$ ($P=0.95$)} \\
    \rowcolor{roweven}
                                               & \ce & {\scriptsize $[-0.60, +0.35]$} & \ce {\scriptsize $[-0.33, +0.45]$} & {\scriptsize $[-0.17, +2.45]$} \\
    \hline
    \rowcolor{rowodd}
    \textbf{11. OECD High-Tuition High-Sub.} & \co 11 (5) & $-0.16$ ($P=0.24$) & \co $+0.05$ ($P=0.57$) & \textbf{$+1.14$ ($P=0.94$)} \\
    \rowcolor{rowodd}
                                               & \co & {\scriptsize $[-0.63, +0.39]$} & \co {\scriptsize $[-0.34, +0.50]$} & {\scriptsize $[-0.16, +2.55]$} \\
    \hline
    \rowcolor{roweven}
    \textbf{12. OECD High-Tuition Low-Sub.} & \ce 13 (4) & $-0.14$ ($P=0.27$) & \ce $+0.09$ ($P=0.64$) & \textbf{$+1.09$ ($P=0.94$)} \\
    \rowcolor{roweven}
                                               & \ce & {\scriptsize $[-0.59, +0.37]$} & \ce {\scriptsize $[-0.31, +0.49]$} & {\scriptsize $[-0.30, +2.46]$} \\
    \bottomrule
  \end{tabular}%
  }
\end{table}

Table~\ref{tab:benchmarks} compares our primary joint four-outcome continuous MV-BCF model at the $+10$-year horizon ($N = 283$, $Z_1$) against a specification that omits follow-up macroeconomic growth controls (Total Effect \texttt{M1b}), three Bayesian tree ablations (\texttt{B1} separate univariate BCFs, \texttt{B2} MV-BCF without predicted funding $\hat{z}(X)$, and \texttt{B3} single-forest BART), and four frequentist estimators (\texttt{B4} linear local projections, \citealt{jorda2005estimation}; \texttt{B5} country-and-wave two-way fixed effects; \texttt{B6} 5-fold country-blocked cross-fitted double machine learning, \citealt{chernozhukov2018double}; and \texttt{B7} a frequentist nonparametric orthogonal causal forest / R-learner fitted on $W_{i,t}$ using 500-tree random forests with country-clustered standard errors, \citealt{athey2019generalized}). Three methodological lessons stand out from Table~\ref{tab:benchmarks}. First, omitting follow-up growth controls strengthens the 10-year slopes on total non-resource tax/GDP ($Y_1 = +0.30$ pp, $P = 0.95$ vs $+0.24$ pp, $P = 0.89$) and direct income tax/GDP ($Y_2 = +0.16$ pp, $P = 0.96$ vs $+0.12$ pp, $P = 0.90$) whilst preserving a positive shift in the direct income tax share ($Y_3 = +0.54$ pp, $P = 0.92$ vs $+0.71$ pp, $P = 0.96$). Second, amongst the Bayesian tree models (\texttt{B1}--\texttt{B3}), univariate BCFs (\texttt{B1}) and MV-BCF without $\hat{z}(X)$ (\texttt{B2}) yield positive slopes closely aligned with our primary model (with the joint covariance matrix $\Sigma$ in our primary model improving posterior precision on $Y_3$: $\text{SD} = 0.385$ vs $0.462$ pp), whereas single-forest BART (\texttt{B3}) shrinks the treatment slopes toward zero ($+0.03$ pp on $Y_2$ and $+0.34$ pp on $Y_3$), illustrating regularisation-induced confounding when $\mu$ and $\tau$ do not receive separate priors \citep{hahn2020bayesian}. Third, across the frequentist estimators (\texttt{B4}--\texttt{B7}), the nonparametric orthogonal causal forest (\texttt{B7}) closely matches our Bayesian MV-BCF slopes ($+0.25$ pp on $Y_1$, $P_{\text{freq}} = 0.92$; $+0.11$ pp on $Y_2$, $P_{\text{freq}} = 0.87$; $+0.50$ pp on $Y_3$, $P_{\text{freq}} = 0.90$; and $-0.44$ pp on $Y_4$, $P_{\text{freq}} = 0.15$), whereas linear two-way fixed effects (\texttt{B5}) is noisy on $Y_1$ ($-0.21$ pp) and linear ridge-based double machine learning (\texttt{B6}) attenuates $Y_3$ ($+0.07$ pp) because it imposes a constant linear outcome model across heterogeneous development tiers.

\begin{table}[htbp]
  \centering
  \caption{Method benchmark and ablation comparison for a $+10$ pp university spending increase ($Z_1$, Scaling A) at the $+10$-year horizon on the identical common sample ($N = 283$, 112 countries). Posterior means, 95\% Bayesian CIs, and posterior probabilities $P$ are shown for the Bayesian tree specifications; frequentist point estimates, 95\% confidence intervals, and one-sided $p$-value complements $P_{\text{freq}} = 1 - p_{\text{one-sided}}$ are shown for \texttt{B4}--\texttt{B7}.}
  \label{tab:benchmarks}
  \resizebox{\textwidth}{!}{%
  \begin{tabular}{l|c|c|c|c}
    \toprule
    \rowcolor{tabhead}
    \textbf{Estimator / Specification ($+10\text{y}$, $Z_1$)} & \textbf{Total Tax} & \textbf{Direct Income} & \textbf{Direct Income} & \textbf{Goods Tax} \\
    \rowcolor{tabhead}
    & \textbf{/ GDP ($Y_1$, pp)} & \textbf{Tax / GDP ($Y_2$, pp)} & \textbf{Tax Share ($Y_3$, pp)} & \textbf{Share ($Y_4$, pp)} \\
    \midrule
    \rowcolor{rowodd}
    \textbf{Primary: Joint MV-BCF (Full Controls, \texttt{M2b})} & \co $+0.24$ ($P=0.89$) & \textbf{$+0.12$ ($P=0.90$)} & \co \textbf{$+0.71$ ($P=0.96$)} & $-0.42$ ($P=0.14$) \\
    \rowcolor{rowodd}
                                                                & \co {\scriptsize $[-0.15, +0.70]$} & {\scriptsize $[-0.08, +0.33]$} & \co {\scriptsize $[-0.08, +1.42]$} & {\scriptsize $[-1.19, +0.34]$} \\
    \hline
    \rowcolor{roweven}
    \textbf{MV-BCF Without Growth Controls (\texttt{M1b})} & \ce \textbf{$+0.30$ ($P=0.95$)} & \textbf{$+0.16$ ($P=0.96$)} & \ce \textbf{$+0.54$ ($P=0.92$)} & $-0.65$ ($P=0.04$) \\
    \rowcolor{roweven}
                                                                & \ce {\scriptsize $[-0.05, +0.65]$} & {\scriptsize $[-0.02, +0.34]$} & \ce {\scriptsize $[-0.19, +1.33]$} & {\scriptsize $[-1.45, +0.10]$} \\
    \hline
    \rowcolor{rowodd}
    \textbf{\texttt{B1}: Univariate BCFs ($D=1$)} & \co \textbf{$+0.45$ ($P=0.97$)} & \textbf{$+0.17$ ($P=0.94$)} & \co \textbf{$+0.71$ ($P=0.93$)} & $-0.47$ ($P=0.15$) \\
    \rowcolor{rowodd}
                                                                & \co {\scriptsize $[-0.004, +0.86]$} & {\scriptsize $[-0.03, +0.39]$} & \co {\scriptsize $[-0.21, +1.63]$} & {\scriptsize $[-1.38, +0.44]$} \\
    \hline
    \rowcolor{roweven}
    \textbf{\texttt{B2}: MV-BCF Without $\hat{z}(X)$} & \ce \textbf{$+0.34$ ($P=0.95$)} & \textbf{$+0.14$ ($P=0.90$)} & \ce \textbf{$+0.59$ ($P=0.94$)} & $-0.51$ ($P=0.10$) \\
    \rowcolor{roweven}
                                                                & \ce {\scriptsize $[-0.05, +0.73]$} & {\scriptsize $[-0.08, +0.35]$} & \ce {\scriptsize $[-0.11, +1.35]$} & {\scriptsize $[-1.23, +0.34]$} \\
    \hline
    \rowcolor{rowodd}
    \textbf{\texttt{B3}: Single-Forest BART} & \co $+0.15$ ($P=0.67$) & $+0.03$ ($P=0.57$) & \co $+0.34$ ($P=0.68$) & $-0.09$ ($P=0.43$) \\
    \rowcolor{rowodd}
                                                                & \co {\scriptsize $[-0.53, +0.85]$} & {\scriptsize $[-0.30, +0.38]$} & \co {\scriptsize $[-0.93, +1.62]$} & {\scriptsize $[-1.39, +1.08]$} \\
    \hline
    \rowcolor{roweven}
    \textbf{\texttt{B4}: Linear Local Projections} & \ce \textbf{$+0.59$ ($P_{\text{freq}}=0.99$)} & \textbf{$+0.18$ ($P_{\text{freq}}=0.90$)} & \ce $+0.51$ ($P_{\text{freq}}=0.82$) & $+0.45$ ($P_{\text{freq}}=0.78$) \\
    \rowcolor{roweven}
                                                                & \ce {\scriptsize $[+0.07, +1.12]$} & {\scriptsize $[-0.09, +0.44]$} & \ce {\scriptsize $[-0.58, +1.59]$} & {\scriptsize $[-0.70, +1.59]$} \\
    \hline
    \rowcolor{rowodd}
    \textbf{\texttt{B5}: Two-Way Fixed Effects} & \co $-0.21$ ($P_{\text{freq}}=0.30$) & $+0.08$ ($P_{\text{freq}}=0.61$) & \co $+0.50$ ($P_{\text{freq}}=0.70$) & $-0.32$ ($P_{\text{freq}}=0.37$) \\
    \rowcolor{rowodd}
                                                                & \co {\scriptsize $[-1.01, +0.59]$} & {\scriptsize $[-0.50, +0.67]$} & \co {\scriptsize $[-1.40, +2.41]$} & {\scriptsize $[-2.27, +1.63]$} \\
    \hline
    \rowcolor{roweven}
    \textbf{\texttt{B6}: Cross-Fitted Double ML} & \ce \textbf{$+0.41$ ($P_{\text{freq}}=0.95$)} & $+0.06$ ($P_{\text{freq}}=0.69$) & \ce $+0.07$ ($P_{\text{freq}}=0.56$) & $+0.56$ ($P_{\text{freq}}=0.80$) \\
    \rowcolor{roweven}
                                                                & \ce {\scriptsize $[-0.09, +0.91]$} & {\scriptsize $[-0.18, +0.29]$} & \ce {\scriptsize $[-0.94, +1.08]$} & {\scriptsize $[-0.73, +1.84]$} \\
    \hline
    \rowcolor{rowodd}
    \textbf{\texttt{B7}: Orthogonal Causal Forest (R-Learner)} & \co \textbf{$+0.25$ ($P_{\text{freq}}=0.92$)} & $+0.11$ ($P_{\text{freq}}=0.87$) & \co \textbf{$+0.50$ ($P_{\text{freq}}=0.90$)} & $-0.44$ ($P_{\text{freq}}=0.15$) \\
    \rowcolor{rowodd}
                                                                & \co {\scriptsize $[-0.11, +0.62]$} & {\scriptsize $[-0.08, +0.30]$} & \co {\scriptsize $[-0.26, +1.26]$} & {\scriptsize $[-1.26, +0.38]$} \\
    \bottomrule
  \end{tabular}%
  }
\end{table}

Table~\ref{tab:robust} reports 17 sensitivity and falsification checks (\texttt{RC1} to \texttt{RC17}) designed to test every data-construction and modelling decision in our pipeline (Note also that under five-year within-country funding changes $Z_2$ in Section~\ref{sec:unbundling}, where baseline tax levels average $\bar{L}_1 = 18.04\%$ of GDP and $\bar{L}_3 = 34.68\%$ of revenue across $N = 176$, the first-order level differential $\tau_{Y_2} \approx (\bar{L}_1 \tau_{Y_3} + \bar{L}_3 \tau_{Y_1}) / 100$ explains why a $+1.56$ pp shift in $Y_3$ combined with positive within-country covariance between baseline tax capacity $L_{1,i,t}$ and $\tau_{Y_3,i}$ raises $Y_2$ by $+0.37$ pp even when $\tau_{Y_1} = -0.29$ pp). First, the $+10$-year positive slopes on $Y_1$, $Y_2$, and $Y_3$ hold when we use raw unadjusted WDI tax-to-GDP without resource rent deductions (\texttt{RC1}, and $+0.24$ pp on $Y_1$, $P = 0.89$, and $+0.70$ pp on $Y_3$, $P = 0.96$ when the $150\%$ pre-winsorisation cap is omitted in \texttt{RC1b}), collapse the panel into non-overlapping 10-year decadal waves (\texttt{RC2}, $+0.82$ pp on $Y_3$, $P = 0.93$), sweep out time-invariant country fixed effects via within-country demeaning (\texttt{RC3a} at $+5$ years, $+0.44$ pp on $Y_3$, $P = 0.97$; and \texttt{RC3b} at $+10$ years, $+0.12$ pp on $Y_2$, $P = 0.94$, and $+0.62$ pp on $Y_3$, 95\% CI $[+0.01, +1.31]$, $P = 0.98$), or drop all observations with any missing baseline covariate (\texttt{RC4}, $+0.26$ pp on $Y_1$, $P = 0.91$, and $+0.66$ pp on $Y_3$, $P = 0.96$). Second, our pre-treatment falsification test (\texttt{RC5}), which checks whether university funding $Z_1$ at wave $t$ ``predicts'' prior tax changes over $t-1 \to t$, yields near-zero slopes ($+0.04$ pp on $Y_2$, $P = 0.67$, and $+0.001$ pp on $Y_3$, $P = 0.51$), reassuring us that treatment is not tracking pre-existing fiscal booms. Third, positive 10-year effects hold when we use a more diffuse $\tau$ tree prior (\texttt{RC6}, $+0.30$ pp on $Y_1$, $P = 0.96$, and $+0.16$ pp on $Y_2$, $P = 0.97$), perturb SWIID Gini inequality with its standard errors (\texttt{RC7}), use row-wise out-of-bag random forest predictions for $\hat{z}(X)$ (\texttt{RC8}), apply two-stage sequential g-estimation (\texttt{RC9}; \citealt{acharya2016explaining}, where Stage 1 regresses $Y_{i,t+2}$ on $M_{i,t\to t+2}$, $Z_{1,i,t}$, and $X_{i,t}$ to demediate the outcome $\tilde{Y}_{i,t+2} = Y_{i,t+2} - M_{i,t\to t+2}\hat{\gamma}$ before fitting MV-BCF in Stage 2, yielding $+0.46$ pp on $Y_1$, $P = 0.99$, $+0.20$ pp on $Y_2$, $P = 0.97$, and $+0.85$ pp on $Y_3$, $P = 0.98$), control for lagged pre-treatment tax trends $\Delta Y_{t-1\to t}$ (\texttt{RC10}), or replace the goods tax share with goods tax as a percentage of GDP (\texttt{RC11}). Fourth, restricting the sample to pre-COVID waves before 2020 (\texttt{RC12}), fitting strictly on the 80\% training countries (\texttt{RC13}, $+0.42$ pp on $Y_1$, $P = 0.99$, and $+0.22$ pp on $Y_2$, $P = 0.98$), using the legacy binary median cutoff (\texttt{RC14}, $Z \in \{0, 1\}$), feeding the identical covariate matrix into both $\mu(\cdot)$ and $\tau(\cdot)$ (\texttt{RC15}, $W = X$, $+0.33$ pp on $Y_1$, $P = 0.97$, and $+0.51$ pp on $Y_3$, $P = 0.91$), centring treatment at zero in $[-1.5, +1.5]$ (\texttt{RC16}, $+0.32$ pp on $Y_1$, $P = 0.93$, $+0.14$ pp on $Y_2$, $P = 0.90$, and $+0.50$ pp on $Y_3$, $P = 0.89$), or estimating the aggregate university budget share $Z_4$ on the exact overlapping sample with $Z_1$ and $Z_3$ (\texttt{RC17}, $N = 272$, which still yields negative tax-to-GDP slopes of $-0.34$ pp on $Y_1$ and $-0.15$ pp on $Y_2$, confirming that $Z_4$'s weak tax-to-GDP return is not an artefact of sample composition) all reinforce our conclusions.

\begin{table}[htbp]
  \centering
  \caption{Robustness, identification, and falsification checks across the primary tax outcomes ($Y_1$, $Y_2$, and $Y_3$, with 95\% CIs on the second line of each cell and posterior probabilities $P = \Pr(\tau > 0 \mid \text{data})$). Checks \texttt{RC1}--\texttt{RC13}, \texttt{RC15}, and \texttt{RC16} report continuous slopes per $+10$ pp funding increase ($Z_1$); \texttt{RC14} reports the legacy binary median-cutoff contrast ($Z \in \{0, 1\}$); and \texttt{RC17} reports the aggregate university budget share ($Z_4$) on the common sample.}
  \label{tab:robust}
  \resizebox{\textwidth}{!}{%
  \renewcommand{\arraystretch}{1.04}
  \begin{tabular}{l|c|c|c|c}
    \toprule
    \rowcolor{tabhead}
    \textbf{Sensitivity Check} & \textbf{Horizon ($N$, Countries)} & \textbf{Total Tax} & \textbf{Direct Income} & \textbf{Direct Income} \\
    \rowcolor{tabhead}
    & & \textbf{/ GDP ($Y_1$, pp)} & \textbf{Tax / GDP ($Y_2$, pp)} & \textbf{Tax Share ($Y_3$, pp)} \\
    \midrule
    \rowcolor{rowodd}
    \textbf{\texttt{RC1}. Unadjusted WDI Tax / GDP} & \co $+10\text{y}$ (283, 112) & $+0.25$ ($P=0.89$) & \co \textbf{$+0.15$ ($P=0.92$)} & $+0.45$ ($P=0.84$) \\
    \rowcolor{rowodd}
                                                      & \co & {\scriptsize $[-0.16, +0.63]$} & \co {\scriptsize $[-0.07, +0.38]$} & {\scriptsize $[-0.34, +1.28]$} \\
    \hline
    \rowcolor{roweven}
    \textbf{\texttt{RC2}. Non-Overlapping 10y Waves} & \ce $+10\text{y}$ (154, 105) & $+0.32$ ($P=0.89$) & \ce $+0.12$ ($P=0.79$) & \textbf{$+0.82$ ($P=0.93$)} \\
    \rowcolor{roweven}
                                                      & \ce & {\scriptsize $[-0.17, +0.85]$} & \ce {\scriptsize $[-0.16, +0.40]$} & {\scriptsize $[-0.30, +1.83]$} \\
    \hline
    \rowcolor{rowodd}
    \textbf{\texttt{RC3a}. Country Fixed Effects ($+5\text{y}$)} & \co $+5\text{y}$ (371, 121) & $+0.01$ ($P=0.53$) & \co $+0.06$ ($P=0.86$) & \textbf{$+0.44$ ($P=0.97$)} \\
    \rowcolor{rowodd}
                                                      & \co & {\scriptsize $[-0.21, +0.23]$} & \co {\scriptsize $[-0.05, +0.17]$} & {\scriptsize $[-0.03, +0.90]$} \\
    \hline
    \rowcolor{roweven}
    \textbf{\texttt{RC3b}. Country Fixed Effects ($+10\text{y}$)} & \ce $+10\text{y}$ (283, 112) & $+0.18$ ($P=0.86$) & \ce \textbf{$+0.12$ ($P=0.94$)} & \textbf{$+0.62$ ($P=0.98$)} \\
    \rowcolor{roweven}
                                                      & \ce & {\scriptsize $[-0.17, +0.50]$} & \ce {\scriptsize $[-0.04, +0.26]$} & {\scriptsize $[+0.01, +1.31]$} \\
    \hline
    \rowcolor{rowodd}
    \textbf{\texttt{RC4}. Strictly Observed Covariates}& \co $+10\text{y}$ (238, 107) & \textbf{$+0.26$ ($P=0.91$)} & \co $+0.11$ ($P=0.84$) & \textbf{$+0.66$ ($P=0.96$)} \\
    \rowcolor{rowodd}
                                                      & \co & {\scriptsize $[-0.11, +0.63]$} & \co {\scriptsize $[-0.09, +0.35]$} & {\scriptsize $[-0.06, +1.47]$} \\
    \hline
    \rowcolor{roweven}
    \textbf{\texttt{RC5}. Placebo ($\Delta Y_{t-1\to t}$)}& \ce Pre-$t$ (347, 117) & $+0.09$ ($P=0.74$) & \ce $+0.04$ ($P=0.67$) & $+0.00$ ($P=0.51$) \\
    \rowcolor{roweven}
                                                      & \ce & {\scriptsize $[-0.19, +0.36]$} & \ce {\scriptsize $[-0.11, +0.18]$} & {\scriptsize $[-0.59, +0.56]$} \\
    \hline
    \rowcolor{rowodd}
    \textbf{\texttt{RC6}. Diffuse $\tau$ Prior} & \co $+10\text{y}$ (283, 112) & \textbf{$+0.30$ ($P=0.96$)} & \co \textbf{$+0.16$ ($P=0.97$)} & \textbf{$+0.61$ ($P=0.94$)} \\
    \rowcolor{rowodd}
                                                      & \co & {\scriptsize $[-0.02, +0.67]$} & \co {\scriptsize $[-0.01, +0.34]$} & {\scriptsize $[-0.17, +1.42]$} \\
    \hline
    \rowcolor{roweven}
    \textbf{\texttt{RC7}. SWIID Error Perturbation} & \ce $+10\text{y}$ (283, 112) & \textbf{$+0.33$ ($P=0.95$)} & \ce $+0.14$ ($P=0.89$) & \textbf{$+0.55$ ($P=0.93$)} \\
    \rowcolor{roweven}
                                                      & \ce & {\scriptsize $[-0.06, +0.68]$} & \ce {\scriptsize $[-0.09, +0.35]$} & {\scriptsize $[-0.20, +1.29]$} \\
    \hline
    \rowcolor{rowodd}
    \textbf{\texttt{RC8}. Row-Wise OOB RF $\hat{z}(X)$}& \co $+10\text{y}$ (283, 112) & \textbf{$+0.31$ ($P=0.93$)} & \co \textbf{$+0.13$ ($P=0.90$)} & \textbf{$+0.62$ ($P=0.94$)} \\
    \rowcolor{rowodd}
                                                      & \co & {\scriptsize $[-0.10, +0.69]$} & \co {\scriptsize $[-0.08, +0.33]$} & {\scriptsize $[-0.19, +1.40]$} \\
    \hline
    \rowcolor{roweven}
    \textbf{\texttt{RC9}. Sequential G-Estimation} & \ce $+10\text{y}$ (283, 112) & \textbf{$+0.46$ ($P=0.99$)} & \ce \textbf{$+0.20$ ($P=0.97$)} & \textbf{$+0.85$ ($P=0.98$)} \\
    \rowcolor{roweven}
                                                      & \ce & {\scriptsize $[+0.09, +0.85]$} & \ce {\scriptsize $[-0.01, +0.41]$} & {\scriptsize $[+0.05, +1.66]$} \\
    \hline
    \rowcolor{rowodd}
    \textbf{\texttt{RC10}. Pre-Trend Controls ($\Delta Y_{t-1\to t}$)} & \co $+10\text{y}$ (283, 112) & \textbf{$+0.28$ ($P=0.96$)} & \co \textbf{$+0.16$ ($P=0.94$)} & \textbf{$+0.58$ ($P=0.93$)} \\
    \rowcolor{rowodd}
                                                      & \co & {\scriptsize $[-0.03, +0.61]$} & \co {\scriptsize $[-0.03, +0.36]$} & {\scriptsize $[-0.18, +1.42]$} \\
    \hline
    \rowcolor{roweven}
    \textbf{\texttt{RC11}. Goods Tax / GDP Check} & \ce $+10\text{y}$ (283, 112) & \textbf{$+0.30$ ($P=0.95$)} & \ce \textbf{$+0.15$ ($P=0.92$)} & \textbf{$+0.50$ ($P=0.90$)} \\
    \rowcolor{roweven}
                                                      & \ce & {\scriptsize $[-0.04, +0.63]$} & \ce {\scriptsize $[-0.06, +0.35]$} & {\scriptsize $[-0.30, +1.37]$} \\
    \hline
    \rowcolor{rowodd}
    \textbf{\texttt{RC12}. Pre-COVID Sample ($<2020$)} & \co $+10\text{y}$ (187, 90) & $+0.23$ ($P=0.82$) & \co $+0.08$ ($P=0.70$) & $+0.46$ ($P=0.82$) \\
    \rowcolor{rowodd}
                                                      & \co & {\scriptsize $[-0.27, +0.70]$} & \co {\scriptsize $[-0.20, +0.36]$} & {\scriptsize $[-0.49, +1.44]$} \\
    \hline
    \rowcolor{roweven}
    \textbf{\texttt{RC13}. 80\% Training Set Only} & \ce $+10\text{y}$ (229, 90) & \textbf{$+0.42$ ($P=0.99$)} & \ce \textbf{$+0.22$ ($P=0.98$)} & \textbf{$+0.64$ ($P=0.92$)} \\
    \rowcolor{roweven}
                                                      & \ce & {\scriptsize $[+0.06, +0.79]$} & \ce {\scriptsize $[+0.001, +0.44]$} & {\scriptsize $[-0.31, +1.63]$} \\
    \hline
    \rowcolor{rowodd}
    \textbf{\texttt{RC14}. Legacy Binary Cutoff ($Z \in \{0,1\}$)} & \co $+10\text{y}$ (283, 112) & \textbf{$+0.58$ ($P=0.93$)} & \co \textbf{$+0.27$ ($P=0.92$)} & \textbf{$+1.18$ ($P=0.95$)} \\
    \rowcolor{rowodd}
                                                      & \co & {\scriptsize $[-0.22, +1.37]$} & \co {\scriptsize $[-0.10, +0.68]$} & {\scriptsize $[-0.25, +2.62]$} \\
    \hline
    \rowcolor{roweven}
    \textbf{\texttt{RC15}. Same Covariates in $\mu$ and $\tau$ ($W = X$)} & \ce $+10\text{y}$ (283, 112) & \textbf{$+0.33$ ($P=0.97$)} & \ce $+0.12$ ($P=0.85$) & \textbf{$+0.51$ ($P=0.91$)} \\
    \rowcolor{roweven}
                                                      & \ce & {\scriptsize $[-0.01, +0.68]$} & \ce {\scriptsize $[-0.09, +0.33]$} & {\scriptsize $[-0.23, +1.23]$} \\
    \hline
    \rowcolor{rowodd}
    \textbf{\texttt{RC16}. Zero-Centred Scaling ($Z_1 - 1.5$)} & \co $+10\text{y}$ (283, 112) & \textbf{$+0.32$ ($P=0.93$)} & \co \textbf{$+0.14$ ($P=0.90$)} & $+0.50$ ($P=0.89$) \\
    \rowcolor{rowodd}
                                                      & \co & {\scriptsize $[-0.09, +0.71]$} & \co {\scriptsize $[-0.06, +0.33]$} & {\scriptsize $[-0.31, +1.26]$} \\
    \hline
    \rowcolor{roweven}
    \textbf{\texttt{RC17}. Common-Sample Budget Share ($Z_4$)} & \ce $+10\text{y}$ (272, 109) & $-0.34$ ($P=0.07$) & \ce $-0.15$ ($P=0.10$) & \textbf{$+0.59$ ($P=0.91$)} \\
    \rowcolor{roweven}
                                                      & \ce & {\scriptsize $[-0.72, +0.08]$} & \ce {\scriptsize $[-0.37, +0.08]$} & {\scriptsize $[-0.26, +1.43]$} \\
    \bottomrule
  \end{tabular}%
  }
\end{table}

Finally, we evaluate all 44 countries from our 10-year modelling sample that sit below their income-group training median ($Z_{1,i,t} < 0$) in the most recent treatment wave (\texttt{2015--2019}, available in full in our online GitHub repository\footnote{\raggedright Complete 44-country funding-gap table: \url{https://github.com/andrewcparnell/education-tax-bcf/blob/main/outputs/tables/tableA4_latest_z0_countries_costs_and_returns.csv}.} alongside the cohort-level lookup in Appendix~\ref{app:policymakers}, Table~\ref{tab:policymakers}). Six countries (New Zealand, Iceland, Uruguay, Romania, Italy, and Slovenia) sit within $2.5$ pp of their income-group median ($26.4\%$ for high income), illustrating why our continuous MV-BCF specification is much more realistic for country-level policy evaluation than a binary cutoff model. Because the continuous model scales expected returns smoothly with the size of the funding increase, closing a tiny gap to the median (such as $+0.31$ pp in New Zealand costing \texteuro$0.04$ billion per year or $+0.35$ pp in Iceland costing \texteuro$0.005$ billion per year) yields a proportionally tiny single-year return at Year 10 (\texteuro$+0.004$ billion and \texteuro$+0.001$ billion per year), whereas a full $+10$ pp expansion costs \texteuro$1.16$ billion in New Zealand and \texteuro$0.13$ billion in Iceland and yields $+0.06$ to $+0.09$ pp of GDP in direct income tax at Year 10 (\texteuro$+0.14$ billion and \texteuro$+0.02$ billion per year) alongside a $+0.72$ to $+0.77$ pp shift in $Y_3$ (note that non-OECD high-income city-states Singapore and Macao SAR fall into a distinct terminal leaf in $W_{i,t}$ with log PPP GDP per person above $11.4$ and working-age population share above $73\%$, yielding higher $Y_2$ slopes of $+0.61$ and $+0.38$ pp). For countries sitting roughly $10$ pp below their income-group median (such as Ireland at $17.3\%$ with a $9.1$ pp gap costing $0.39\%$ of GDP or \texteuro$2.16$ billion per year to close vs \texteuro$2.37$ billion for a $+10$ pp increase, South Korea at $14.9\%$, Argentina at $16.7\%$, and Belarus at $17.0\%$), closing the gap and adding $+10$ pp represent comparable cohort investments whose full 40-year career returns are detailed in Appendix~\ref{app:policymakers}.
\FloatBarrier

\section{Continuous Treatment Orthogonalisation, Residual Covariance, and MCMC Diagnostics}
\label{app:diagnostics}

Table~\ref{tab:balance} summarises how our random forest predicted funding level $\hat{z}(X_{i,t}) = \mathbb{E}[Z_{i,t} \mid X_{i,t}]$ orthogonalises the continuous university funding variable against all 19 pre-treatment covariates across the $+10$-year modelling sample ($N = 283$, with missing auxiliary covariates imputed via training-fold medians). Before conditioning on $\hat{z}(X_{i,t})$, country-waves with above-median university funding ($Z_{1,i,t} \ge 0$) have higher total education spending as a share of GDP ($5.26\%$ vs $3.84\%$, raw correlation $r = +0.540$), higher baseline non-resource tax-to-GDP ($18.57\%$ vs $15.65\%$, $r = +0.325$), higher baseline direct income tax share ($36.57\%$ vs $32.93\%$, $r = +0.196$), and slower lagged GDP per person growth ($1.53\%$ vs $3.49\%$, $r = -0.297$) than country-waves below the median ($Z_{1,i,t} < 0$). After residualising $Z_{i,t}$ on the random forest prediction $\hat{z}(X_{i,t})$, the mean absolute correlation between the residual funding level $Z_{i,t} - \hat{z}(X_{i,t})$ and all 19 starting covariates drops by more than 60\% from $0.157$ (maximum $0.540$) to $0.061$ (maximum $0.213$ for Scaling A, and from $0.140$ to $0.056$ for Scaling B; note that because the random forest orthogonalises $Z_{i,t}$ jointly against all 19 correlated predictors rather than one by one, a single variable with a small raw correlation such as gross tertiary enrolment can shift slightly from $-0.090$ to $-0.131$ while the strong confounders are sharply reduced). This confirms that the random forest first stage captures the systematic pre-treatment predictors of funding intensity and that including $\hat{z}(X_{i,t})$ inside $\mu(\cdot)$ orthogonalises the treatment variable against baseline confounders.

\begin{table}[htbp]
  \centering
  \caption{Continuous treatment covariate correlations before and after residualising on the random forest predicted funding level $\hat{z}(X_{i,t}) = \mathbb{E}[Z_{i,t} \mid X_{i,t}]$ across six representative baseline predictors and the 19-covariate summary ($+10$-year sample, $N = 283$). Columns 2 and 3 report subgroup means below and above the Scaling A median ($Z_1 = 0$), Columns 4 and 5 report raw and residual correlations with $Z_1$ (Scaling A), and Columns 6 and 7 report raw and residual correlations with $Z^{(\text{B})}$ (Scaling B).}
  \label{tab:balance}
  \resizebox{\textwidth}{!}{%
  \footnotesize
  \setlength{\tabcolsep}{4.2pt}
  \begin{tabular}{l|c|c|c|c|c|c}
    \toprule
    \rowcolor{tabhead}
    \textbf{Baseline Pre-Treatment Covariate} & \textbf{Below Med.} & \textbf{Above Med.} & \textbf{Raw Corr.} & \textbf{Resid. Corr.} & \textbf{Raw Corr.} & \textbf{Resid. Corr.} \\
    \rowcolor{tabhead}
    & \textbf{($Z_1 < 0$)} & \textbf{($Z_1 \ge 0$)} & \textbf{(Scaling A)} & \textbf{(Scaling A)} & \textbf{(Scaling B)} & \textbf{(Scaling B)} \\
    \midrule
    \rowcolor{rowodd}
    \textbf{Total Education Spend (\% GDP)} & \co 3.84 & 5.26 & \co $+0.540$ & $+0.166$ & \co $+0.516$ & $+0.179$ \\
    \rowcolor{roweven}
    \textbf{Baseline Non-Resource Tax / GDP (\%)} & \ce 15.65 & 18.57 & \ce $+0.325$ & $+0.044$ & \ce $+0.275$ & $+0.037$ \\
    \rowcolor{rowodd}
    \textbf{Lagged GDP per Person Growth (\%)} & \co 3.49 & 1.53 & \co $-0.297$ & $-0.134$ & \co $-0.317$ & $-0.138$ \\
    \rowcolor{roweven}
    \textbf{Baseline Direct Income Tax Share (\%)} & \ce 32.93 & 36.57 & \ce $+0.196$ & $+0.028$ & \ce $+0.176$ & $+0.015$ \\
    \rowcolor{rowodd}
    \textbf{Log Real GDP per Person (PPP)} & \co 9.75 & 9.74 & \co $+0.142$ & $+0.025$ & \co $+0.074$ & $-0.015$ \\
    \rowcolor{roweven}
    \textbf{Gross Tertiary Enrolment (\%)} & \ce 45.33 & 36.23 & \ce $-0.090$ & $-0.131$ & \ce $-0.119$ & $-0.127$ \\
    \midrule
    \rowcolor{tabhead}
    \textbf{All 19 Covariates (Mean / Max $\lvert r \rvert$)} & \textbf{--} & \textbf{--} & \textbf{0.157 / 0.540} & \textbf{0.061 / 0.213} & \textbf{0.140 / 0.516} & \textbf{0.056 / 0.202} \\
    \bottomrule
  \end{tabular}%
  }
\end{table}

Table~\ref{tab:corr} reports the posterior mean residual standard deviations and all six cross-outcome residual correlations from the continuous-treatment MV-BCF model. These posterior correlations show why modelling all four tax outcomes ($Y_1$ to $Y_4$) jointly in a Multivariate BCF is preferable to fitting four independent regressions. Unobserved fiscal shocks to $Y_1$ (total non-resource tax/GDP) and $Y_2$ (direct income tax/GDP) are strongly positively correlated ($r = +0.62$ at $+5$ years and $+0.65$ at $+10$ years), as are shocks to $Y_2$ and $Y_3$ (the direct income tax share, $r = +0.71$ and $+0.75$). Conversely, shocks to $Y_3$ and $Y_4$ (the goods and services tax share) are strongly negatively correlated ($r = -0.46$ at $+5$ years and $-0.59$ at $+10$ years) because the two revenue shares compete within the same national budget. Estimating the full $4 \times 4$ covariance matrix $\Sigma$ via the Inverse-Wishart Gibbs step allows the continuous tree updates to borrow strength across these correlated fiscal innovations.

\begin{table}[htbp]
  \centering
  \caption{Posterior mean cross-outcome residual correlations (with 95\% CIs) and residual standard deviations from the four-outcome continuous-treatment Multivariate Bayesian Causal Forest ($Z_1$, Scaling A).}
  \label{tab:corr}
  \small
  \setlength{\tabcolsep}{5.0pt}
  \resizebox{\textwidth}{!}{%
  \begin{tabular}{l|c|c}
    \toprule
    \rowcolor{tabhead}
    \textbf{Residual Parameter Pair (MV-BCF, $Z_1$)} & \textbf{$+5\text{y}$ Mean [95\% CI]} & \textbf{$+10\text{y}$ Mean [95\% CI]} \\
    \midrule
    \rowcolor{rowodd}
    \textbf{1. Non-Res. Tax / GDP ($Y_1$) $\leftrightarrow$ Income Tax / GDP ($Y_2$)} & \co \textbf{$+0.62$ $[+0.54, +0.69]$} & \textbf{$+0.65$ $[+0.56, +0.73]$} \\
    \rowcolor{roweven}
    \textbf{2. Income Tax / GDP ($Y_2$) $\leftrightarrow$ Income Tax Share ($Y_3$)} & \ce \textbf{$+0.71$ $[+0.65, +0.77]$} & \textbf{$+0.75$ $[+0.67, +0.81]$} \\
    \rowcolor{rowodd}
    \textbf{3. Income Tax Share ($Y_3$) $\leftrightarrow$ Goods Tax Share ($Y_4$)} & \co \textbf{$-0.46$ $[-0.56, -0.36]$} & \textbf{$-0.59$ $[-0.69, -0.48]$} \\
    \rowcolor{roweven}
    \textbf{4. Income Tax / GDP ($Y_2$) $\leftrightarrow$ Goods Tax Share ($Y_4$)} & \ce \textbf{$-0.24$ $[-0.35, -0.12]$} & \textbf{$-0.25$ $[-0.40, -0.11]$} \\
    \rowcolor{rowodd}
    \textbf{5. Non-Res. Tax / GDP ($Y_1$) $\leftrightarrow$ Goods Tax Share ($Y_4$)} & \co $+0.20$ $[+0.07, +0.32]$ & $+0.33$ $[+0.19, +0.46]$ \\
    \rowcolor{roweven}
    \textbf{6. Non-Res. Tax / GDP ($Y_1$) $\leftrightarrow$ Income Tax Share ($Y_3$)} & \ce $-0.03$ $[-0.14, +0.09]$ & $+0.08$ $[-0.07, +0.23]$ \\
    \rowcolor{rowodd}
    \textbf{7. Residual SDs ($\sigma_1, \sigma_2, \sigma_3, \sigma_4$)}& \co 1.80, 1.00, 4.08, 4.21 pp & 2.51, 1.39, 5.14, 4.80 pp \\
    \bottomrule
  \end{tabular}%
  }
\end{table}

\begin{figure}[htbp]
  \centering
  \includegraphics[width=\textwidth]{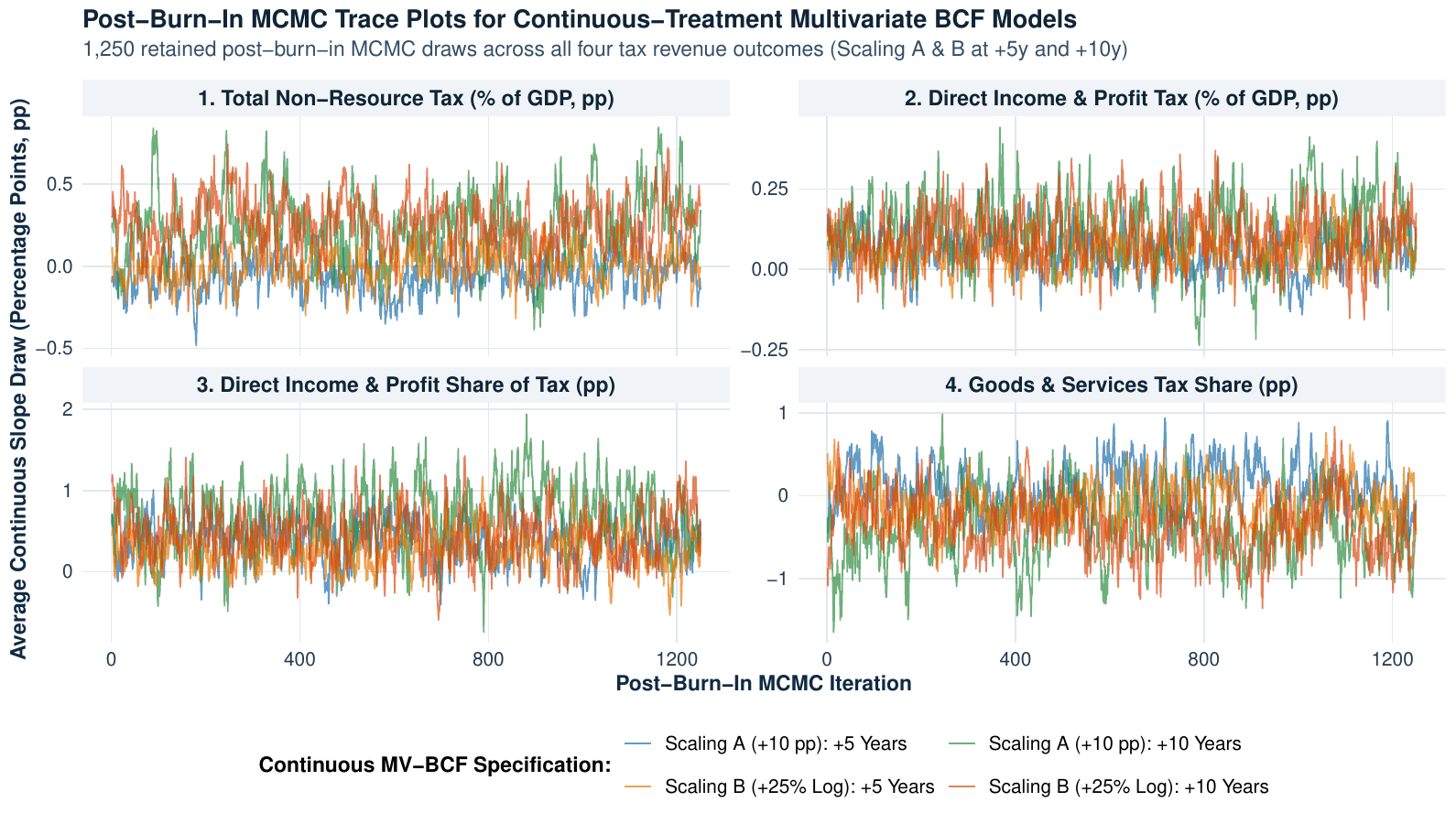}
  \caption{Post-burn-in MCMC trace plots (1,250 retained draws) across all four tax outcomes ($Y_1$ to $Y_4$) from the $+5\text{y}$ and $+10\text{y}$ continuous-treatment Multivariate Bayesian Causal Forest sampler under both Linear Ratio Scaling A ($Z_1$, $+10$ pp) and Relative Logarithmic Scaling B ($+25\%$ log).}
  \label{fig:traces}
\end{figure}

Table~\ref{tab:mcmc} and Figure~\ref{fig:traces} present the post-burn-in MCMC convergence diagnostics and in-sample posterior predictive fit metrics ($R^2_{\text{in}}$) across the $+5$-year and $+10$-year continuous-treatment MV-BCF models (1,250 retained draws per baseline run, supplemented by our four-chain dispersed verification in Footnote~\ref{fn:multichain}). Across all eight horizon-by-outcome parameters in Table~\ref{tab:mcmc}, single-chain split $\hat{R}$ values lie between $0.999$ and $1.030$ with effective sample sizes (ESS) of $65$ to $120$ and Monte Carlo standard errors (MCSE) of $0.006$ to $0.046$ pp (accounting for within-country clustering across repeated waves yields $P_{\text{clu}} = 0.896$ on $Y_1$, $0.896$ on $Y_2$, and $0.961$ on $Y_3$ at $+10$ years). When we pool four independent chains initialised from dispersed seeds ($4 \times 3,000$ iterations, $8,000$ retained draws), between-chain Gelman-Rubin $\hat{R}$ falls to $1.000\text{--}1.006$ with combined $\text{ESS} = 767\text{--}1,932$ ($\text{MCSE}_P \le 0.011$), yielding $+10$-year slopes of $+0.31$ pp on $Y_1$ ($P = 0.94$), $+0.15$ pp on $Y_2$ ($P = 0.93$), $+0.71$ pp on $Y_3$ ($P = 0.96$), and $-0.43$ pp on $Y_4$ ($P = 0.15$). In-sample fit values ($R^2_{\text{in}} = 0.247\text{--}0.641$, with in-sample 95\% predictive interval coverage \texttt{Cov95} of $95.4\%\text{--}97.9\%$ across both horizons and $93.8\%\text{--}95.8\%$ of standardised residuals within $\pm 2\text{ SD}$) reflect the flexibility of the 75 prognostic trees in absorbing baseline variation rather than out-of-sample forecasting power (which is evaluated separately via $R^2_{\text{OOS}}$ in Table~\ref{tab:scaling_ab}). Extended residual checks\footnote{\raggedright Extended wave, autocorrelation, and income-group residual diagnostics in our GitHub repository: \url{https://github.com/andrewcparnell/education-tax-bcf/blob/main/outputs/tables/tableA6_extended_residual_diagnostics.csv}.} further verify that $+10$-year wave-specific mean residuals are close to zero across all five-year periods including the 2008 financial crisis and 2020 COVID-19 window (ANOVA $p > 0.99$ on $Y_1$, $0.47$ on $Y_2$, $0.22$ on $Y_3$, and $0.27$ on $Y_4$), that within-country lag-1 residual autocorrelation is slightly negative at the non-overlapping $+5$-year horizon ($\hat{\rho}_1 = -0.14\text{ to }-0.03$) and $+0.28\text{ to }+0.32$ at the overlapping $+10$-year horizon (addressed via non-overlapping Check \texttt{RC2}), and that $+10$-year residual standard deviations are comparable across low-, lower-middle-, upper-middle-, and high-income groups ($1.74, 2.46, 2.43, 2.28$ pp on $Y_1$; $0.85, 1.19, 1.36, 1.35$ pp on $Y_2$; $4.63, 4.88, 4.71, 4.76$ pp on $Y_3$; and $5.14, 4.22, 4.61, 4.07$ pp on $Y_4$).

\begin{table}[htbp]
  \centering
  \caption{Post-burn-in MCMC convergence diagnostics (split-chain $\hat{R}$, effective sample size ESS, Monte Carlo standard error MCSE, and Geweke $z$-score) and in-sample posterior predictive fit ($R^2_{\text{in}}$, 95\% CI coverage, and share of standardised residuals within $\pm 2$ SD) across the $+5$-year and $+10$-year continuous MV-BCF models ($Z_1$, Scaling A; see Table~\ref{tab:scaling_ab} for out-of-sample $R^2_{\text{OOS}}$).}
  \label{tab:mcmc}
  \resizebox{\textwidth}{!}{%
  \begin{tabular}{l|c|c|c|c|c|c|c|c|c}
    \toprule
    \rowcolor{tabhead}
    \textbf{Model \& Outcome ($Z_1$)} & \textbf{Horizon ($N$)} & \textbf{Mean (SD)} & \textbf{$\hat{R}$} & \textbf{ESS} & \textbf{MCSE} & \textbf{Geweke} & \textbf{$R^2_{\text{in}}$} & \textbf{Cov95} & \textbf{$\lvert e^*\rvert \le 2\text{SD}$} \\
    \midrule
    \rowcolor{rowodd}
    \textbf{1. MV-BCF Tax / GDP ($Y_1$)} & \co $+5\text{y}$ (371) & $-0.055$ (0.113) & \co \textbf{1.026} & 91 & \co 0.012 & $-1.67$ & \co 0.396 & 96.2\% & \co 93.8\% \\
    \rowcolor{roweven}
    \textbf{2. MV-BCF Inc. Tax / GDP ($Y_2$)} & \ce $+5\text{y}$ (371) & $+0.053$ (0.061) & \ce \textbf{1.030} & 92 & \ce 0.006 & $+1.80$ & \ce 0.247 & 96.2\% & \ce 95.4\% \\
    \rowcolor{rowodd}
    \textbf{3. MV-BCF Inc. Tax Share ($Y_3$)} & \co $+5\text{y}$ (371) & $+0.365$ (0.244) & \co \textbf{1.023} & 120 & \co 0.022 & $+0.64$ & \co 0.332 & 95.4\% & \co 94.1\% \\
    \rowcolor{roweven}
    \textbf{4. MV-BCF Goods Share ($Y_4$)} & \ce $+5\text{y}$ (371) & $+0.098$ (0.293) & \ce \textbf{1.011} & 75 & \ce 0.034 & $+0.79$ & \ce 0.426 & 97.0\% & \ce 94.6\% \\
    \rowcolor{rowodd}
    \textbf{5. MV-BCF Tax / GDP ($Y_1$)} & \co $+10\text{y}$ (283) & $+0.236$ (0.200) & \co \textbf{1.003} & 65 & \co 0.025 & $-0.29$ & \co 0.423 & 95.8\% & \co 94.0\% \\
    \rowcolor{roweven}
    \textbf{6. MV-BCF Inc. Tax / GDP ($Y_2$)} & \ce $+10\text{y}$ (283) & $+0.124$ (0.103) & \ce \textbf{1.002} & 109 & \ce 0.010 & $-0.34$ & \ce 0.299 & 96.5\% & \ce 95.4\% \\
    \rowcolor{rowodd}
    \textbf{7. MV-BCF Inc. Tax Share ($Y_3$)} & \co $+10\text{y}$ (283) & $+0.707$ (0.385) & \co \textbf{1.024} & 100 & \co 0.038 & $-1.24$ & \co 0.460 & 96.8\% & \co 94.7\% \\
    \rowcolor{roweven}
    \textbf{8. MV-BCF Goods Share ($Y_4$)} & \ce $+10\text{y}$ (283) & $-0.422$ (0.390) & \ce \textbf{0.999} & 72 & \ce 0.046 & $-1.75$ & \ce 0.641 & 97.9\% & \ce 95.8\% \\
    \bottomrule
  \end{tabular}%
  }
\end{table}

\FloatBarrier

\section{Comparison of Linear Ratio Scaling A and Relative Logarithmic Scaling B}
\label{app:scaling_comparison}

As discussed in Section~\ref{sec:unesco_trap}, the UNESCO per-student university spending ratio can be transformed either on a linear percentage-point scale within each World Bank income group ($Z_1$ or Scaling A: $\Delta Z_1 = +1.0$ per $+10$ pp of GDP per person per student around the training-set group median $m_{g(i,t)}$, winsorised at $\pm 1.5$ units so $Z_{1,i,t} \in [-1.5, +1.5]$ with $\text{SD}(Z_1) = 1.05$) or on a concave relative logarithmic scale (Scaling B: $Z^{(\text{B})}_{i,t} = \text{clip}\bigl(\log(\text{Spend}_{i,t} / m_{g(i,t)}) / \log(1.25), -1.5, +1.5\bigr) \in [-1.5, +1.5]$ with $\text{SD}(Z^{(\text{B})}) = 1.21$, where $\Delta Z^{(\text{B})} = +1.0$ represents a $+25\%$ relative increase in per-student funding). In high- and upper-middle-income countries (where median per-student spending is $26.4\%$ and $27.6\%$ of GDP per person), a $+25\%$ relative increase equals roughly $+6.6$ to $+6.9$ pp of GDP per person. In lower-middle- and low-income countries (where raw medians are $68.9\%$ and $183.8\%$), a $+25\%$ relative increase corresponds to $+17.2$ to $+45.9$ pp and compresses the right tail of the UNESCO ratio distribution via the logarithm, directly embedding diminishing marginal returns to additional percentage points of spending. In the main text we present Scaling A ($Z_1$) because percentage-point increments of GDP per person translate directly into national budget costs; here we compare Scaling A and Scaling B head to head across posterior slopes, in-sample residual diagnostics across $Z \in [-1.5, +1.5]$, 5-fold country-blocked cross-validation, and out-of-sample holdout predictions on the 22 unseen countries in Fold 5.

\begin{table}[htbp]
  \centering
  \caption{Head-to-head comparison of Linear Ratio Scaling A ($Z_1$, per $+10$ pp of GDP per person per student), Relative Logarithmic Scaling B (per $+25\%$ log per-student increase), and the $\tau \equiv 0$ prognostic-only ablation across $+5\text{y}$ and $+10\text{y}$ posterior slopes (with 95\% CIs), in-sample posterior predictive fit ($R^2_{\text{in}}$, CRPS, and 95\% coverage), 5-fold country-blocked CV across all 112 countries ($R^2_{\text{OOS}} = 1 - \text{MSE}/\text{MSE}_{\text{mean}}$, RMSE, and CRPS), and locked holdout validation on the 22 unseen countries in Fold 5 ($R^2_{\text{OOS}}$, RMSE, CRPS, and 95\% coverage).}
  \label{tab:scaling_ab}
  \resizebox{\textwidth}{!}{%
  \renewcommand{\arraystretch}{1.04}
  \footnotesize
  \setlength{\tabcolsep}{3.2pt}
  \begin{tabular}{l|c|c|c|c}
    \toprule
    \rowcolor{tabhead}
    \textbf{Comparison Metric \& Horizon} & \textbf{1. Total Non-Res.} & \textbf{2. Direct Income} & \textbf{3. Direct Income} & \textbf{4. Goods \& Serv.} \\
    \rowcolor{tabhead}
    & \textbf{Tax / GDP ($Y_1$, pp)} & \textbf{Tax / GDP ($Y_2$, pp)} & \textbf{Tax Share ($Y_3$, pp)} & \textbf{Tax Share ($Y_4$, pp)} \\
    \midrule
    \rowcolor{rowodd}
    \textbf{A1. Scaling A Direct ($+5\text{y}$, \texttt{M2a})} & \co $-0.06$ ($P=0.31$) {\scriptsize $[-0.28, +0.18]$} & $+0.05$ ($P=0.81$) {\scriptsize $[-0.07, +0.17]$} & \co \textbf{$+0.36$ ($P=0.93$)} {\scriptsize $[-0.13, +0.81]$} & $+0.10$ ($P=0.62$) {\scriptsize $[-0.47, +0.67]$} \\
    \rowcolor{roweven}
    \textbf{B1. Scaling B Direct ($+5\text{y}$, \texttt{M2c})} & \ce $+0.004$ ($P=0.50$) {\scriptsize $[-0.22, +0.22]$} & $+0.06$ ($P=0.86$) {\scriptsize $[-0.05, +0.17]$} & \ce \textbf{$+0.27$ ($P=0.89$)} {\scriptsize $[-0.15, +0.73]$} & $-0.10$ ($P=0.34$) {\scriptsize $[-0.58, +0.39]$} \\
    \hline
    \rowcolor{rowodd}
    \textbf{A2. Scaling A Direct ($+10\text{y}$, \texttt{M2b})} & \co $+0.24$ ($P=0.89$) {\scriptsize $[-0.15, +0.70]$} & \textbf{$+0.12$ ($P=0.90$)} {\scriptsize $[-0.08, +0.33]$} & \co \textbf{$+0.71$ ($P=0.96$)} {\scriptsize $[-0.08, +1.42]$} & $-0.42$ ($P=0.14$) {\scriptsize $[-1.19, +0.34]$} \\
    \rowcolor{roweven}
    \textbf{B2. Scaling B Direct ($+10\text{y}$, \texttt{M2d})} & \ce \textbf{$+0.26$ ($P=0.95$)} {\scriptsize $[-0.04, +0.56]$} & \textbf{$+0.11$ ($P=0.90$)} {\scriptsize $[-0.06, +0.27]$} & \ce \textbf{$+0.47$ ($P=0.93$)} {\scriptsize $[-0.16, +1.09]$} & $-0.38$ ($P=0.13$) {\scriptsize $[-0.99, +0.33]$} \\
    \hline
    \rowcolor{rowodd}
    \textbf{A3. Scaling A Total ($+10\text{y}$, \texttt{M1b})} & \co \textbf{$+0.30$ ($P=0.95$)} & \textbf{$+0.16$ ($P=0.96$)} & \co \textbf{$+0.54$ ($P=0.92$)} & $-0.65$ ($P=0.04$) \\
    \rowcolor{roweven}
    \textbf{B3. Scaling B Total ($+10\text{y}$, \texttt{M1d})} & \ce \textbf{$+0.21$ ($P=0.93$)} & \textbf{$+0.11$ ($P=0.90$)} & \ce $+0.41$ ($P=0.89$) & $-0.42$ ($P=0.08$) \\
    \hline
    \rowcolor{rowodd}
    \textbf{A4. Scaling A In-Sample ($+10\text{y}$)} & \co $R^2_{\text{in}}=0.42$, $\text{CRPS}=1.27$ ($95.8\%$) & $R^2_{\text{in}}=0.30$, $\text{CRPS}=0.72$ ($96.5\%$) & \co $R^2_{\text{in}}=0.46$, $\text{CRPS}=2.67$ ($96.8\%$) & $R^2_{\text{in}}=0.64$, $\text{CRPS}=2.38$ ($97.9\%$) \\
    \rowcolor{roweven}
    \textbf{B4. Scaling B In-Sample ($+10\text{y}$)} & \ce $R^2_{\text{in}}=0.49$, $\text{CRPS}=1.20$ ($95.8\%$) & $R^2_{\text{in}}=0.29$, $\text{CRPS}=0.72$ ($95.8\%$) & \ce $R^2_{\text{in}}=0.47$, $\text{CRPS}=2.64$ ($97.2\%$) & $R^2_{\text{in}}=0.63$, $\text{CRPS}=2.42$ ($97.9\%$) \\
    \hline
    \rowcolor{rowodd}
    \textbf{A5. Scaling A 5-Fold CV ($N=283$)} & \co $R^2_{\text{OOS}}=-0.07$, $\text{RMSE}=3.15$, $\text{CRPS}=1.64$ & $R^2_{\text{OOS}}=+0.02$, $\text{RMSE}=1.54$, $\text{CRPS}=0.85$ & \co \textbf{$R^2_{\text{OOS}}=+0.10$}, $\text{RMSE}=6.11$, $\text{CRPS}=3.41$ & \textbf{$R^2_{\text{OOS}}=+0.18$}, $\text{RMSE}=6.48$, $\text{CRPS}=3.43$ \\
    \rowcolor{roweven}
    \textbf{B5. Scaling B 5-Fold CV ($N=283$)} & \ce $R^2_{\text{OOS}}=-0.07$, $\text{RMSE}=3.16$, $\text{CRPS}=1.64$ & $R^2_{\text{OOS}}=-0.01$, $\text{RMSE}=1.56$, $\text{CRPS}=0.86$ & \ce $R^2_{\text{OOS}}=+0.04$, $\text{RMSE}=6.32$, $\text{CRPS}=3.51$ & \textbf{$R^2_{\text{OOS}}=+0.15$}, $\text{RMSE}=6.59$, $\text{CRPS}=3.47$ \\
    \rowcolor{rowodd}
    \textbf{C5. $\tau \equiv 0$ Prognostic CV ($N=283$)} & \co $R^2_{\text{OOS}}=-0.08$, $\text{RMSE}=3.16$, $\text{CRPS}=1.64$ & $R^2_{\text{OOS}}=+0.02$, $\text{RMSE}=1.54$, $\text{CRPS}=0.85$ & \co \textbf{$R^2_{\text{OOS}}=+0.10$}, $\text{RMSE}=6.11$, $\text{CRPS}=3.41$ & \textbf{$R^2_{\text{OOS}}=+0.19$}, $\text{RMSE}=6.46$, $\text{CRPS}=3.37$ \\
    \hline
    \rowcolor{roweven}
    \textbf{A6. Scaling A Holdout ($N=54$)} & \ce $R^2_{\text{OOS}}=-0.34$, $\text{RMSE}=3.19$ {\scriptsize ($\text{CRPS}=1.73, 88.9\%$)} & $R^2_{\text{OOS}}=-0.15$, $\text{RMSE}=1.53$ {\scriptsize ($\text{CRPS}=0.86, 94.4\%$)} & \ce \textbf{$R^2_{\text{OOS}}=+0.14$}, $\text{RMSE}=5.78$ {\scriptsize ($\text{CRPS}=3.24, 94.4\%$)} & $R^2_{\text{OOS}}=+0.02$, $\text{RMSE}=5.48$ {\scriptsize ($\text{CRPS}=3.08, 96.3\%$)} \\
    \rowcolor{rowodd}
    \textbf{B6. Scaling B Holdout ($N=54$)} & \co $R^2_{\text{OOS}}=-0.27$, $\text{RMSE}=3.10$ {\scriptsize ($\text{CRPS}=1.68, 85.2\%$)} & $R^2_{\text{OOS}}=-0.11$, $\text{RMSE}=1.50$ {\scriptsize ($\text{CRPS}=0.83, 94.4\%$)} & \co \textbf{$R^2_{\text{OOS}}=+0.16$}, $\text{RMSE}=5.69$ {\scriptsize ($\text{CRPS}=3.15, 96.3\%$)} & $R^2_{\text{OOS}}=-0.13$, $\text{RMSE}=5.88$ {\scriptsize ($\text{CRPS}=3.26, 92.6\%$)} \\
    \bottomrule
  \end{tabular}%
  }
\end{table}

\begin{figure}[htbp]
  \centering
  \includegraphics[width=\textwidth]{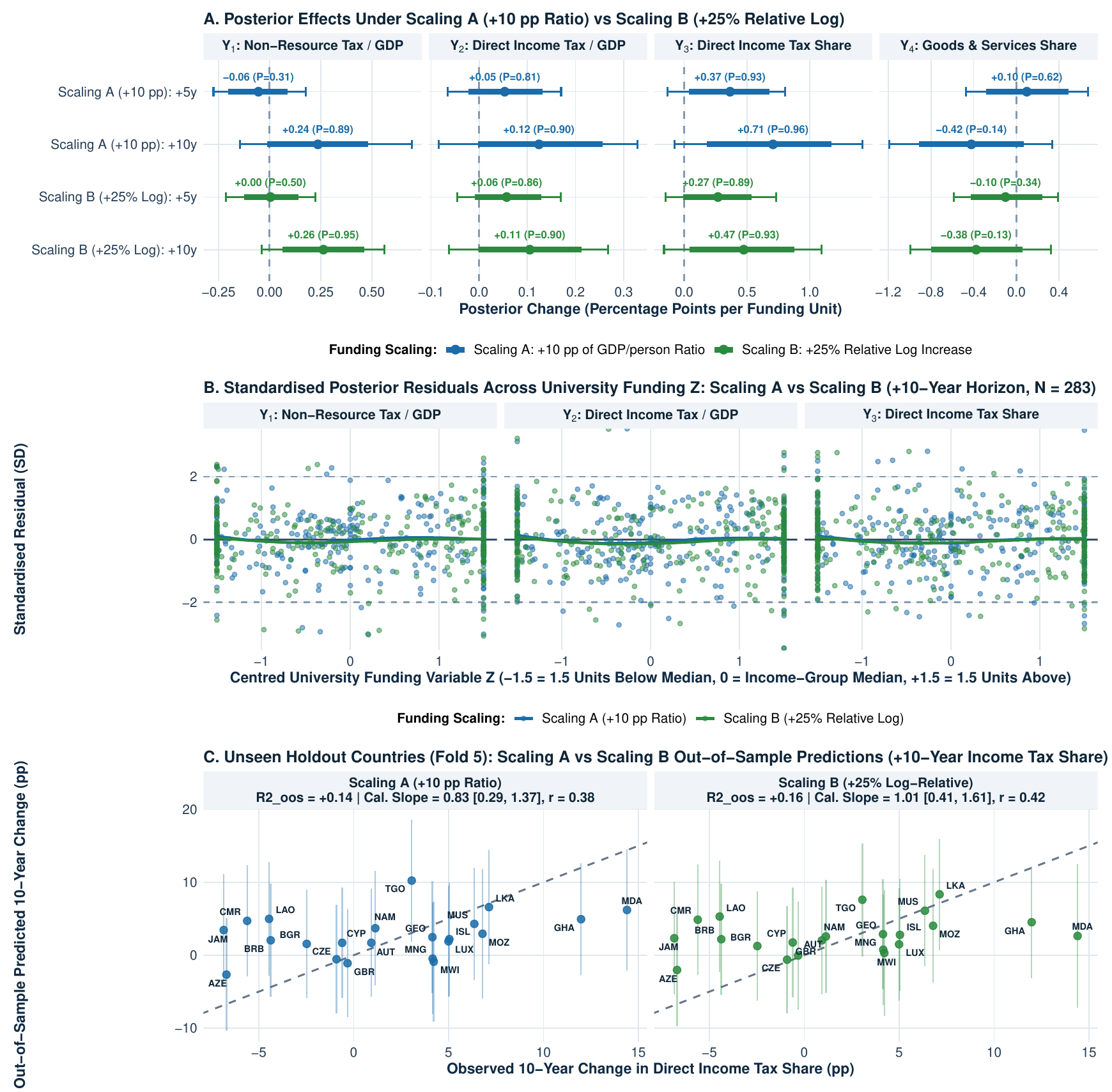}
  \caption{Head-to-head comparison of Linear Ratio Scaling A ($Z_1$, per $+10$ pp of GDP per person per student) vs Relative Logarithmic Scaling B (per $+25\%$ relative per-student increase): (A) $+5\text{y}$ and $+10\text{y}$ posterior slopes with 80\% and 95\% CIs, (B) standardised $+10\text{y}$ posterior residuals across $Z \in [-1.5, +1.5]$ with LOESS smooths for both scalings, and (C) observed (horizontal axis) vs out-of-sample predicted (vertical axis, with vertical 80\% predictive interval bars, $R^2_{\text{OOS}}$, Mincer-Zarnowitz calibration slopes, and Pearson $r$) 10-year changes in the direct income tax share ($Y_3$) across the 22 unseen holdout countries in Fold 5.}
  \label{fig:scaling_ab}
\end{figure}
\FloatBarrier

Table~\ref{tab:scaling_ab} and Figure~\ref{fig:scaling_ab} highlight three insights from comparing the two scalings. First, both scalings tell the exact same substantive story across horizons and outcomes (Rows A1--B3 and Figure~\ref{fig:scaling_ab}A): under both Scaling A and Scaling B, the 5-year effect appears in the direct income tax share ($Y_3 = +0.36$ pp, $P = 0.93$ under Scaling A; $+0.27$ pp, $P = 0.89$ under Scaling B, and $+0.40$ pp, $P = 0.97$ for the total effect \texttt{M1c}), whilst the 10-year horizon shows gains across $Y_1$ ($+0.24$ pp, $P = 0.89$ under Scaling A; $+0.26$ pp, $P = 0.95$ under Scaling B), $Y_2$ ($+0.12$ pp, $P = 0.90$ under Scaling A; $+0.11$ pp, $P = 0.90$ under Scaling B), and $Y_3$ ($+0.71$ pp, $P = 0.96$ under Scaling A; $+0.47$ pp, $P = 0.93$ under Scaling B). Second, in Figure~\ref{fig:scaling_ab}B, the LOESS residual curves across $Z \in [-1.5, +1.5]$ are flat and centred on zero for both scalings, and because a $+10$ pp increment represents a larger proportional step in high- and upper-middle-income countries, Scaling A achieves slightly lower 5-fold cross-validation error across all 112 countries (Rows A5 vs B5: mean CRPS $2.33$ vs $2.37$ pp). Third, because the logarithm in Scaling B compresses the right tail of the UNESCO ratio in low- and lower-middle-income countries, Scaling B achieves higher $+10$-year posterior precision on $Y_1$ ($\text{SD} = 0.154$ vs $0.200$ pp, raising $P$ from $0.89$ to $0.95$) and slightly lower holdout prediction error on the 22 unseen countries in Fold 5 for $Y_1$, $Y_2$, and $Y_3$ (Row B6 and Figure~\ref{fig:scaling_ab}C: CRPS $1.68$ vs $1.73$ pp on $Y_1$, $0.83$ vs $0.86$ pp on $Y_2$, and $3.15$ vs $3.24$ pp on $Y_3$, with Diebold-Mariano $p = 0.82\text{--}0.84$). At the same time, Rows A6 and B6 show that whilst 95\% holdout interval coverage is well calibrated on $Y_2$, $Y_3$, and $Y_4$ ($92.6\%$ to $96.3\%$), it dips to $88.9\%$ under Scaling A and $85.2\%$ under Scaling B on $Y_1$, confirming that unseen countries face wider aggregate fiscal shocks and that nominal $Y_1$ return-on-investment probabilities (such as $P_{\text{net}} = 0.77$ for Ireland) are somewhat optimistic.

\section{Assistance for Policy Makers}
\label{app:policymakers}

\begin{figure}[htbp]
  \centering
  \includegraphics[width=\textwidth]{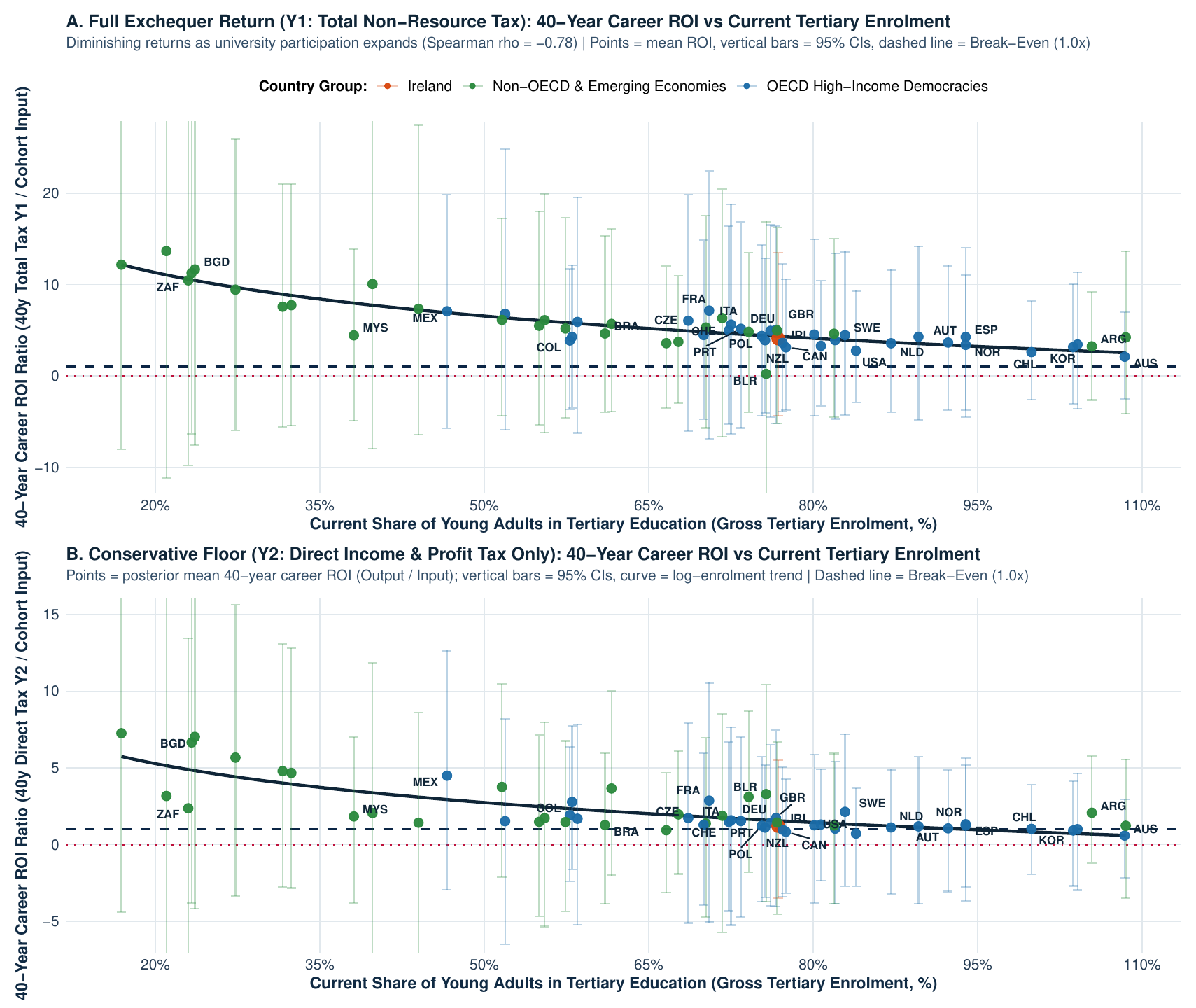}
  \caption{Current share of young adults in tertiary education (gross tertiary enrolment \%, horizontal axis; note UNESCO gross enrolment can slightly exceed $100\%$ where mature, postgraduate, or international students enrol alongside the five-year post-secondary age cohort) vs estimated 40-year career Return on Investment ratio ($\text{ROI}$, 40-year career tax output divided by cohort degree input, vertical axis, with points showing posterior means, vertical bars showing 95\% credible intervals, and solid curves showing the log-enrolment diminishing-returns trend) for (A) Total Non-Resource Tax ($Y_1$, full Exchequer return) and (B) Direct Income and Profit Tax ($Y_2$, conservative floor). The horizontal dashed line marks break-even ($\text{ROI} = 1.0\times$), and the dotted line marks zero return ($0.0\times$). Very low-enrolment ($<12\%$) or extreme-enrolment ($>112\%$) cases and city-state outliers are omitted for legibility and reported in full in the online 70-country GitHub table (Footnote~\ref{fn:policymakers}).}
  \label{fig:roi_scatter}
\end{figure}

To supplement the cohort return-on-investment analysis in Section~\ref{sec:roi}, Figure~\ref{fig:roi_scatter} plots each country's current share of young adults in tertiary education (gross tertiary enrolment, horizontal axis) against its estimated 40-year career return-on-investment ratio ($\text{ROI}$, defined as 40-year career tax output divided by cohort degree input, with vertical bars indicating 95\% credible intervals and a log-enrolment trend curve illustrating diminishing returns, Spearman $\rho = -0.78$) for both Total Non-Resource Tax ($Y_1$, Panel A) and Direct Income Tax ($Y_2$, Panel B), while Table~\ref{tab:policymakers} provides a compact country-by-country lookup guide for 20 representative high-, middle-, and low-income economies from the 70 countries with complete \texttt{2020--2024} macroeconomic, demographic, enrolment, and fiscal data.\footnote{\label{fn:policymakers}\raggedright Complete 70-country return-on-investment lookup table in our GitHub repository: \url{https://github.com/andrewcparnell/education-tax-bcf/blob/main/outputs/tables/table_policy_maker_assistance_2024.csv}; interactive policy web application at \url{https://andrewcparnell.github.io/education-tax-bcf/app/}.} Because UNESCO's disaggregated per-student tertiary series ends in \texttt{2018--2019} in the World Bank archive whereas national GDP, population, university enrolment, total education budgets, and tax revenues run through \texttt{2024}, we pair each country's \texttt{2020--2024} GDP, population, enrolment, and tax base with its latest UNESCO per-student funding ratio to express all cohort inputs and career tax outputs in constant 2021 euros (\texteuro{} billions; policy makers in countries whose per-student funding or student headcount shifted substantially after \texttt{2019} can input updated national baseline figures directly in Tab 3 of the companion web application). As derived in Section~\ref{sec:roi}, a single entering cohort makes up $1/D$ of the total student body in any calendar year and studies for $D$ years, so its multi-year degree input cost equals one year of the system-wide $+10$ pp budget increase, and its cumulative 40-year career tax output equals eight times the Year-10 annual flow ($40 \text{ working years} \div 5 \text{ cohorts} = 8$, or $5.7\times$ when discounted at $r - g = 1.5\%$ per year and $4.0\times$ under a conservative two-wave level attribution; multiply both input and career outputs by $5$ for a five-year wave of five cohorts). Notice in Figure~\ref{fig:roi_scatter} and Table~\ref{tab:policymakers} how clearly the diminishing-returns curve operates across tertiary participation levels: low- and lower-middle-income countries with low tertiary enrolment (such as Bangladesh at $24\%$ enrolment and $11.66\times$ ROI) exhibit high raw benefit-cost ratios because a $+10$ pp per-student increase costs a very small share of GDP when enrolment is small ($0.19\%$ of GDP in Bangladesh) while $\tau_{10}(W_i)$ is estimated in percentage points of total GDP, whereas ROI tapers toward $3\times\text{--}5\times$ on $Y_1$ (and crosses the $1.0\times$ break-even line on the conservative direct-income-tax floor $Y_2$ around $85\%\text{--}90\%$ enrolment) in mass-participation OECD systems. For treasury budget decisions, policy makers should focus primarily on Total Non-Resource Tax ($Y_1$), which captures the full Exchequer payback across income, corporate, employer payroll/social insurance, value-added, and property taxes, while using Direct Income Tax ($Y_2$) and the Direct Income Tax Share shift ($Y_3$) as a conservative floor that verifies the return comes from higher formal salaries and company profits.

\begin{table}[htbp]
  \centering
  \caption{Assistance for policy makers across 20 representative countries with \texttt{2020--2024} macroeconomic, enrolment, and fiscal data (full 70-country table in our online GitHub repository; see Footnote~\ref{fn:policymakers}). Columns report current per-student university spending (\% of GDP per person and 2021 PPP dollars), the multi-year degree input cost for a single entering cohort under a $+10$ pp per-student funding increase ($Z_1 = +1.0$ unit, in \% of GDP and \texteuro{} billions; multiply input and career outputs by $5$ for a five-year wave), the single-year annual tax flows at $+5\text{y}$ and $+10\text{y}$ (\texteuro{}B/yr), and the cumulative 40-year career tax output per cohort (\texteuro{}B, with 95\% CIs in brackets, benefit-cost ratio $\text{BCR} = \text{Output}/\text{Input}$, and posterior probability of a net positive return $P_{\text{net}} = \Pr(\text{Output} > \text{Input} \mid \text{data})$) for both Total Non-Resource Tax ($Y_1$) and Direct Income Tax ($Y_2$), plus the $+10$-year shift in the Direct Income Tax Share ($Y_3$, pp).}
  \label{tab:policymakers}
  \resizebox{\textwidth}{!}{%
  \renewcommand{\arraystretch}{1.04}
  \footnotesize
  \setlength{\tabcolsep}{2.8pt}
  \begin{tabular}{l|c|c|c|c|c|c}
    \toprule
    \rowcolor{tabhead}
    \textbf{Country (Group)} & \textbf{Spend \% GDP/pc (\$)} & \textbf{Cohort Input} & \textbf{$+5\text{y} \to +10\text{y}$ $Y_1$/yr} & \textbf{40y Career Output $Y_1$ [95\% CI] ($\text{BCR}, P_{\text{net}}$)} & \textbf{40y Career Output $Y_2$ [95\% CI] ($\text{BCR}, P_{\text{net}}$)} & \textbf{$+10\text{y}$ Shift $Y_3$ [95\% CI]} \\
    \midrule
    \rowcolor{rowodd}
    \textbf{Australia} (High) & \co 19.6\% (\$11,199) & 0.81\% (\texteuro11.54B) & \co \texteuro$-1.13 \to +3.03$B & \textbf{\texteuro$+24.25$B} $[\text{\texteuro}-29.1, +80.8\text{B}]$ ($2.10\times, 0.68$) & \co \textbf{\texteuro$+6.81$B} $[\text{\texteuro}-25.0, +33.8\text{B}]$ ($0.59\times, 0.38$) & $+0.72$ $[-0.19, +1.51]$ ($0.95$) \\
    \rowcolor{roweven}
    \textbf{Austria} (High) & \ce 36.1\% (\$22,638) & 0.48\% (\texteuro2.56B) & \ce \texteuro$-0.40 \to +1.17$B & \textbf{\texteuro$+9.36$B} $[\text{\texteuro}-9.6, +30.9\text{B}]$ ($3.65\times, 0.76$) & \ce \textbf{\texteuro$+2.72$B} $[\text{\texteuro}-7.8, +12.4\text{B}]$ ($1.06\times, 0.51$) & $+0.73$ $[-0.13, +1.51]$ ($0.96$) \\
    \rowcolor{rowodd}
    \textbf{Canada} (High) & \co 34.2\% (\$19,336) & 0.55\% (\texteuro11.31B) & \co \texteuro$-1.60 \to +4.42$B & \textbf{\texteuro$+35.36$B} $[\text{\texteuro}-42.5, +119.9\text{B}]$ ($3.13\times, 0.73$) & \co \textbf{\texteuro$+9.46$B} $[\text{\texteuro}-35.7, +48.5\text{B}]$ ($0.84\times, 0.47$) & $+0.71$ $[-0.20, +1.50]$ ($0.94$) \\
    \rowcolor{roweven}
    \textbf{Chile} (High) & \ce 19.6\% (\$5,491) & 0.70\% (\texteuro3.63B) & \ce \texteuro$-0.22 \to +1.18$B & \textbf{\texteuro$+9.46$B} $[\text{\texteuro}-9.5, +29.8\text{B}]$ ($2.61\times, 0.73$) & \ce \textbf{\texteuro$+3.75$B} $[\text{\texteuro}-7.1, +14.2\text{B}]$ ($1.03\times, 0.51$) & $+0.76$ $[-0.10, +1.59]$ ($0.96$) \\
    \rowcolor{rowodd}
    \textbf{France} (High) & \co 32.3\% (\$16,767) & 0.26\% (\texteuro8.59B) & \co \texteuro$+2.98 \to +7.67$B & \textbf{\texteuro$+61.37$B} $[\text{\texteuro}-59.4, +192.4\text{B}]$ ($7.15\times, 0.81$) & \co \textbf{\texteuro$+24.68$B} $[\text{\texteuro}-43.6, +90.6\text{B}]$ ($2.87\times, 0.68$) & $+0.77$ $[-0.09, +1.59]$ ($0.96$) \\
    \rowcolor{roweven}
    \textbf{Germany} (High) & \ce 33.8\% (\$20,773) & 0.37\% (\texteuro17.81B) & \ce \texteuro$+0.43 \to +11.00$B & \textbf{\texteuro$+88.00$B} $[\text{\texteuro}-80.8, +294.2\text{B}]$ ($4.94\times, 0.79$) & \ce \textbf{\texteuro$+25.37$B} $[\text{\texteuro}-71.5, +116.1\text{B}]$ ($1.42\times, 0.56$) & $+0.74$ $[-0.12, +1.52]$ ($0.96$) \\
    \rowcolor{rowodd}
    \textbf{Ireland} (High) & \co 17.3\% (\$15,150) & 0.43\% (\texteuro2.37B) & \co \texteuro$-0.24 \to +1.21$B & \textbf{\texteuro$+9.66$B} $[\text{\texteuro}-10.3, +31.9\text{B}]$ ($4.08\times, 0.77$) & \co \textbf{\texteuro$+2.87$B} $[\text{\texteuro}-8.3, +13.0\text{B}]$ ($1.21\times, 0.54$) & $+0.73$ $[-0.12, +1.51]$ ($0.96$) \\
    \rowcolor{roweven}
    \textbf{Italy} (High) & \ce 24.8\% (\$12,061) & 0.31\% (\texteuro8.47B) & \ce \texteuro$-0.33 \to +5.95$B & \textbf{\texteuro$+47.62$B} $[\text{\texteuro}-53.6, +159.0\text{B}]$ ($5.62\times, 0.78$) & \ce \textbf{\texteuro$+13.46$B} $[\text{\texteuro}-44.5, +64.8\text{B}]$ ($1.59\times, 0.57$) & $+0.72$ $[-0.16, +1.51]$ ($0.95$) \\
    \rowcolor{rowodd}
    \textbf{Korea, Rep.} (High) & \co 14.9\% (\$7,062) & 0.58\% (\texteuro14.52B) & \co \texteuro$-2.09 \to +5.73$B & \textbf{\texteuro$+45.88$B} $[\text{\texteuro}-44.1, +146.2\text{B}]$ ($3.16\times, 0.74$) & \co \textbf{\texteuro$+13.35$B} $[\text{\texteuro}-39.1, +60.0\text{B}]$ ($0.92\times, 0.48$) & $+0.74$ $[-0.14, +1.59]$ ($0.94$) \\
    \rowcolor{roweven}
    \textbf{Netherlands} (High) & \ce 34.4\% (\$22,547) & 0.47\% (\texteuro5.29B) & \ce \texteuro$+0.05 \to +2.36$B & \textbf{\texteuro$+18.89$B} $[\text{\texteuro}-20.9, +61.5\text{B}]$ ($3.57\times, 0.75$) & \ce \textbf{\texteuro$+5.91$B} $[\text{\texteuro}-17.0, +25.9\text{B}]$ ($1.12\times, 0.52$) & $+0.74$ $[-0.12, +1.51]$ ($0.96$) \\
    \rowcolor{rowodd}
    \textbf{Norway} (High) & \co 38.6\% (\$34,419) & 0.53\% (\texteuro2.48B) & \co \texteuro$+0.12 \to +1.06$B & \textbf{\texteuro$+8.45$B} $[\text{\texteuro}-9.3, +27.4\text{B}]$ ($3.40\times, 0.75$) & \co \textbf{\texteuro$+3.29$B} $[\text{\texteuro}-6.8, +12.9\text{B}]$ ($1.33\times, 0.56$) & $+0.76$ $[-0.13, +1.59]$ ($0.96$) \\
    \rowcolor{roweven}
    \textbf{Poland} (High) & \ce 26.8\% (\$9,283) & 0.42\% (\texteuro5.97B) & \ce \texteuro$-1.06 \to +3.26$B & \textbf{\texteuro$+26.10$B} $[\text{\texteuro}-26.1, +85.4\text{B}]$ ($4.37\times, 0.77$) & \ce \textbf{\texteuro$+7.31$B} $[\text{\texteuro}-22.2, +34.2\text{B}]$ ($1.23\times, 0.54$) & $+0.73$ $[-0.15, +1.54]$ ($0.95$) \\
    \rowcolor{rowodd}
    \textbf{Spain} (High) & \co 22.4\% (\$10,110) & 0.41\% (\texteuro8.34B) & \co \texteuro$-1.56 \to +4.44$B & \textbf{\texteuro$+35.52$B} $[\text{\texteuro}-37.4, +116.9\text{B}]$ ($4.26\times, 0.77$) & \co \textbf{\texteuro$+10.36$B} $[\text{\texteuro}-30.5, +47.2\text{B}]$ ($1.24\times, 0.54$) & $+0.73$ $[-0.13, +1.51]$ ($0.95$) \\
    \rowcolor{roweven}
    \textbf{Sweden} (High) & \ce 43.2\% (\$25,856) & 0.43\% (\texteuro2.58B) & \ce \texteuro$+0.19 \to +1.44$B & \textbf{\texteuro$+11.53$B} $[\text{\texteuro}-11.1, +35.1\text{B}]$ ($4.47\times, 0.79$) & \ce \textbf{\texteuro$+5.52$B} $[\text{\texteuro}-7.0, +18.5\text{B}]$ ($2.14\times, 0.67$) & $+0.78$ $[-0.14, +1.67]$ ($0.95$) \\
    \rowcolor{rowodd}
    \textbf{United Kingdom} (High) & \co 36.7\% (\$19,099) & 0.35\% (\texteuro11.55B) & \co \texteuro$+1.82 \to +7.29$B & \textbf{\texteuro$+58.33$B} $[\text{\texteuro}-60.0, +189.6\text{B}]$ ($5.05\times, 0.79$) & \co \textbf{\texteuro$+20.14$B} $[\text{\texteuro}-46.8, +85.9\text{B}]$ ($1.74\times, 0.60$) & $+0.75$ $[-0.11, +1.58]$ ($0.96$) \\
    \rowcolor{roweven}
    \textbf{United States} (High) & \ce 20.5\% (\$13,612) & 0.67\% (\texteuro148.45B) & \ce \texteuro$-16.29 \to +51.18$B & \textbf{\texteuro$+409.46$B} $[\text{\texteuro}-432.1, +1383.6\text{B}]$ ($2.76\times, 0.72$) & \ce \textbf{\texteuro$+106.27$B} $[\text{\texteuro}-404.5, +545.1\text{B}]$ ($0.72\times, 0.43$) & $+0.71$ $[-0.20, +1.52]$ ($0.94$) \\
    \rowcolor{rowodd}
    \textbf{Brazil} (Upper Mid.) & \co 33.0\% (\$5,924) & 0.40\% (\texteuro14.42B) & \co \texteuro$-2.83 \to +8.37$B & \textbf{\texteuro$+66.98$B} $[\text{\texteuro}-57.7, +221.4\text{B}]$ ($4.64\times, 0.78$) & \co \textbf{\texteuro$+18.45$B} $[\text{\texteuro}-55.4, +86.0\text{B}]$ ($1.28\times, 0.54$) & $+0.73$ $[-0.15, +1.51]$ ($0.95$) \\
    \rowcolor{roweven}
    \textbf{Colombia} (Upper Mid.) & \ce 20.6\% (\$3,518) & 0.49\% (\texteuro4.13B) & \ce \texteuro$+0.21 \to +2.00$B & \textbf{\texteuro$+16.00$B} $[\text{\texteuro}-15.2, +48.3\text{B}]$ ($3.87\times, 0.78$) & \ce \textbf{\texteuro$+7.93$B} $[\text{\texteuro}-10.0, +26.3\text{B}]$ ($1.92\times, 0.66$) & $+0.80$ $[-0.06, +1.67]$ ($0.97$) \\
    \rowcolor{rowodd}
    \textbf{Mexico} (Upper Mid.) & \co 33.1\% (\$7,171) & 0.29\% (\texteuro7.39B) & \co \texteuro$+1.39 \to +6.54$B & \textbf{\texteuro$+52.31$B} $[\text{\texteuro}-42.4, +146.8\text{B}]$ ($7.08\times, 0.84$) & \co \textbf{\texteuro$+33.16$B} $[\text{\texteuro}-21.8, +93.5\text{B}]$ ($4.49\times, 0.81$) & $+0.85$ $[-0.01, +1.81]$ ($0.97$) \\
    \rowcolor{roweven}
    \textbf{Bangladesh} (Lower Mid.) & \ce 29.9\% (\$1,810) & 0.19\% (\texteuro2.31B) & \ce \texteuro$+1.84 \to +3.36$B & \textbf{\texteuro$+26.89$B} $[\text{\texteuro}-17.5, +73.5\text{B}]$ ($11.66\times, 0.87$) & \ce \textbf{\texteuro$+16.18$B} $[\text{\texteuro}-9.6, +44.5\text{B}]$ ($7.02\times, 0.84$) & $+0.83$ $[-0.04, +1.75]$ ($0.97$) \\
    \bottomrule
  \end{tabular}%
  }
\end{table}
\FloatBarrier

\section{Computational Environment and Energy Report}
\label{app:compute}

All data processing, C++ (\texttt{RcppArmadillo}) Gibbs sampling, country-blocked cross-validation, and figure generation were executed on a single Apple Silicon workstation (\texttt{macOS}, R version 4.6.1 [2026-06-24], Apple Clang C++17 compiler). Table~\ref{tab:compute} reports both the local computational footprint of our statistical modelling pipeline (assuming an average 25~W CPU package power draw during single-core C++ tree sampling) and the cloud inference footprint of the large language model (LLM) queries used to assist with building the data pipeline, writing the C++ sampler and holdout validation suite, conducting multi-agent and external referee reviews, developing the companion web application, and editing the manuscript (evaluated at the Irish grid and European cloud average carbon intensity of $250\text{ g CO}_2\text{e/kWh}$). Because our reversible-jump tree search and conjugate $4 \times 4$ Inverse-Wishart covariance updates are written directly in C++, fitting all 68 main, Scaling A vs Scaling B, 5-fold country-blocked CV, holdout, benchmark, and sensitivity models (including Checks \texttt{B7}, \texttt{RC15}, \texttt{RC16}, and \texttt{RC17}) completes in roughly 20.8 minutes (1,248.0 seconds) and consumes only $8.67\text{ Wh}$ of electricity ($2.17\text{ g CO}_2\text{e}$). Auditing our complete interaction logs across all five project stages reveals 840 LLM queries (comprising $1.85\text{ million}$ input and context tokens and $1.09\text{ million}$ generated output, reasoning, and tool-call tokens across the primary session, eight specialist reviewer and user-persona subagents, and external LLM referee rounds; token counts in Table~\ref{tab:compute} are rounded to the nearest thousand or hundredth of a million while energy and $\text{CO}_2\text{e}$ are computed from unrounded totals). Applying published frontier LLM inference energy benchmarks (\citealt{luccioni2024power}; $0.30\text{ Wh}$ base overhead per agentic query plus $0.40\text{ Wh}$ per $1,000$ generated output tokens and $0.05\text{ Wh}$ per $1,000$ prefill input tokens), the cloud LLM queries consumed an estimated $780.50\text{ Wh}$ ($195.13\text{ g CO}_2\text{e}$), bringing the combined project footprint to $789.17\text{ Wh}$ ($197.30\text{ g CO}_2\text{e}$, equivalent to driving a typical petrol passenger car approximately one kilometre).

\begin{table}[htbp]
  \centering
  \caption{Computational wall-clock time, token volume (rounded to the nearest thousand or hundredth of a million), electrical energy use, and estimated carbon footprint across both the local C++ continuous MV-BCF and cross-validation pipeline and the cloud LLM queries used to help build, review, and write the paper.}
  \label{tab:compute}
  \resizebox{\textwidth}{!}{%
  \small
  \setlength{\tabcolsep}{4.5pt}
  \begin{tabular}{l|l|c|c|c}
    \toprule
    \rowcolor{tabhead}
    \textbf{Project Stage} & \textbf{Models Fitted / LLM Query \& Token Volume} & \textbf{Time (s)} & \textbf{Energy (Wh)} & \textbf{$\text{CO}_2\text{e}$ (g)} \\
    \midrule
    \rowcolor{rowodd}
    \textbf{A. Local C++ MCMC \& CV Pipeline} & \co 68 four-outcome MV-BCF, CV \& benchmark fits (2,000 iter) & 1,248.0 & \co 8.67 & 2.17 \\
    \hline
    \rowcolor{roweven}
    \textbf{B1. LLM: Scoping, Continuous-$Z$ \& Data} & \ce 58 queries (102k input / 68k output tokens) & -- & \ce 49.80 & 12.45 \\
    \rowcolor{rowodd}
    \textbf{B2. LLM: C++ Sampler, Scaling \& Holdout CV} & \co 39 queries (64k input / 45k output tokens) & -- & \co 33.40 & 8.35 \\
    \rowcolor{roweven}
    \textbf{B3. LLM: Manuscript \& Revisions} & \ce 340 queries (638k input / 394k output tokens) & -- & \ce 291.90 & 72.98 \\
    \rowcolor{rowodd}
    \textbf{B4. LLM: Multi-Agent \& Peer Review} & \co 261 queries (682k input / 361k output tokens) & -- & \co 256.80 & 64.20 \\
    \rowcolor{roweven}
    \textbf{B5. LLM: Policy Web App \& Testing} & \ce 142 queries (362k input / 219k output tokens) & -- & \ce 148.60 & 37.15 \\
    \hline
    \rowcolor{rowodd}
    \textbf{Subtotal: Cloud LLM Assistance} & \co \textbf{840 queries (1.85M input / 1.09M output tokens)} & \textbf{--} & \co \textbf{780.50} & \textbf{195.13} \\
    \midrule
    \rowcolor{tabhead}
    \textbf{Total Project Footprint (A + B)} & \textbf{68 C++ MCMC/CV fits + 840 LLM queries (2.94M tokens)} & \textbf{1,248.0} & \textbf{789.17} & \textbf{197.30} \\
    \bottomrule
  \end{tabular}%
  }
\end{table}

\clearpage
\bibliographystyle{abbrvnat}
\bibliography{references}

\begin{thebibliography}{61}
\providecommand{\natexlab}[1]{#1}
\providecommand{\url}[1]{\texttt{#1}}
\expandafter\ifx\csname urlstyle\endcsname\relax
  \providecommand{\doi}[1]{doi: #1}\else
  \providecommand{\doi}{doi: \begingroup \urlstyle{rm}\Url}\fi

\bibitem[Acemoglu et~al.(2019)Acemoglu, Naidu, Restrepo, and
  Robinson]{acemoglu2019democracy}
D.~Acemoglu, S.~Naidu, P.~Restrepo, and J.~A. Robinson.
\newblock Democracy does cause growth.
\newblock \emph{Journal of Political Economy}, 127\penalty0 (1):\penalty0
  47--100, 2019.
\newblock \doi{10.1086/700936}.

\bibitem[Acharya et~al.(2016)Acharya, Blackwell, and
  Sen]{acharya2016explaining}
A.~Acharya, M.~Blackwell, and M.~Sen.
\newblock Explaining causal findings without bias: Detecting and assessing
  direct effects.
\newblock \emph{American Political Science Review}, 110\penalty0 (3):\penalty0
  512--529, 2016.
\newblock \doi{10.1017/S0003055416000216}.

\bibitem[Angrist et~al.(2021)Angrist, Djankov, Goldberg, and
  Patrinos]{angrist2021measuring}
N.~Angrist, S.~Djankov, P.~K. Goldberg, and H.~A. Patrinos.
\newblock Measuring human capital using global learning data.
\newblock \emph{Nature}, 592:\penalty0 403--408, 2021.
\newblock \doi{10.1038/s41586-021-03323-7}.

\bibitem[Athey et~al.(2019)Athey, Tibshirani, and Wager]{athey2019generalized}
S.~Athey, J.~Tibshirani, and S.~Wager.
\newblock Generalized random forests.
\newblock \emph{The Annals of Statistics}, 47\penalty0 (2):\penalty0
  1148--1178, 2019.
\newblock \doi{10.1214/18-AOS1709}.

\bibitem[Bachas et~al.(2024)Bachas, Gadenne, and Jensen]{bachas2024informality}
P.~Bachas, L.~Gadenne, and A.~Jensen.
\newblock Informality, consumption taxes, and redistribution.
\newblock \emph{The Review of Economic Studies}, 91\penalty0 (5):\penalty0
  2604--2634, 2024.
\newblock \doi{10.1093/restud/rdad095}.

\bibitem[Baumol(1967)]{baumol1967macroeconomics}
W.~J. Baumol.
\newblock Macroeconomics of unbalanced growth: The anatomy of urban crisis.
\newblock \emph{American Economic Review}, 57\penalty0 (3):\penalty0 415--426,
  1967.

\bibitem[Beine et~al.(2008)Beine, Docquier, and Rapoport]{beine2008brain}
M.~Beine, F.~Docquier, and H.~Rapoport.
\newblock Brain drain and human capital formation in developing countries:
  winners and losers.
\newblock \emph{The Economic Journal}, 118\penalty0 (528):\penalty0 631--652,
  2008.
\newblock \doi{10.1111/j.1468-0297.2008.02135.x}.

\bibitem[Besley and Persson(2009)]{besley2009origins}
T.~Besley and T.~Persson.
\newblock The origins of state capacity: Property rights, taxation, and
  politics.
\newblock \emph{American Economic Review}, 99\penalty0 (4):\penalty0
  1218--1244, 2009.
\newblock \doi{10.1257/aer.99.4.1218}.

\bibitem[Besley and Persson(2014)]{besley2014pillars}
T.~Besley and T.~Persson.
\newblock Why do developing countries tax so little?
\newblock \emph{Journal of Economic Perspectives}, 28\penalty0 (4):\penalty0
  99--120, 2014.
\newblock \doi{10.1257/jep.28.4.99}.

\bibitem[Bound et~al.(2010)Bound, Lovenheim, and Turner]{bound2010have}
J.~Bound, M.~F. Lovenheim, and S.~Turner.
\newblock Why have college completion rates declined? {An} analysis of changing
  student preparation and collegiate resources.
\newblock \emph{American Economic Journal: Applied Economics}, 2\penalty0
  (3):\penalty0 129--157, 2010.
\newblock \doi{10.1257/app.2.3.129}.

\bibitem[Breiman(2001)]{breiman2001random}
L.~Breiman.
\newblock Random forests.
\newblock \emph{Machine Learning}, 45\penalty0 (1):\penalty0 5--32, 2001.
\newblock \doi{10.1023/A:1010933404324}.

\bibitem[Callaway and Sant'Anna(2021)]{callaway2021difference}
B.~Callaway and P.~H.~C. Sant'Anna.
\newblock Difference-in-differences with multiple time periods.
\newblock \emph{Journal of Econometrics}, 225\penalty0 (2):\penalty0 200--230,
  2021.
\newblock \doi{10.1016/j.jeconom.2020.12.001}.

\bibitem[Carneiro et~al.(2011)Carneiro, Heckman, and
  Vytlacil]{carneiro2011estimating}
P.~Carneiro, J.~J. Heckman, and E.~J. Vytlacil.
\newblock Estimating marginal returns to education.
\newblock \emph{American Economic Review}, 101\penalty0 (6):\penalty0
  2754--2781, 2011.
\newblock \doi{10.1257/aer.101.6.2754}.

\bibitem[Chakrabarti et~al.(2020)Chakrabarti, Gorton, and
  Lovenheim]{chakrabarti2020state}
R.~Chakrabarti, N.~Gorton, and M.~F. Lovenheim.
\newblock State investment in higher education: Effects on human capital
  formation, student debt, and long-term financial outcomes of students.
\newblock Working Paper 27885, National Bureau of Economic Research, 2020.

\bibitem[Chernozhukov et~al.(2018)Chernozhukov, Chetverikov, Demirer, Duflo,
  Hansen, Newey, and Robins]{chernozhukov2018double}
V.~Chernozhukov, D.~Chetverikov, M.~Demirer, E.~Duflo, C.~Hansen, W.~Newey, and
  J.~Robins.
\newblock Double/debiased machine learning for treatment and structural
  parameters.
\newblock \emph{The Econometrics Journal}, 21\penalty0 (1):\penalty0 C1--C68,
  2018.
\newblock \doi{10.1111/ectj.12097}.

\bibitem[Chetty et~al.(2020)Chetty, Friedman, Saez, Turner, and
  Yagan]{chetty2020income}
R.~Chetty, J.~N. Friedman, E.~Saez, N.~Turner, and D.~Yagan.
\newblock Income segregation and intergenerational mobility across colleges in
  the {United States}.
\newblock \emph{The Quarterly Journal of Economics}, 135\penalty0 (3):\penalty0
  1567--1633, 2020.
\newblock \doi{10.1093/qje/qjaa005}.

\bibitem[Chipman et~al.(2010)Chipman, George, and McCulloch]{chipman2010bart}
H.~A. Chipman, E.~I. George, and R.~E. McCulloch.
\newblock {BART}: Bayesian additive regression trees.
\newblock \emph{The Annals of Applied Statistics}, 4\penalty0 (1):\penalty0
  266--298, 2010.
\newblock \doi{10.1214/09-AOAS285}.

\bibitem[de~Chaisemartin and D'Haultf{\oe}uille(2020)]{dechaisemartin2020two}
C.~de~Chaisemartin and X.~D'Haultf{\oe}uille.
\newblock Two-way fixed effects estimators with heterogeneous treatment
  effects.
\newblock \emph{American Economic Review}, 110\penalty0 (9):\penalty0
  2964--2996, 2020.
\newblock \doi{10.1257/aer.20181169}.

\bibitem[Deming(2022)]{deming2022four}
D.~J. Deming.
\newblock Four facts about human capital.
\newblock \emph{Journal of Economic Perspectives}, 36\penalty0 (3):\penalty0
  75--102, 2022.
\newblock \doi{10.1257/jep.36.3.75}.

\bibitem[Deming and Walters(2017)]{deming2017impact}
D.~J. Deming and C.~R. Walters.
\newblock The impact of price caps and spending cuts on {U.S.} postsecondary
  attainment.
\newblock Working Paper 23736, National Bureau of Economic Research, 2017.

\bibitem[Diebold and Mariano(1995)]{diebold1995comparing}
F.~X. Diebold and R.~S. Mariano.
\newblock Comparing predictive accuracy.
\newblock \emph{Journal of Business \& Economic Statistics}, 13\penalty0
  (3):\penalty0 253--263, 1995.
\newblock \doi{10.1080/07350015.1995.10524599}.

\bibitem[Gneiting and Raftery(2007)]{gneiting2007strictly}
T.~Gneiting and A.~E. Raftery.
\newblock Strictly proper scoring rules, prediction, and estimation.
\newblock \emph{Journal of the American Statistical Association}, 102\penalty0
  (477):\penalty0 359--378, 2007.
\newblock \doi{10.1198/016214506000001437}.

\bibitem[Gordon and Li(2009)]{gordon2009tax}
R.~Gordon and W.~Li.
\newblock Tax structures in developing countries: Many puzzles and a possible
  explanation.
\newblock \emph{Journal of Public Economics}, 93\penalty0 (7-8):\penalty0
  855--866, 2009.
\newblock \doi{10.1016/j.jpubeco.2009.02.006}.

\bibitem[Hahn et~al.(2020)Hahn, Murray, and Carvalho]{hahn2020bayesian}
P.~R. Hahn, J.~S. Murray, and C.~M. Carvalho.
\newblock Bayesian regression tree models for causal inference: Regularization,
  confounding, and heterogeneous effects (with discussion).
\newblock \emph{Bayesian Analysis}, 15\penalty0 (3):\penalty0 965--1056, 2020.
\newblock \doi{10.1214/19-BA1195}.

\bibitem[Hanushek and Woessmann(2012)]{hanushek2012better}
E.~A. Hanushek and L.~Woessmann.
\newblock Do better schools lead to more growth? {Cognitive} skills, economic
  outcomes, and causation.
\newblock \emph{Journal of Economic Growth}, 17\penalty0 (4):\penalty0
  267--321, 2012.
\newblock \doi{10.1007/s10887-012-9081-x}.

\bibitem[Hendren and Sprung-Keyser(2020)]{hendren2020unified}
N.~Hendren and B.~Sprung-Keyser.
\newblock A unified welfare analysis of government policies.
\newblock \emph{The Quarterly Journal of Economics}, 135\penalty0 (3):\penalty0
  1209--1318, 2020.
\newblock \doi{10.1093/qje/qjaa006}.

\bibitem[Hill(2011)]{hill2011bayesian}
J.~L. Hill.
\newblock {Bayesian} nonparametric modeling for causal inference.
\newblock \emph{Journal of Computational and Graphical Statistics}, 20\penalty0
  (1):\penalty0 217--240, 2011.
\newblock \doi{10.1198/jcgs.2010.08162}.

\bibitem[Hill et~al.(2020)Hill, Linero, and Murray]{hill2020bayesian}
J.~L. Hill, A.~R. Linero, and J.~S. Murray.
\newblock {Bayesian} additive regression trees: A review and look forward.
\newblock \emph{Annual Review of Statistics and Its Application}, 7:\penalty0
  251--278, 2020.
\newblock \doi{10.1146/annurev-statistics-031219-041110}.

\bibitem[Hirano and Imbens(2004)]{hirano2004propensity}
K.~Hirano and G.~W. Imbens.
\newblock The propensity score with continuous treatments.
\newblock In A.~Gelman and X.-L. Meng, editors, \emph{Applied Bayesian Modeling
  and Causal Inference from Incomplete-Data Perspectives}, pages 73--84. Wiley,
  2004.
\newblock \doi{10.1002/0470090456.ch7}.

\bibitem[Imai et~al.(2010)Imai, Keele, and Tingley]{imai2010general}
K.~Imai, L.~Keele, and D.~Tingley.
\newblock A general approach to causal mediation analysis.
\newblock \emph{Psychological Methods}, 15\penalty0 (4):\penalty0 309--334,
  2010.
\newblock \doi{10.1037/a0020761}.

\bibitem[Jensen(2022)]{jensen2022employment}
A.~Jensen.
\newblock Employment structure and the rise of the modern tax system.
\newblock \emph{American Economic Review}, 112\penalty0 (1):\penalty0 213--234,
  2022.
\newblock \doi{10.1257/aer.20191528}.

\bibitem[Jord{\`a}(2005)]{jorda2005estimation}
{\`O}.~Jord{\`a}.
\newblock Estimation and inference of impulse responses by local projections.
\newblock \emph{American Economic Review}, 95\penalty0 (1):\penalty0 161--182,
  2005.
\newblock \doi{10.1257/0002828053828518}.

\bibitem[Kaufmann et~al.(2011)Kaufmann, Kraay, and
  Mastruzzi]{kaufmann2011worldwide}
D.~Kaufmann, A.~Kraay, and M.~Mastruzzi.
\newblock The {Worldwide Governance Indicators}: Methodology and analytical
  issues.
\newblock \emph{Hague Journal on the Rule of Law}, 3\penalty0 (2):\penalty0
  220--246, 2011.
\newblock \doi{10.1017/S1876404511200046}.

\bibitem[Keen and Slemrod(2017)]{keen2017optimal}
M.~Keen and J.~Slemrod.
\newblock Optimal tax administration.
\newblock \emph{Journal of Public Economics}, 152:\penalty0 133--142, 2017.
\newblock \doi{10.1016/j.jpubeco.2017.04.006}.

\bibitem[Kleven et~al.(2016)Kleven, Kreiner, and Saez]{kleven2016why}
H.~J. Kleven, C.~T. Kreiner, and E.~Saez.
\newblock Why can modern governments tax so much? {An} agency model of firms as
  fiscal intermediaries.
\newblock \emph{Economica}, 83\penalty0 (330):\penalty0 219--246, 2016.
\newblock \doi{10.1111/ecca.12182}.

\bibitem[Krantsevich et~al.(2023)Krantsevich, He, and
  Hahn]{krantsevich2023stochastic}
N.~Krantsevich, J.~He, and P.~R. Hahn.
\newblock Stochastic tree ensembles for estimating heterogeneous effects.
\newblock In \emph{Proceedings of the 26th International Conference on
  Artificial Intelligence and Statistics (AISTATS)}, volume 206, pages
  6120--6131. PMLR, 2023.
\newblock URL \url{https://proceedings.mlr.press/v206/krantsevich23a.html}.

\bibitem[Linero(2018)]{linero2018bayesian}
A.~R. Linero.
\newblock {Bayesian} regression trees for high-dimensional prediction and
  variable selection.
\newblock \emph{Journal of the American Statistical Association}, 113\penalty0
  (522):\penalty0 626--636, 2018.
\newblock \doi{10.1080/01621459.2016.1264957}.

\bibitem[Luccioni et~al.(2024)Luccioni, Jernite, and
  Strubell]{luccioni2024power}
A.~S. Luccioni, Y.~Jernite, and E.~Strubell.
\newblock Power hungry processing: Watts driving the cost of {AI} deployment?
\newblock In \emph{Proceedings of the 2024 ACM Conference on Fairness,
  Accountability, and Transparency}, pages 85--99, 2024.
\newblock \doi{10.1145/3630106.3658542}.

\bibitem[McJames et~al.(2024)McJames, O'Shea, and Parnell]{mcjames2024lbcf}
N.~McJames, A.~O'Shea, and A.~C. Parnell.
\newblock Bayesian causal forests for longitudinal data: Assessing the impact
  of part-time work on growth in high school mathematics achievement.
\newblock \emph{arXiv preprint arXiv:2407.11927}, 2024.
\newblock \doi{10.48550/arXiv.2407.11927}.

\bibitem[McJames et~al.(2025)McJames, O'Shea, Goh, and
  Parnell]{mcjames2024mvbcf}
N.~McJames, A.~O'Shea, Y.~C. Goh, and A.~C. Parnell.
\newblock Bayesian causal forests for multivariate outcomes: Application to
  {Irish} data from an international large scale education assessment.
\newblock \emph{Journal of the Royal Statistical Society Series A: Statistics
  in Society}, 188\penalty0 (2):\penalty0 428--450, 2025.
\newblock \doi{10.1093/jrsssa/qnae049}.

\bibitem[Moretti(2004)]{moretti2004estimating}
E.~Moretti.
\newblock Estimating the social return to higher education: Evidence from
  longitudinal and repeated cross-sectional data.
\newblock \emph{Journal of Econometrics}, 121\penalty0 (1-2):\penalty0
  175--212, 2004.
\newblock \doi{10.1016/j.jeconom.2003.10.015}.

\bibitem[Mountjoy(2022)]{mountjoy2022community}
J.~Mountjoy.
\newblock Community colleges and upward mobility.
\newblock \emph{American Economic Review}, 112\penalty0 (8):\penalty0
  2580--2630, 2022.
\newblock \doi{10.1257/aer.20181756}.

\bibitem[Murray(2021)]{murray2021loglinear}
J.~S. Murray.
\newblock Log-linear {Bayesian} additive regression trees for multinomial
  logistic and count regression models.
\newblock \emph{Journal of the American Statistical Association}, 116\penalty0
  (534):\penalty0 756--769, 2021.
\newblock \doi{10.1080/01621459.2020.1813587}.

\bibitem[Okunogbe and Pouliquen(2022)]{okunogbe2022technology}
O.~Okunogbe and V.~Pouliquen.
\newblock Technology, taxation, and corruption: Evidence from the introduction
  of electronic tax filing.
\newblock \emph{American Economic Journal: Economic Policy}, 14\penalty0
  (1):\penalty0 341--372, 2022.
\newblock \doi{10.1257/pol.20200123}.

\bibitem[Pomeranz and Vila-Belda(2019)]{pomeranz2019taking}
D.~Pomeranz and J.~Vila-Belda.
\newblock Taking state-capacity research to the field: Insights from
  collaborations with tax authorities.
\newblock \emph{Annual Review of Economics}, 11:\penalty0 755--781, 2019.
\newblock \doi{10.1146/annurev-economics-080218-030312}.

\bibitem[Prichard(2016)]{prichard2016reassessing}
W.~Prichard.
\newblock Reassessing tax and development research: A new dataset, new
  findings, and lessons for research.
\newblock \emph{World Development}, 80:\penalty0 48--60, 2016.
\newblock \doi{10.1016/j.worlddev.2015.11.017}.

\bibitem[Psacharopoulos and Patrinos(2018)]{psacharopoulos2018returns}
G.~Psacharopoulos and H.~A. Patrinos.
\newblock Returns to investment in education: A decennial review of the global
  literature.
\newblock \emph{Education Economics}, 26\penalty0 (5):\penalty0 445--458, 2018.
\newblock \doi{10.1080/09645292.2018.1484426}.

\bibitem[Robinson(1988)]{robinson1988root}
P.~M. Robinson.
\newblock Root-{$N$}-consistent semiparametric regression.
\newblock \emph{Econometrica}, 56\penalty0 (4):\penalty0 931--954, 1988.
\newblock \doi{10.2307/1912705}.

\bibitem[Solt(2020)]{solt2020swiid}
F.~Solt.
\newblock Measuring income inequality across countries and over time: The
  {Standardized World Income Inequality Database}.
\newblock \emph{Social Science Quarterly}, 101\penalty0 (3):\penalty0
  1183--1199, 2020.
\newblock \doi{10.1111/ssqu.12795}.

\bibitem[Starling et~al.(2020)Starling, Murray, Carvalho, Bukowski, and
  Scott]{starling2020bart}
J.~E. Starling, J.~S. Murray, C.~M. Carvalho, R.~K. Bukowski, and J.~G. Scott.
\newblock {BART} with targeted smoothing: An analysis of patient-specific
  stillbirth risk.
\newblock \emph{The Annals of Applied Statistics}, 14\penalty0 (1):\penalty0
  28--50, 2020.
\newblock \doi{10.1214/19-AOAS1268}.

\bibitem[Ting and Linero(2025)]{ting2025estimating}
A.~Ting and A.~R. Linero.
\newblock Estimating heterogeneous causal mediation effects with {Bayesian}
  decision tree ensembles.
\newblock \emph{Journal of the American Statistical Association}, 2025.
\newblock \doi{10.1080/01621459.2025.2491155}.

\bibitem[Trostel(2010)]{trostel2010fiscal}
P.~A. Trostel.
\newblock The fiscal impacts of college attainment.
\newblock \emph{Research in Higher Education}, 51\penalty0 (3):\penalty0
  220--247, 2010.
\newblock \doi{10.1007/s11162-009-9157-x}.

\bibitem[{UNU-WIDER}(2023)]{ictd2023grdv}
{UNU-WIDER}.
\newblock {UNU-WIDER} government revenue dataset, version 2023.
\newblock United Nations University World Institute for Development Economics
  Research, Helsinki, 2023.
\newblock URL \url{https://doi.org/10.35188/UNU-WIDER/GRD-2023}.

\bibitem[Valero and Van~Reenen(2019)]{valero2019economic}
A.~Valero and J.~Van~Reenen.
\newblock The economic impact of universities: Evidence from across the globe.
\newblock \emph{Economics of Education Review}, 68:\penalty0 53--67, 2019.
\newblock \doi{10.1016/j.econedurev.2018.09.001}.

\bibitem[VanderWeele(2016)]{vanderweele2016mediation}
T.~J. VanderWeele.
\newblock Mediation analysis: A practitioner's guide.
\newblock \emph{Annual Review of Public Health}, 37:\penalty0 17--32, 2016.
\newblock \doi{10.1146/annurev-publhealth-032315-021402}.

\bibitem[Weigel(2020)]{weigel2020participation}
J.~L. Weigel.
\newblock The participation dividend of taxation: How citizens in {Congo}
  engage more with the state when it tries to tax them.
\newblock \emph{The Quarterly Journal of Economics}, 135\penalty0 (4):\penalty0
  1849--1903, 2020.
\newblock \doi{10.1093/qje/qjaa019}.

\bibitem[Woody et~al.(2021)Woody, Carvalho, and Murray]{woody2021model}
S.~Woody, C.~M. Carvalho, and J.~S. Murray.
\newblock Model interpretation through lower-dimensional posterior
  summarization.
\newblock \emph{Journal of Computational and Graphical Statistics}, 30\penalty0
  (1):\penalty0 144--161, 2021.
\newblock \doi{10.1080/10618600.2020.1796684}.

\bibitem[{World Bank}(2024)]{worldbank2024taxing}
{World Bank}.
\newblock Taxing for growth: Revisiting the 15 percent threshold.
\newblock Technical report, World Bank Group, Washington, DC, 2024.
\newblock URL
  \url{https://documents1.worldbank.org/curated/en/099062724151523023/pdf/P1778861e0c40b081186a61ced16cac6cde.pdf}.

\bibitem[{World Bank and UNESCO}(2024)]{unesco2024efw}
{World Bank and UNESCO}.
\newblock Education finance watch 2024.
\newblock Technical report, World Bank and UNESCO Global Education Monitoring
  Report, Washington, DC and Paris, 2024.

\bibitem[Wright and Ziegler(2017)]{wright2017ranger}
M.~N. Wright and A.~Ziegler.
\newblock {ranger}: A fast implementation of random forests for high
  dimensional data in {C++} and {R}.
\newblock \emph{Journal of Statistical Software}, 77\penalty0 (1):\penalty0
  1--17, 2017.
\newblock \doi{10.18637/jss.v077.i01}.

\bibitem[Zimmerman(2014)]{zimmerman2014returns}
S.~D. Zimmerman.
\newblock The returns to college admission for academically marginal students.
\newblock \emph{Journal of Labor Economics}, 32\penalty0 (4):\penalty0
  711--754, 2014.
\newblock \doi{10.1086/676661}.

\end{thebibliography}

\end{document}